\documentclass[journal]{IEEEtran}
\usepackage{amsmath,amsfonts,amsthm}
\usepackage{algorithmic}
\usepackage{algorithm}
\usepackage{array}
\usepackage{subfig}
\usepackage{textcomp}
\usepackage{stfloats}
\usepackage{xurl}
\usepackage{graphicx}
\usepackage{cite}
\usepackage{acro}
\usepackage{xcolor}
\usepackage{enumitem}
\usepackage{thm-restate}
\usepackage{float}
\usepackage{hyperref}
\newtheorem{definition}{Definition}
\newtheorem{theorem}{Theorem}
\newtheorem{lemma}{Lemma}
\newtheorem{remark}{Remark}

\DeclareMathOperator*{\argmin}{arg\,min}
\DeclareMathOperator*{\argmax}{arg\,max}

\DeclareAcronym{cdf}{short = CDF, long  = cumulative distribution function}
\DeclareAcronym{ecdf}{short = eCDF, long  = empirical cumulative distribution function}
\DeclareAcronym{dp}{short = DP, long  = differential privacy}
\DeclareAcronym{dkw}{short = DKW, long  = Dvoretzky–Kiefer–Wolfowitz}
\DeclareAcronym{hq}{
  short = HQ,
  long  = histogram queries
}
\DeclareAcronym{aq}{
  short = AQ,
  long  = adaptive quantiles
}
\DeclareAcronym{pp}{
  short = PP,
  long  = polynomial projection
}
\DeclareAcronym{mp}{
  short = MP,
  long  = matching pursuit
}
\DeclareAcronym{rnm}{
  short = RNM,
  long  = report noisy max
}

\newcommand{\truecdf}{F}
\newcommand{\opttruecdf}{\hat{F}}
\newcommand{\empcdf}{F_n}
\newcommand{\optempcdf}{\hat{F}_n}
\newcommand{\dpempcdf}{\tilde{F}_n}

\usepackage[dvipsnames]{xcolor}
\usepackage{changes}

\begin{document}

\title{Functional Approximation Methods for Differentially Private Distribution Estimation}

\author{Ye Tao,~\IEEEmembership{Member,~IEEE} and Anand D. Sarwate,~\IEEEmembership{Senior Member,~IEEE} 
\thanks{The authors are with the Department of Electrical and Computer Engineering, Rutgers, The State University of New Jersey, Piscataway, NJ 08854 USA. \texttt{(yt371, ads221)@rutgers.edu}} 
\thanks{An earlier version of this paper was presented in part at the 2025 IEEE International Conference on Acoustics, Speech and Signal Processing (ICASSP)~\cite{tao2025differentially}, DOI: 10.1109/ICASSP49660.2025.10890461. 
This work is funded in part by the US National Institutes of Health under award 2R01DA040487 and the National Science Foundation under award CNS-2148104.}}



\maketitle



\begin{abstract}
The \ac{cdf} is fundamental for characterizing random variables, making it essential in applications that require privacy-preserving data analysis. This paper introduces a novel framework for constructing differentially private \ac{cdf}s inspired by functional analysis and the functional mechanism. We develop two variants: a polynomial projection method, which projects the empirical \ac{cdf} into a polynomial space, and a sparse approximation method via matching pursuit, which projects it into arbitrary function spaces constructed from dictionaries. In both cases, the empirical \ac{cdf} is approximated within the chosen space, and the corresponding coefficients are privatized to guarantee differential privacy. Compared with existing approaches such as histogram queries, tree-based methods, and adaptive quantiles, our approach achieves comparable or superior performance. Our methods are particularly well-suited to decentralized settings and scenarios where \ac{cdf}s must be efficiently updated with newly collected or streaming data. They are also readily extensible to multivariate distributions while remaining effective. In addition, we investigate the influence of parameters such as dictionary size and systematically evaluate different dictionary constructions, including Legendre polynomials, B-splines, and distribution-based functions. Overall, our contributions advance the development of practical and reliable methods for privacy-preserving \ac{cdf} estimation.
\end{abstract}

\section{Introduction}
\IEEEPARstart{T}{he} cumulative distribution function (CDF) is a fundamental object in statistical analysis, serving as a cornerstone in both classical statistics and modern machine learning. For instance, in hypothesis testing, test statistics are compared against critical values derived from the CDF to guide decisions about hypotheses. Furthermore, CDFs play a central role in risk assessment and decision-making under uncertainty, as they provide a comprehensive characterization of the distribution of possible outcomes. When a distribution is not known and only sampled observed data are available, we typically use the \ac{ecdf} to estimate the unknown true CDF.

In this paper we provide new methods for approximating a CDF when the data is sensitive or private. More specifically, we design new approaches for estimating CDFs under \ac{dp} constraints~\cite{DworkMNS:06sensitivity}. This paper is motivated by two related application scenarios. In federated learning, the sensitive data is spread across many data holders. For example, different groups may have subject data from human health research, and summary statistics such as a CDF can reveal aggregate features of the population distribution. An aggregator collects a DP estimate from each group's local data to form this population summary.  The second is when data comes in over time in batches. A DP CDF estimate is updated and published as more data is collected. Central to both of these tasks is DP CDF estimation at a single site: ideally we want an estimation method which can be aggregated or updated over time.

The simplest method for approximating a CDF under DP is to histogram the data using a fixed binning~\cite{DworkRoth}. This effectively quantizes the data: a histogram is a collection of count queries for each bin. Adding noise to the bin counts guarantees DP, and we can compute an eCDF using the quantized values. However, fixed binning may lead to suboptimal approximation accuracy due to the discretization of the data domain. To alleviate this limitation, adaptive histogram boundaries~\cite{XuZXYYW:13boundaries} dynamically adjust bin widths based on data density. This approach assumes that the data are pre-quantized or that the data domain is a subset of the integers. Hierarchical histograms~\cite{QardajiYL:13hierarchical,CormodePSSY:12spatial} improve the resolution of the binning by building a tree structure on a fine partitioning of the data, thereby mitigating the limitations of fixed binning. Additional discretization-based methods have been studied under various settings. For instance, in large-scale federated settings with a massive number of sites or users, random subsampling has been used to construct histograms efficiently~\cite{CormodeB:22sample}. Other work has considered multidimensional data using transforms such as wavelets~\cite{BCDKMT07,XiaoWG:11wavelet}. Related distribution estimation problems have also been investigated under alternative privacy models, including local and shuffle models~\cite{BassilyS:15succinct,AcharyaSZ:19hadamard,WangBLJ:17local}. Beyond histogram-based approaches, adaptive quantile mechanisms~\cite{mckenna2021estimating} provide an alternative approach to CDF estimation. Instead of relying on histogram counts, these methods estimate private quantiles. However, these adaptive methods typically incur high communication costs due to their iterative nature, which limits their applicability in decentralized settings. While differentially private estimation of probability density functions (PDFs) has also been studied~\cite{kroll2021density, wagner2023fast, liu2024differentially}, we focus on the CDF, as it allows direct access to ranks, quantiles, thresholds, and other statistics: estimating these from DP PDFs may involve integrating the private PDF, making it less straightforward to provide approximation guarantees.

Given the vast work on DP statistical estimation, why do we need new methods for DP CDF estimation? Existing methods lack flexibility or efficiency in certain scenarios, such as decentralized settings or streaming data updates. For example, in decentralized settings, approaches like adaptive quantiles involve multiple communication rounds, while a more efficient alternative is to allow each site to transmit its information once for centralized aggregation into a global DP CDF. In streaming scenarios, methods such as adaptive quantiles require accessing old data when incorporating new samples, leading to repeated noise addition and increased privacy loss. Histogram-based methods may require recomputing the bin counts or rebuilding the histogram structure when new data arrive, making them inefficient for continuous updates. Motivated by these limitations, in this work we propose a novel framework for constructing differentially private CDFs in which the empirical CDF is projected into an appropriate function space and approximated in a manner analogous to standard signal decomposition techniques~\cite{cooley1965algorithm, oppenheim1999discrete, tolstov2012fourier}. The coefficients of these functions are then privatized to ensure differential privacy. Within this framework, we introduce two approaches. The first approach, called the polynomial projection method, projects the eCDF into a polynomial space using families of orthogonal polynomials commonly employed in functional analysis~\cite{luenberger1997optimization, kantorovich2014functional, yosida2012functional, alda2017bernstein}, followed by application of the functional mechanism to ensure differential privacy~\cite{zhang2012functional}. The second approach, a sparse approximation method via matching pursuit, constructs arbitrary function spaces from dictionaries instead of relying on polynomials, enabling flexible approximation of more complex CDF shapes. Our primary focus is one-dimensional CDF estimation, although the proposed framework naturally extends to high-dimensional CDF estimation, as demonstrated in our experiments.

The contributions of this work are:
    \begin{itemize}
        \item We develop two methods for approximating empirical CDFs via projections into function spaces: a polynomial projection method based on a fixed polynomial function space, and a sparse approximation method via matching pursuit that generalizes this idea by allowing flexible function spaces, together forming a general framework for privacy-preserving CDF estimation and offering a novel perspective on the problem.
        \item We provide theoretical analysis for approximation using orthonormal families, including upper bounds on the estimation error between the DP CDFs and true CDFs, and investigate the role of post-processing, showing that it preserves the validity of the CDF without compromising the approximation performance. 
        \item We demonstrate that our methods achieve comparable or superior performance to existing techniques across a range of scenarios. In particular, our approaches are well-suited for decentralized settings, and when updating the CDF with newly collected data, they outperform alternatives by avoiding the need to access previously collected data, thereby conserving the privacy budget. We further evaluate the applicability of our framework to high-dimensional CDF estimation, demonstrating the effectiveness of the sparse approximation method in such settings.
        \item We explore the effects of key parameters, such as dictionary size, and systematically examine different dictionary constructions, including polynomial, B-spline, and parametric distribution-based bases for function spaces, demonstrating the flexibility and robustness of our methods.
    \end{itemize}

\section{Background}

\subsection{Differential Privacy}
Differential privacy has been developed to enable statistical analyses on datasets while preserving the privacy of individuals included in the data. It is defined in terms of neighboring databases; two sets are considered neighboring if they differ by a single entry.

\begin{definition}[$(\epsilon, \delta)$-DP~\cite{dwork2014algorithmic}]
Let $\epsilon, \delta \geq 0$, a randomized algorithm $M: \mathcal{D} \rightarrow \mathcal{Y}$ is $(\epsilon, \delta)$-differentially private if for all $Y \subseteq  \mathcal{Y}$ and for all neighboring datasets $D, D' \in \mathcal{D}$, 
\begin{align*}
\mathbb{P}[M(D) \in Y] \leq e^\epsilon \mathbb{P}[M(D') \in Y] + \delta.
\end{align*}
\end{definition}

The parameter $\epsilon$ represents the privacy budget, where a smaller value offers stronger privacy but may result in less accurate responses. The parameter $\delta$ indicates the probability of information leakage. To design a $(\epsilon, \delta)$-DP algorithm that approximates a desired function $f: \mathcal{D} \rightarrow \mathbb{R}^d$, a common approach is to add random noise of appropriate magnitude to the output of $f(D)$~\cite{mcsherry2007mechanism, geng2016optimal, geng2016optimal2, liu2018generalized, alghamdi2022cactus, geng2015staircase, kairouz2014extremal, balle2018improving}. For example, when $\epsilon, \delta \in (0,1)$, the Gaussian mechanism is defined as $M(D) = f(D) + \mathcal{N}(0, \sigma^2 \mathbf{I})$, where $\sigma = \Delta \sqrt{2 \log(1.25/\delta)}/\epsilon$ and $\Delta = \underset {D \sim D'}{\max} \|f(D)-f(D')\|_2$ represents $\ell_2$ sensitivity of function $f$.

Differential privacy has several properties that make it particularly useful in applications. The composition property~\cite{dwork2014algorithmic, kairouz2015composition, mironov2017renyi, dong2019gaussian, gopi2021numerical, murtagh2015complexity} allows for a modular design of mechanisms: if each component of a mechanism is differentially private, then their composition also preserves differential privacy. The post-processing property ensures the safety of conducting arbitrary computations on the output of a differentially private mechanism: because of a data processing inequality~\cite{beaudry2011intuitive}, post-processing can only reduce the privacy risk provided by the mechanism.

\subsection{CDF Estimation}
In this work, we focus on the problem of estimating a cumulative distribution function from observations of a random variable. Accurate CDF estimation under privacy constraints is crucial in many applications, and various techniques have been developed to achieve this goal while preserving differential privacy. Existing approaches can be broadly categorized into histogram-based methods, hierarchical tree-based methods, and quantile-based methods. These methods typically introduce noise into data-dependent representations, such as histogram counts, hierarchical structures, or estimated quantiles, to achieve differential privacy. In our evaluation, we compare our methods with three representative baselines from these categories:


\subsubsection{Histogram Queries}
The \ac{hq} method partitions the data into a fixed number of uniformly spaced bins. The number of observations in each bin is counted, and noise is added to protect the privacy of these counts. The resulting noisy histogram can be used to estimate the CDF in a differentially private manner. To enhance accuracy, post-processing is applied by setting any negative noisy counts to 0, ensuring the estimated CDF is non-decreasing and non-negative.

\subsubsection{Tree-Based Histogram}
The tree-based (TB) histogram method extends standard flat histograms by organizing the data domain into a hierarchical structure, where each node stores the count of observations within a specific range. Noisy counts are added to the tree nodes to guarantee differential privacy, and CDF values are estimated by aggregating noisy counts from selected nodes in the hierarchy. A common construction is based on a balanced binary tree over the data domain. Recently, Rameshwar et al.~\cite{rameshwar2026optimal} advanced this paradigm by introducing optimized tree-based mechanisms and identifying near-optimal tree heights and branching factors. In our framework, we use this tree-based mechanism as a benchmarking baseline.

\subsubsection{Adaptive Quantiles}
The \ac{aq} method provides a more refined approach by leveraging known quantiles to guide the estimation process. Initially, a set of known quantiles $Q = \{0:a, 1:b\}$ is established, where $a$ and $b$ represent the lower and upper bounds of the data range. The method proceeds by selecting an initial value $x_0 = (a+b)/2$, counting the number of observations above and below $x_0$, and adding noise to these counts to preserve privacy. The corresponding quantile $q$ is then computed and stored as $Q[q] = x_0$. This process is iteratively repeated, with each new candidate $x_i$ selected as the midpoint of the largest interval between known quantiles until the privacy budget is exhausted. Due to the noise added during the process, situations may arise where $x_1 > x_2$ but $q_1 < q_2$. Post-processing is necessary, and one approach is to reorder the $x$ and $q$ values.

\begin{table}[h]
\caption{Key Notations}
\label{table:notation}
\centering
\begin{tabular}{ll}
\hline
Symbol & Definition \\
\hline
$\truecdf$ & true CDF of the distribution \\
$F_n$ & eCDF of samples $\{x_k\}_{k=1}^n$ \\
$\hat{F}$ & optimal projection of $F$ onto predefined space \\ 
$\hat{F}_n$ & optimal projection of $F_n$ onto predefined space \\
$\tilde{F}_n$ & privacy-preserving approximation of $F_n$ \\
$\hat{F}_n^s$ & optimal projection of $F_n$ with sparsity level $s$ \\
$\tilde{F}_n^s$ & privacy-preserving approximation of $\hat{F}_n^s$ \\
\hline
\end{tabular}
\end{table}

\section{CDF Approximation as Function Approximation}
\label{sec:method_overview}

\begin{figure}[t]
\centering
\includegraphics[width=0.85\linewidth]{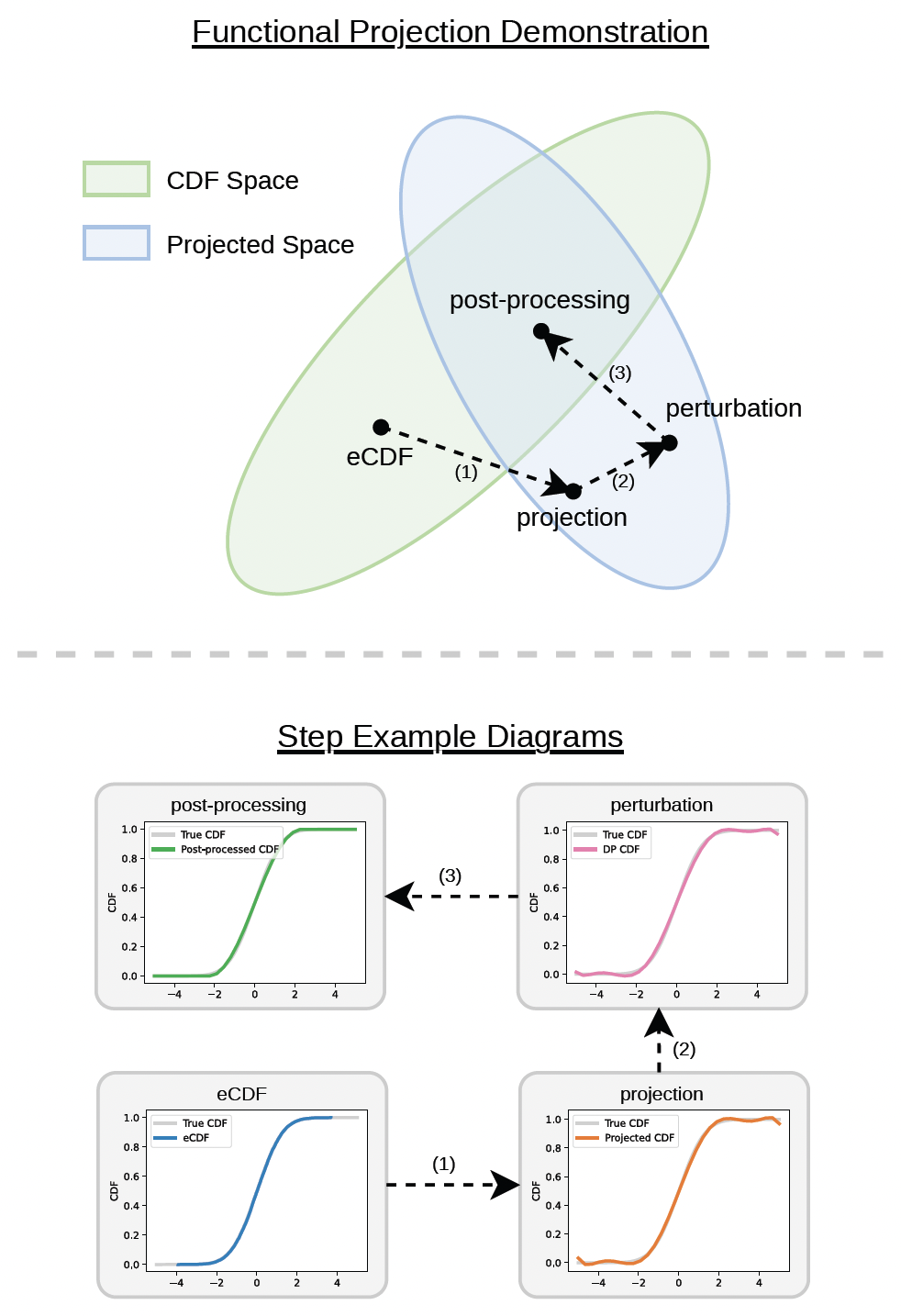}
\caption{Illustration of our proposed method. The top panel shows the overview: the \ac{ecdf} is projected onto the predefined function space, perturbed for differential privacy, and post-processed to obtain a valid \ac{dp} \ac{cdf}. The bottom panel presents step-by-step example results for each stage.}
\label{fig:method_overview}
\end{figure}

Our approach first approximates the \ac{ecdf} using a weighted combination of predefined functions and subsequently applies a differential privacy technique to protect the coefficients. With the perturbed coefficients and the associated predefined functions, we are able to reconstruct a privacy-preserving \ac{ecdf} (see Figure~\ref{fig:method_overview}). To formalize this, we begin by defining the relevant quantities. Key notations are summarized in Table~\ref{table:notation}. Throughout this work, we assume that the data range is known and approximate the empirical CDF in the Hilbert space $L^2([-A, A])$. When the data range is unknown, existing differentially private techniques can be applied prior to CDF construction, such as privately estimating the domain bounds or truncating the data under approximate differential privacy~\cite{KorolovaKMN:09clicks,BunNS:16concepts,BalcerV:19finite}. 

Let $\truecdf$ be the \ac{cdf} of a distribution supported on the interval $[-A,A]$, and let $\{x_k\}_{k=1}^n$ be i.i.d. samples drawn from $\truecdf$, ordered such that $x_1 \leq x_2 \leq \cdots \leq x_n$. The \ac{ecdf} is defined as
\[
\empcdf(x) = \frac{1}{n} \sum_{k=1}^n \mathbb{I}(x_k \leq x),
\] 
where $\mathbb{I}(x_k \leq x)$ is the indicator function, equal to $1$ if $x_k \leq x$, and $0$ otherwise. Let $\{ \phi_i \}_{i=1}^m$ be an orthonormal basis for an $m$-dimensional subspace of $L^2([-A, A])$. 
Define the function space:
\[
\mathcal{F} = \operatorname{span}\{\phi_1, \phi_2, \ldots,\phi_m\}.
\]
The orthogonal projection of $\empcdf$ onto $\mathcal{F}$ is
\[
\optempcdf = \argmin_{f \in \mathcal{F}} \| f - \empcdf \|_2,
\]
which admits the representation 
\[
\optempcdf(x) = \sum_{i=1}^m c_i \phi_i(x).
\]
The standard approach to guaranteeing $(\epsilon,\delta)$-\ac{dp} is to perturb the coefficients with noise:
\[
\tilde{c}_i = c_i + z_i,
\]
where $\{z_i\}_{i=1}^m$ are i.i.d. random variables drawn from an $(\epsilon,\delta)$-\ac{dp} mechanism such as Gaussian or Laplace mechanism. The privacy-preserving approximation of the \ac{ecdf} is then given by:
\[
\dpempcdf(x) = \sum_{i=1}^m \tilde{c}_i \phi_i(x).
\]
Assuming that $\opttruecdf$ is the optimal approximation of $\truecdf$ in $\mathcal{F}$ satisfying $\|\truecdf - \opttruecdf\|_2 \leq \alpha$. The total error of the privacy-preserving estimator $\dpempcdf$ can be decomposed as:
\begin{align*}
\|\truecdf-\dpempcdf\|_2 &\leq \|\truecdf-\empcdf\|_2 + \|\empcdf-\optempcdf\|_2 + \|\optempcdf-\dpempcdf\|_2
\\ &\leq \|\truecdf-\empcdf\|_2 + \|\empcdf-\opttruecdf\|_2 + \|\optempcdf-\dpempcdf\|_2
\\ &\leq \underbrace{\|\truecdf-\opttruecdf\|_2}_{\text{approximation error}} + 2\underbrace{\|\truecdf-\empcdf\|_2}_{\text{empirical error}} + \underbrace{\|\optempcdf-\dpempcdf\|_2}_{\text{privacy error}}.
\end{align*}
The first and third inequalities follow from the triangle inequality. The second inequality uses the optimality of $\optempcdf$, which ensures that $\|\empcdf - \optempcdf\|_2 \leq \|\empcdf - \opttruecdf\|_2$. This decomposition separates the total error into three components: the approximation error from projecting the true \ac{cdf} onto the function subspace, the empirical error due to finite sampling, and the privacy error introduced by perturbation. The empirical error can be further bounded as
\begin{align*}
\|\truecdf - \empcdf\|_2 &= \left( \int_{-A}^A |\truecdf(x) - \empcdf(x)|^2 \, dx \right)^{1/2} 
\\ &\leq \left( \int_{-A}^A \|\truecdf - \empcdf\|_\infty^2 \, dx \right)^{1/2} 
\\ &= (2A)^{1/2} \|\truecdf - \empcdf\|_\infty,
\end{align*}
where the inequality holds because $|\truecdf(x) - \empcdf(x)| \leq \|\truecdf - \empcdf\|_\infty$ for all $x \in [-A, A]$. The privacy error term can be explicitly bounded in terms of the noise magnitude and the properties of the basis functions:
\begin{align*}
\|\optempcdf - \dpempcdf\|_2 &= \left\| \sum_{i=1}^m z_i \phi_i \right\|_2 \leq \sum_{i=1}^m \| z_i \phi_i \|_2
= \sum_{i=1}^m |z_i| \|\phi_i\|_2,
\end{align*}
where the inequality follows from triangle inequality. 
We define the total error threshold $\tau$ as a weighted sum of the individual error components:
\[
\tau = \alpha + 2\beta + \eta,
\]
where $\alpha$ bounds the approximation error, $\beta$ accounts for the empirical error, and $\eta$ quantifies the privacy error. The total error satisfies the following tail bound:
\begin{align*}
&\mathbb{P}(\|\truecdf-\dpempcdf\|_2 > \tau)
\\ \leq & \mathbb{P}\left( \|\truecdf-\opttruecdf\|_2 + 2\|\truecdf-\empcdf\|_2 + \|\optempcdf-\dpempcdf\|_2 > \tau \right) 
\\ \leq &\mathbb{P}\left(2(2A)^{1/2} \|\truecdf - \empcdf\|_\infty + \sum_{i=1}^m |z_i| \|\phi_i\|_2 > 2\beta + \eta \right) 
\\ \leq &\mathbb{P}\left(\|\truecdf-\empcdf\|_\infty > \beta/(2A)^{1/2}\right) + \mathbb{P}\left(\sum_{i=1}^m |z_i| \|\phi_i\|_2 > \eta \right)
\\ \leq &2 \exp\left(-2n \beta^2/2A \right) + \mathbb{P}\left(\sum_{i=1}^m |z_i| > \eta/\max_{1 \leq i \leq m} \|\phi_i\|_2\right).
\end{align*}
The last inequality follows from the \ac{dkw} inequality~\cite{dvoretzky1956asymptotic}, which provides a finite-sample bound on the deviation between the empirical distribution function $\empcdf$ and the true distribution function $\truecdf$. The bound for privacy error depends on the differential privacy mechanism. For instance, if the Gaussian mechanism is applied, where the noise vector $\mathbf{z} = [z_1, z_2, \ldots, z_m]^\top$ is drawn from $\mathcal{N}(0, \sigma^2 \mathbf{I})$, we have $\mathbb{E}[|z_i|]=\sigma \sqrt{2/\pi}$. By the linearity of expectation, we know:
\[
\mathbb{E}\left[\sum_{i=1}^m |z_i|\right] = \sum_{i=1}^m \mathbb{E}[|z_i|] = m \sigma \sqrt{2/\pi}.
\]
Define $\eta' = \eta/\max_{1 \leq i \leq m} \|\phi_i\|_2$. Applying Markov's inequality yields:
\begin{align*}
\mathbb{P}\left(\sum_{i=1}^m |z_i| \geq \eta'\right) \leq \frac{\mathbb{E}\left[\sum_{i=1}^m |z_i|\right]}{\eta'} = \frac{m \sigma \sqrt{2/\pi}}{\eta'}.
\end{align*}
An even tighter bound can be obtained by applying the Chernoff bound and leveraging the moment generating function of the random variable $\|\mathbf{z}\|_1$~\footnote{A detailed derivation can be found at \url{https://www.ryanhmckenna.com/2021/12/tail-bounds-on-sum-of-half-normal.html}.}.

For more general $L^p$ approximation spaces (e.g., $L^1$), the orthogonal projection framework available in the Hilbert space $L^2$ is generally unavailable. Nevertheless, convergence and approximation error bounds for polynomial approximation in $L^p$ norms can still be established through classical approximation theory in Banach spaces~\cite{devore1993constructive}. Extending the proposed framework and its theoretical analysis to more general function spaces remains an interesting direction for future work.

\section{Polynomial Projection}
\label{sec:polynomial_projection}

Building on the theoretical intuition from Section~\ref{sec:method_overview}, we propose a \ac{pp} method to obtain a differentially private approximation of the \ac{ecdf}. Without loss of generality, we assume $A = 1$. If the sample range lies outside $[-1, 1]$, we first scale the data to fit within this interval. For example, if the data range is $[-A, A]$, each $x_k$ is divided by $A$. The bounds can be determined either through prior knowledge or by applying differentially private estimation techniques~\cite{liu2019differential, tao2024privacy}, which require additional privacy budget. After obtaining the \ac{dp} \ac{ecdf}, the scaling can be reversed to map the result back to the original data range. Section~\ref{sec:nondp_pp} presents the polynomial projection approximation in the non-private setting, while Section~\ref{sec:dp_pp} introduces its differentially private counterpart.

\subsection{Non-private Polynomial Projection}
\label{sec:nondp_pp}
In this work, we use the Legendre polynomials, although other polynomial families (e.g., Legendre, Chebyshev) can also be employed. The first $m$ functions from any such family span the same subspace of polynomials of degree less than $m$. Thus, the choice of polynomial family does not affect the projection space itself, but may impact numerical stability and computational efficiency~\cite{boyd2001chebyshev}. Appendix~\ref{appendix:alternative_family} provides a brief discussion on the selection of polynomial families and other function families, such as Fourier-based approaches.

We leverage the projection theorem~\cite{luenberger1997optimization} (see Theorem~\ref{theorem:projection} in Appendix~\ref{appendix:proofs}) to identify the optimal approximation of the \ac{ecdf} within the selected polynomial space $\mathcal{P}$. This formulation allows straightforward computation of the projection coefficients, which are related to the moments of the variables. These moment-based coefficients facilitate sensitivity analysis and the calibration of noise for privacy protection.

In the context of the Hilbert space $L^2[-1,1]$, we establish a polynomial space $\mathcal{P}$ with the basis $\{P_0, P_1,\ldots,P_m\}$, where each $P_i$ is a Legendre polynomial of degree $i$ over the interval $[-1,1]$. This space $\mathcal{P}$ is equipped with an inner product defined as $\langle f_1, f_2 \rangle =\int_{-1}^1 f_1(x) f_2(x) \, dx$ and a norm given by $\|f\|_2 = \left(\int_{-1}^1 f(x)^2 \, dx \right)^{1/2}$. The Legendre polynomials are orthogonal to each other. Since every finite-dimensional inner product space is also a Hilbert space, and the Legendre polynomials $\{P_i\}_{i=0}^\infty$ form a complete orthogonal sequence in $L^2[-1,1]$, $\mathcal{P}$ is a closed subspace of $L^2[-1,1]$. Given that $\int_{-1}^1 |\empcdf(x)|^2 \, dx < \infty$, the \ac{ecdf} $\empcdf(x)$ belongs to the $L^2[-1,1]$ space. According to the projection theorem (Theorem~\ref{theorem:projection}), we can identify a unique vector $\optempcdf$ in the space $\mathcal{P}$, which is the optimal approximation of the \ac{ecdf}, as demonstrated in Theorem~\ref{theorem:nondp_approximation} (see Appendix~\ref{appendix:proofs} for proof).

\begin{restatable}[Optimal Approximation of \ac{ecdf}]{theorem}{ecdfapprox}
\label{theorem:nondp_approximation}
Consider the polynomial space $\mathcal{P}$ spanned by the Legendre polynomials $\{P_0, P_1,\ldots,P_m\}$, where each $P_i$ is of degree $i$. The optimal approximation $\optempcdf$ of the \ac{ecdf} within $\mathcal{P}$ is given by
\begin{align*}
\optempcdf(x)= \sum_{i=0}^m \langle \empcdf, e_i \rangle e_i =  \sum_{i=0}^m \alpha_i \sum_{j=0}^i \left(\beta_{i,j} \left(1-\mu_{j+1}\right)\right) e_i, 
\end{align*}
where $\alpha_i = 2^i \sqrt{\frac{2i+1}{2}}$, $\beta_{i,j} = \frac{1}{j+1} \binom{i}{j} \binom{\frac{i+j-1}{2}}{i}$, $\mu_{j+1} = \frac{1}{n}\sum_{k=1}^n x_k^{j+1}$ represents mean of the $(j+1)$-th moment of the data, and $e_i=\sqrt{\frac{2i+1}{2}} P_i$ is the orthonormal basis of $\mathcal{P}$.
\end{restatable}

\subsection{Privacy-preserving Polynomial Projection}
\label{sec:dp_pp}

To achieve a privacy-preserving \ac{ecdf}, we protect the coefficients $\langle \empcdf , e_i \rangle$ by adding noise to the variables $\mu_j$ for $j \in [1, m+1]$. The differentially private approximation procedure based on the Analytic Gaussian mechanism is presented in Algorithm~\ref{alg:dp_pp}. Theorem~\ref{theorem:dp_guarantee} guarantees that Algorithm~\ref{alg:dp_pp} satisfies $(\epsilon, \delta)$-\ac{dp}. Moreover, this framework is compatible with a variety of DP mechanisms and noise distributions beyond the Gaussian mechanism, providing flexibility for diverse scenarios.

\begin{algorithm}[H]
\caption{Differentially Private Legendre Polynomial Projection}
\begin{algorithmic}
\STATE {\textbf{Input:}} \ac{ecdf} $\empcdf$, Legendre polynomial basis $\{P_i\}_{i=0}^m$, and the $\ell_2$ sensitivity $\Delta_2 = \sqrt{\frac{5m+8}{2 n^2}}$ of $\boldsymbol{\mu}$.
\STATE {\textbf{Output:}} Privacy-preserving coefficients $\{\tilde{c}_i\}_{i=0}^{m}$ and $\dpempcdf$.
\STATE \textbf{Do:}
\begin{enumerate}[label=\arabic*.]
    \item Compute the moment vector $\boldsymbol{\mu} = [\mu_1, \mu_2, \ldots, \mu_{m+1}]^\top$ where $\mu_j = \frac{1}{n} \sum_{k=1}^n x_k^j$ for $j \in [1, m+1]$.
    \item Add noise $\tilde{\boldsymbol{\mu}} = \boldsymbol{\mu} + \mathbf{z}$, where $\mathbf{z} \sim \mathcal{N}(0, \sigma^2 \mathbf{I})$. The noise scale $\sigma^2$ is computed based on the sensitivity $\Delta_2$ and~\cite[Algorithm 1]{balle2018improving}.
    \item Calculate the \ac{dp} coefficients $\tilde{c}_i = \alpha_i \sum_{j=0}^i \beta_{i,j} (1-\tilde{\mu}_{j+1})$ for $i \in [0, m]$.
    \item Construct the \ac{dp} \ac{ecdf}: $\dpempcdf(x) = \sum_{i=0}^m \tilde{c}_i e_i(x)$.
\end{enumerate}
\STATE {\textbf{End}}
\end{algorithmic}
\label{alg:dp_pp}
\end{algorithm}

\begin{theorem}
\label{theorem:dp_guarantee}
Algorithm~\ref{alg:dp_pp} satisfies $(\epsilon, \delta)$-differential privacy.
\end{theorem}
\begin{proof}
According to Lemma~\ref{lemma:sensitivity} in Appendix~\ref{appendix:proofs}, the $\ell_2$ sensitivity of $\boldsymbol{\mu}$ is $\Delta_2 = \sqrt{\frac{5m+8}{2 n^2}}$. Adding Gaussian noise calibrated via the Analytic Gaussian mechanism~\cite{balle2018improving} with scale based on $\Delta_2$ ensures that Algorithm~\ref{alg:dp_pp} satisfies $(\epsilon, \delta)$-differential privacy.
\end{proof}

To evaluate the noisy approximation $\tilde{F}_n$, Theorem~\ref{theorem:upper_bound_pp} provides an upper bound on its distance from the true \ac{cdf} $F$. Specifically, the theorem shows that if the polynomial space $\mathcal{P}$ is a suitable approximation space, meaning $\hat{F} \in \mathcal{P}$ adequately represents the true \ac{cdf}, then the noisy approximation of the \ac{ecdf} will also remain close to the true \ac{cdf}.

\begin{theorem}[Upper Bound for $\|F-\tilde{F}_n\|_2$]
\label{theorem:upper_bound_pp}
Let $F$ be the true CDF of a random variable with $x \in [-1,1]$. If $\hat{F}$ is the optimal approximation of $F$ in the polynomial space $\mathcal{P}$ and $\|F-\hat{F}\|_2 \leq \alpha$, then with probability at least $1-2 \exp\left(-\frac{n(\tau-\alpha)^2}{16}\right) - 2(m+1) \exp \left(-\frac{(\tau-\alpha)^2}{4(m+1)^4\sigma^2} \right)$, we have $\|F-\tilde{F}_n\|_2 \leq \tau$ for $\tau>\alpha>0$.
\end{theorem}
\begin{proof}
Let $F$ be the true CDF of a random variable with support $x \in [-1,1]$, then $F \in L^{\infty}[-1,1]$ and $F \in L^2[-1,1]$. Let $\hat{F}$ be the optimal approximation of $F$ in the polynomial space $\mathcal{P}$ assuming $\|F-\hat{F}\|_2 \leq \alpha$ for some $\alpha > 0$. The upper bound for $\|F-\tilde{F}_n\|_2$ is then given by
\begin{align}
\|F-\tilde{F}_n\|_2  &\leq \|F-F_n\|_2 + \|F_n-\hat{F}_n\|_2 + \|\hat{F}_n-\tilde{F}_n\|_2\nonumber 
\\ &\leq \|F-F_n\|_2 + \|F_n-\hat{F}\|_2 + \|\hat{F}_n-\tilde{F}_n\|_2 \nonumber 
\\ &\leq \|F-\hat{F}\|_2 + 2\|F-F_n\|_2 + \|\hat{F}_n-\tilde{F}_n\|_2 \nonumber 
\\ &\leq \alpha + 2\sqrt{2} \|F-F_n\|_{\infty} + \sqrt{2}\|\hat{F}_n-\tilde{F}_n\|_{\infty}.
\label{eq:4}
\end{align}
The first and third inequalities follow from the triangle inequality, the second inequality follows from Theorem~\ref{theorem:projection}, and the last inequality uses the bound $\|f\|_2 \leq \sqrt{2} \|f\|_{\infty}$. By leveraging the triangle inequality and the fact that $\hat{F}$ is the optimal approximation of the true distribution, we avoid directly computing the difference between $\hat{F}$ and $\hat{F}_n$. The bound for $\|\hat{F}_n-\tilde{F}_n\|_{\infty}$ is (see Lemma~\ref{lemma:privacy_error_bound} in Appendix~\ref{appendix:proofs})
\begin{align}
\|\hat{F}_n-\tilde{F}_n\|_{\infty} \leq \frac{(m+1)^2}{2} \max_{i \in [1,m+1]} |z_i|.
\label{eq:5}
\end{align}
By combining~\eqref{eq:4} and~\eqref{eq:5}, we obtain the following result:
\begin{align}
&\mathbb{P}(\|F-\tilde{F}_n\|_2 > \tau) \nonumber 
\\ \leq &\mathbb{P}(\alpha + 2\sqrt{2} \|F-F_n\|_{\infty} + \sqrt{2}\|\hat{F}_n-\tilde{F}_n\|_{\infty} > \tau) \nonumber
\\ \leq &\mathbb{P}\left(\|F-F_n\|_{\infty} > \frac{\tau-\alpha}{4\sqrt{2}}\right) + \mathbb{P}\left(\|\hat{F}_n-\tilde{F}_n\|_{\infty} > \frac{\tau-\alpha}{2\sqrt{2}}\right) \nonumber
\\ \leq & 2 \exp\left(-\frac{n(\tau-\alpha)^2}{16}\right) + 2(m+1) \exp \left(-\frac{(\tau-\alpha)^2}{4(m+1)^4\sigma^2} \right). 
\label{eq:bound}
\end{align}
The inequality in the last step is due to the \ac{dkw} inequality~\cite{dvoretzky1956asymptotic} and an upper bound for the the maxima of subgaussian random variables~\cite{wainwright2019high}. 
\end{proof}

\begin{remark}
The magnitude of the approximation term $\alpha$ and its decay with the polynomial degree $m$ depend on the regularity of the underlying CDF. Specifically, approximation results for orthogonal polynomial projections show that, if $F$ belongs to the Sobolev space $H^r([-1,1])$ for some $r > 0$, then $\alpha \leq C_r (m+1)^{-r} \|F\|_{H^r([-1,1])}$, where $C_r > 0$ is independent of $m$~\cite{canuto1982approximation, verfurth1999note}. In particular, $F \in H^1([-1,1])$ whenever the distribution admits a square-integrable density on $[-1,1]$. More generally, if the density $f = F'$ has higher Sobolev regularity, then the CDF $F$ inherits correspondingly higher regularity, leading to a faster decay of the approximation error with $m$. Thus, smoother distributions can be approximated accurately with lower-degree polynomial spaces.
\end{remark}

\begin{remark}
Equation~\eqref{eq:bound} indicates that the difference between the \ac{dp} \ac{ecdf} and the true \ac{cdf} is governed by two bounds. The first term in~\eqref{eq:bound} represents the distance between the true \ac{cdf} and the \ac{ecdf}, while the second term represents the distance between the approximation of the \ac{ecdf} and its noisy counterpart. As the number of data points $n$ increases, the first term decreases due to the convergence of the \ac{ecdf} to the true \ac{cdf}. The second term also decreases because the noise required diminishes as the sensitivity $\Delta_2$ decreases. A larger number of polynomials $m$ does not necessarily lead to a better
result. While a higher $m$ may improves the approximation of the \ac{ecdf}, it also increases the noise due to a higher sensitivity. 
\end{remark}

The proposed polynomial projection framework can also be extended to high-dimensional CDF estimation using tensor-product polynomial spaces. Theorem~\ref{theorem:nondp_approximation_multi} provides a closed-form expression for the corresponding projection coefficients in terms of multivariate empirical moments, while Lemma~\ref{lemma:sensitivity_multi} characterizes their sensitivity for privacy-preserving estimation.

\subsection{Effect of Post-processing}
\label{sec:post_processing}
Similar to the \ac{hq} and \ac{aq} methods discussed earlier, our approach also requires post-processing, since the obtained $\tilde{F}_n$ is not guaranteed to be monotonically non-decreasing on $[0,1]$. We adopt isotonic regression~\cite{lavine1995nonparametric} as the post-processing method. Proposition~\ref{pro:iso} (see Appendix~\ref{appendix:proofs} for proof) shows that this method brings the noisy approximation closer to the true \ac{cdf} without compromising the accuracy of distribution estimation. Although isotonic regression is traditionally formulated under the $L^2$ norm, it can be extended to other $L^p$ norms. 

\begin{restatable}{proposition}{iso}
\label{pro:iso}
Let $\tilde{F}_n \in L^2([-1,1])$, and let $F : [-1,1] \to [0,1]$ be a monotone non-decreasing function. Define
\[
\tilde{F}_n^{\mathrm{iso}} = \argmin_{f \in \mathcal{C}} \|f - \tilde{F}_n\|_2^2,
\]
where the constraint set is $\mathcal{C} = \mathcal{C}_1 \cap \mathcal{C}_2$, with
\begin{align*}
&\mathcal{C}_1 = \left\{ f \in L^2([-1,1]) \mid f(x_1) \leq f(x_2) \ \text{for a.e. } x_1 \leq x_2 \right\}, 
\\ &\mathcal{C}_2 = \left\{ f \in L^2([-1,1]) \mid f(x) \in [0,1] \ \text{for a.e. } x \in[-1,1] \right\}.
\end{align*}
Then the following inequality holds:
\[
\|\tilde{F}_n^{\mathrm{iso}} - F\|_2 \leq \|\tilde{F}_n - F\|_2.
\]
\end{restatable}

At first glance, all this result says is that postprocessing using isotonic regression does not hurt the accuracy. Empirically, however, it can make a significant difference. 


\section{Sparse Approximation via Matching Pursuit}

The \ac{dp} polynomial projection method in the previous question is essentially an ``off-the-shelf'' approach which does not rely on the properties of real \ac{cdf} functions, such as monotonicity and the values at endpoints. Increasing the number of basis functions does not necessarily improve the approximation, since it also amplifies the noise required for privacy. On the other hand, using too few basis functions may lead to a poor approximation of the true \ac{cdf}. To balance this trade-off, we consider a function space $\mathcal{G}$ spanned by a large dictionary of $m$ arbitrary functions, which need not be orthonormal. We then select the top $s$ functions with the largest absolute inner products with the empirical distribution. Our theoretical analysis focuses on the case of orthonormal dictionaries. The experiments consider both orthonormal dictionaries, which align with the theoretical setting, and non-orthogonal dictionaries. The latter are investigated as empirical extensions to explore the effects of different dictionary constructions and potential extensions of the framework beyond the current theoretical setting. The empirical performance of different dictionary constructions is discussed in Section~\ref{sec:dict_compose}.
Lemma~\ref{lemma:4} (see Appendix~\ref{appendix:proofs} for proof) shows that choosing $s$ functions from a larger pool of $m$ orthonormal basis functions achieves a smaller approximation error compared to using a fixed set of $s$ basis functions. Although a larger dictionary increases computational cost, it enhances the expressive power of the representation. 

\begin{restatable}{lemma}{lemmafour}
Let $\mathcal{D}_m = \{\phi_i\}_{i=1}^m$ be an orthonormal basis for a subspace of $L^2([-1,1])$. For any target function $f \in L^2([-1,1])$, define:
\begin{itemize}
    \item[(1)] Fixed basis approximation: $\hat{f} = \sum_{i=1}^s \langle f, \phi_i \rangle \phi_i$, where $\{\phi_i\}_{i=1}^s$ are the first $s$ basis elements in $\mathcal{D}_m$.
    
    \item[(2)] Adaptive selection: $\hat{f}^s = \sum_{i=1}^s \langle f, \phi_{I_i} \rangle \phi_{I_i}$, where $\{\phi_{I_i}\}_{i=1}^s$ corresponds to the $s$ basis functions with the largest absolute inner products $|\langle f, \phi_i \rangle|$.
\end{itemize}
Then the following inequality holds:
\[
\|f - \hat{f}^s\|_2 \leq \|f - \hat{f}\|_2,
\]
with strict inequality unless $\{\phi_{I_i}\}_{i=1}^s = \{\phi_i\}_{i=1}^s$.
\label{lemma:4}
\end{restatable}

To implement this selection process, we adopt a sparse representation approach, namely \ac{mp}~\cite{mallat1993matching}, a classical algorithm widely used in dictionary learning. Formally, given an orthonormal dictionary $\mathcal{D} = \{\phi_i\}_{i=1}^m \subset L^2([-1,1])$, the \ac{mp} algorithm iteratively selects the indices $\{I_i\}_{i=1}^s$ corresponding to the $s$ most relevant basis functions. Relevance is quantified by the inner product between each candidate basis function and the current residual, defined as the unexplained portion of the target function after subtracting the contributions of the previously selected basis functions. This yields the sparse approximation
\[
\hat{F}_n^s(x) = \sum_{i=1}^s c_i \phi_{I_i}(x),
\]
where $\phi_{I_i}$ are dynamically selected from $\mathcal{D}$ and $c_i$ denote the inner products with the residuals at each iteration. Lemma~\ref{lemma:5} (see Appendix~\ref{appendix:proofs} for proof) establishes that, when the dictionary is orthonormal, this procedure is equivalent to selecting the $s$ basis functions with the largest coefficients from the full dictionary. Moreover, our approach naturally extends to dictionaries with non-orthonormal atoms, in which case the \ac{mp} procedure remains applicable.

\begin{restatable}{lemma}{lemmafive}
If $\mathcal{D} = \{\phi_i\}_{i=1}^m$ is an orthonormal basis, then for any target function $f$, \ac{mp} algorithm with sparsity level $s$ selects the $s$ basis functions with the largest absolute inner products $|\langle f, \phi_i \rangle|$. In this case, the result of \ac{mp} is equivalent to selecting the top $s$ basis elements ranked by $|\langle f, \phi_i \rangle|$.
\label{lemma:5}
\end{restatable}

To ensure differential privacy in the selection of $I_i$ and its coefficient $c_i$, we employ the \ac{rnm} mechanism~\cite{ding2021permute}. The resulting differentially private approximation (see Algorithm~\ref{alg:dp_mp}) is
\[
\tilde{F}_n^s(x) = \sum_{i=1}^s \tilde{c}_i \phi_{\tilde{I}_i}(x),
\]
where $\tilde{I}_i$ and $\tilde{c}_i$ denote the privacy-preserving index and coefficient, respectively. Lemma~\ref{lemma:sen_ip} and Lemma~\ref{lemma:sen_aip} (see Appendix~\ref{appendix:proofs} for proof) analyze the sensitivity of the (absolute) coefficients, which is crucial for determining the appropriate noise level required to ensure differential privacy. Based on these results, Theorem~\ref{theorem:privacy} establishes that Algorithm~\ref{alg:dp_mp} satisfies $(2s\epsilon', 0)$-DP under the basic composition rule. Although more advanced composition methods could be applied to relax the privacy cost by introducing a non-zero $\delta$. Finally, the utility of the proposed method is analyzed in Theorem~\ref{theorem:utility}, which provides a bound on the estimation error introduced by the added noise.

\begin{algorithm}[t]
\caption{Differentially private approximation via matching pursuit}
\begin{algorithmic}
\STATE 
\STATE {\textbf{Input:}} eCDF $F_n$, dictionary $\mathcal{D} = \{\phi_i\}_{i=1}^m $, sparsity level $s$, privacy parameter $\epsilon'$ and sensitivity $\Delta_{\text{\ac{mp}}}$.
\STATE {\textbf{Output:}} List of privacy-preserving coefficients $\{\tilde{c}_i\}_{i=1}^s$ and indices for corresponding atoms $\{\tilde{I}_i\}_{i=1}^s$.
\STATE {\textbf{Initialization:}} $\tilde{r}_1 \leftarrow F_n$.
\STATE {\textbf{For $i=1$ to $s$:}}
\begin{enumerate}[label=\arabic*.]
    \item Find $\phi_{\tilde{I}_i} \in \mathcal{D}$ using the \ac{rnm} mechanism with Laplace noise of scale $b = \Delta_{\text{\ac{mp}}}/\epsilon'$: \[\phi_{\tilde{I}_i} = \argmax_{\phi \in \mathcal{D}} \left( \left| \langle \tilde{r}_i, \phi \rangle \right| +  \text{Laplace}(0, b) \right).\]
    \item Compute the noisy coefficient: 
    $\tilde{c}_i \leftarrow \langle \tilde{r}_i, \phi_{\tilde{I}_i} \rangle + \text{Laplace}(0, b)$.
    \item Update the residual: $\tilde{r}_{i+1} \leftarrow \tilde{r}_i - \tilde{c}_i \phi_{\tilde{I}_i}$.
    \item $i \leftarrow i+1$.
\end{enumerate}
\STATE {\textbf{End}}
\end{algorithmic}
\label{alg:dp_mp}
\end{algorithm}

\begin{restatable}[Sensitivity of Inner Product]{lemma}{senip}
Let $\langle \tilde{r}_i, \phi \rangle$ denote the inner product between the residual $\tilde{r}_i$ at the $i$-th step and an arbitrary basis function $\phi$. Assume the residual is defined as
\[
\tilde{r}_i = F_D - \sum_{j=1}^{i-1} \tilde{c}_j \phi_{\tilde{I}_j},
\]
where $F_D$ is the \ac{ecdf} of the dataset $D = \{x_k\}_{k=1}^n$, and $\tilde{c}_j$ are the coefficients computed in previous steps. Then, the sensitivity of this inner product is
\[
\Delta_{\text{\ac{mp}}} = \frac{1}{n} \int \left| \phi(x) \right| \, dx.
\]
\label{lemma:sen_ip}
\end{restatable}

\begin{restatable}[Sensitivity of Absolute Inner Product]{lemma}{senaip}
Let $|\langle \tilde{r}_i, \phi \rangle|$ denote the absolute inner product between the residual $\tilde{r}_i$ at the $i$-th step and an arbitrary basis function $\phi$. Then, the sensitivity of the absolute inner product is
\[\Delta_{\text{A\ac{mp}}} = \Delta_{\text{\ac{mp}}}.\]
\label{lemma:sen_aip}
\end{restatable}

\begin{theorem}
Let $\epsilon'$ denote the privacy budget allocated to each step, and let $\epsilon = 2s\epsilon'$. Then, Algorithm~\ref{alg:dp_mp} satisfies $\epsilon$-DP, where $s$ denotes the sparsity level.
\label{theorem:privacy}
\end{theorem}
\begin{proof}
Each iteration of Algorithm~\ref{alg:dp_mp} involves two operations that require differential privacy: selecting the basis index and releasing the selected coefficient. Each of these operations uses $\epsilon'$-DP. Over $s$ iterations, the total privacy cost is $2s\epsilon'$-DP, according to the basic composition theorem.
\end{proof}

\begin{theorem}[Upper Bound for $\|F-\tilde{F}_n^s\|_2$]
\label{theorem:utility}
Fix a privacy parameter $\epsilon > 0$, sparsity level $s$, and sample size $n$. Let $\mathcal{D} = \{ \phi_i \}_{i=1}^{m}$ denote a family of orthonormal functions on $[-1,1]$, and let $\mathcal{G}$ be the linear span of $\mathcal{D}$. Let $F$ be the true \ac{cdf}, and let $F_n$ denote the \ac{ecdf} formed from $n$ i.i.d. samples from $F$. Let $\hat{F}^s$ and $\hat{F}_n^s$ denote the optimal non-private $s$-sparse approximations of the true CDF $F$ and the empirical CDF $F_n$ in $\mathcal{G}$, respectively, where $\hat{F}_n^s$ has the form 
\[
\hat{F}_n^s(x) = \sum_{i=1}^s c_i \phi_{I_i}(x),
\] 
and the index set $\{I_1, \ldots, I_s\}$ corresponds to the $s$ functions with the largest absolute inner products with $F_n$. Let $\tilde{F}_n^s$ be the output of Algorithm \ref{alg:dp_mp} with input $F_n$, $s$, $\epsilon$, and $\Delta_{\text{\ac{mp}}}$. Finally, define $\beta = \sqrt{2} \sum_{i=1}^{s} |c_i|$. Then if $\|F - \hat{F}^s\|_{2} \le \alpha$, with probability at least $1-2\exp\left(-\frac{n(\tau-\alpha-\beta)^2}{16}\right) - \exp\Big(s - \frac{n \epsilon (\tau-\alpha - \beta)}{2 \|\phi\|_1} \Big) \left(\frac{n\epsilon (\tau-\alpha - \beta)}{2 s \|\phi\|_1}\right)^s$, we have $\|F-\tilde{F}_n^s\|_2 \leq \tau$ for $\tau>\alpha+\beta>0$ and $\tau-\alpha-\beta > 2s\|\phi\|_1/n\epsilon$.
\end{theorem}
\begin{proof}
Let $\check{F}_n^s = \sum_{i=1}^s c_i \phi_{\tilde{I}_i}(x)$, where $\tilde{I}_i$ denotes the index selected via \ac{rnm}. The total error between the true \ac{cdf} and its privacy-preserving counterpart can be decomposed as follows:
\begin{align*}
\|F - \tilde{F}_n^s \|_2 &\leq \|F - F_n \|_2 + \|F_n - \hat{F}_n^s \|_2 + \|\hat{F}_n^s - \tilde{F}_n^s \|_2 
\\ &\leq \|F - F_n \|_2 + \|F_n - \hat{F}^s \|_2 + \|\hat{F}_n^s - \tilde{F}_n^s \|_2
\\ &\leq \underbrace{\|F - \hat{F}^s \|_2}_{\text{approximation error}} + 2\underbrace{\|F - F_n\|_2}_{\text{empirical error}} 
\\ &\quad + \underbrace{\|\hat{F}_n^s - \check{F}_n^s\|_2}_{\text{index error}} + \underbrace{\|\check{F}_n^s - \tilde{F}_n^s\|_2}_{\text{coefficient error}}.
\end{align*}
In the context of the \ac{mp} algorithm, privacy error arises from two sources: the index error and the coefficient error. We now derive worst-case upper bounds for both components. Specifically, we assume that index perturbation occurs at every step, i.e., $I_i \neq \tilde{I}_i$ for all $i$, and we do not assume orthogonality among the perturbed basis functions $\tilde{I}_i$. In this case, we derive the following bounds:
\begin{align*}
\|\hat{F}_n^s - \check{F}_n^s\|_2 &= \left\|\sum_{i=1}^s c_i \phi_{I_i} - \sum_{i=1}^s c_i \phi_{\tilde{I}_i} \right\|_2 
\\ &\leq \sum_{i=1}^s |c_i| \| \phi_{I_i} - \phi_{\tilde{I}_i} \|_2 
\\ &= \sqrt{2} \sum_{i=1}^s |c_i| \cdot \mathbb{I}(I_i \neq \tilde{I}_i) 
\\ &\leq \sqrt{2} \sum_{i=1}^s |c_i|.
\end{align*}
Let $\beta = \sqrt{2} \sum_{i=1}^s |c_i|$ denote an upper bound for the index error.
We now turn to the coefficient error term, which satisfies:
\begin{align*}
\|\check{F}_n^s - \tilde{F}_n^s\|_2 &= \left\|\sum_{i=1}^s c_i \phi_{\tilde{I}_i} - \sum_{i=1}^s \tilde{c}_i \phi_{\tilde{I}_i} \right\|_2 
\\ &\leq \sum_{i=1}^s |c_i - \tilde{c}_i| \| \phi_{\tilde{I}_i} \|_2 \\ &= \sum_{i=1}^s |z_i|,
\end{align*}
where $z_i \sim \text{Laplace}(0, \Delta_{\text{\ac{mp}}}/\epsilon)$ and $\Delta_{\text{\ac{mp}}} = \frac{1}{n} \int \left| \phi(x) \right| \, dx = \frac{1}{n} \|\phi\|_1$. Then for any $\tau > \alpha + \beta$ and $\frac{\tau-\alpha - \beta}{2} > \frac{s\Delta_{\text{\ac{mp}}}}{\epsilon}$, we obtain
\begin{align*}
&\mathbb{P}(\|F-\tilde{F}_n^s\|_2 > \tau) 
\\ 
\leq &\mathbb{P}(\alpha + 2\sqrt{2}\|F - F_n\|_\infty + \beta + \|\check{F}_n^s - \tilde{F}_n^s\|_2 > \tau)
\\ \leq &\mathbb{P}\left(\|F-F_n\|_{\infty} > \frac{\tau-\alpha - \beta}{4\sqrt{2}}\right) 
\\ \quad &+ \mathbb{P}\left(\sum_{i=1}^s |z_i| > \frac{\tau-\alpha - \beta}{2}\right)
\\ 
\leq &2\exp\left(-\frac{n(\tau-\alpha-\beta)^2}{16}\right) 
\\ \quad &+ \exp\Big(s - \frac{n \epsilon (\tau-\alpha - \beta)}{2 \|\phi\|_1} \Big) \left(\frac{n\epsilon (\tau-\alpha - \beta)}{2 s \|\phi\|_1}\right)^s.
\end{align*}
The first term in the last inequality comes from the \ac{dkw} inequality. The second term arises because $z_i \sim \text{Laplace}(0, \Delta_{\text{\ac{mp}}}/\epsilon)$, which implies that $|z_i| \sim \text{Exponential}(\epsilon/\Delta_{\text{\ac{mp}}})$. Consequently, the sum follows an Erlang distribution, i.e., $\sum_{i=1}^s |z_i| \sim \text{Erlang} (s, \epsilon/\Delta_{\text{\ac{mp}}})$, where $\epsilon/\Delta_{\text{\ac{mp}}}$ is the rate parameter. Based on Lemma~\ref{lemma:erlang_bound} in Appendix~\ref{appendix:proofs}, we obtain the second term in the bound.
\end{proof}

\begin{remark}
Theorem~\ref{theorem:utility} shows that increasing the sparsity level $s$, and thus using more functions to approximate the \ac{ecdf}, does not necessarily reduce the distance between the true CDF and its noisy approximation, as a larger $s$ increases the likelihood of incorrect index and coefficient selection. Similarly, enlarging the dictionary size $m$ does not necessarily bring the DP CDF closer to the true CDF, since a richer dictionary reduces the approximation error $\alpha$ but simultaneously increases the noise-induced error $\beta$. By contrast, increasing the sample size $n$ consistently decreases the distance between the CDF and its approximation, because the empirical error diminishes and the magnitude of the added noise is reduced.
\end{remark}

\section{Experiments}
In this experimental section~\footnote{The implementation code is available at \url{https://github.com/yetaoyl/dp-cdf}.}, we investigate the impact of key parameters, including sparsity level and dictionary size, on CDF approximation (see Section~\ref{sec:effect_para}). We then provide a comparative analysis of different methods (see Section~\ref{sec:compare_methods}) and finally discuss the effect of varying dictionary compositions (see Section~\ref{sec:dict_compose}).

\subsection{Effect of Parameters}
\label{sec:effect_para}
To quantitatively study the effect of parameters, we assess the distance between the \ac{cdf} and its (noisy) approximation using three measures: the Kolmogorov-Smirnov distance~\cite{lilliefors1967kolmogorov}, the earth mover's distance~\cite{rubner1998metric}, and the energy distance~\cite{rizzo2016energy}. The Kolmogorov-Smirnov distance is defined as the maximum difference between the two functions. In contrast, the earth mover's distance, given in the univariate case by $W(f_1, f_2) = \int |f_1 - f_2|$, measures the integral of the absolute difference between the \ac{cdf}s. The energy distance, expressed as $E(f_1, f_2) = \sqrt{2\int (f_1 - f_2)^2}$, penalizes larger deviations between the distributions more heavily.

\begin{figure*}[htb!]
\centering
\subfloat[$\mathcal{N}(0,1)$]{\includegraphics[height=3.2cm]{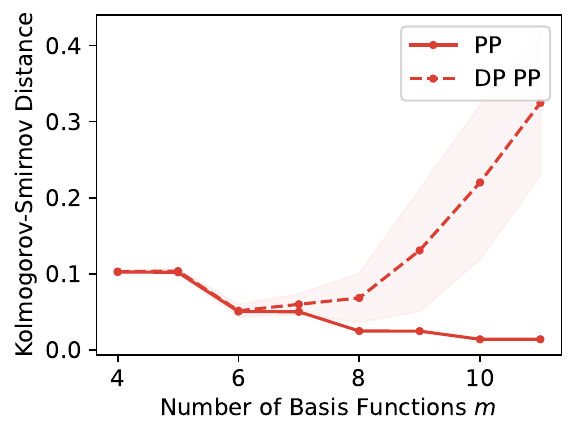}}\hfil
\subfloat[$\mathcal{N}(0,1)$]{\includegraphics[height=3.2cm]{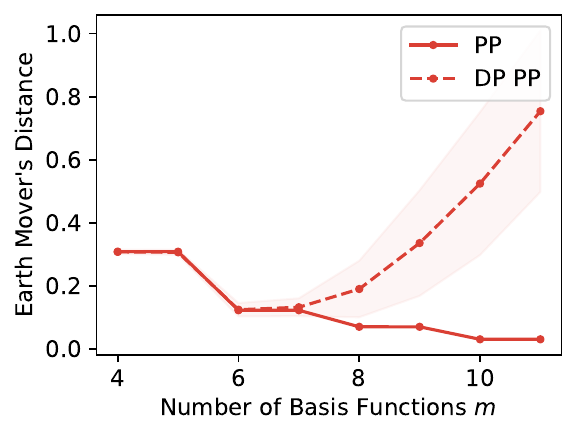}}\hfil
\subfloat[$\mathcal{N}(0,1)$]{\includegraphics[height=3.2cm]{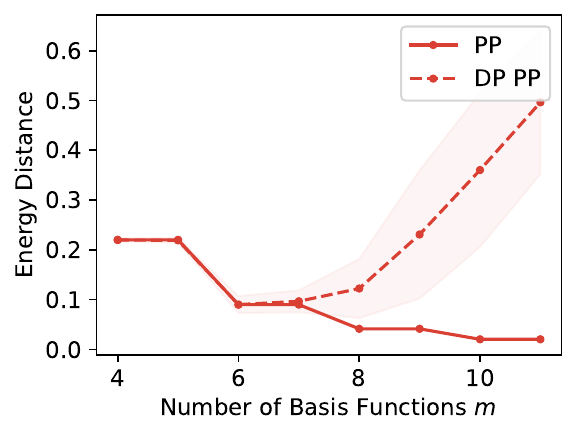}}

\subfloat[$\mathcal{N}(0,1)$]{\includegraphics[height=3.2cm]{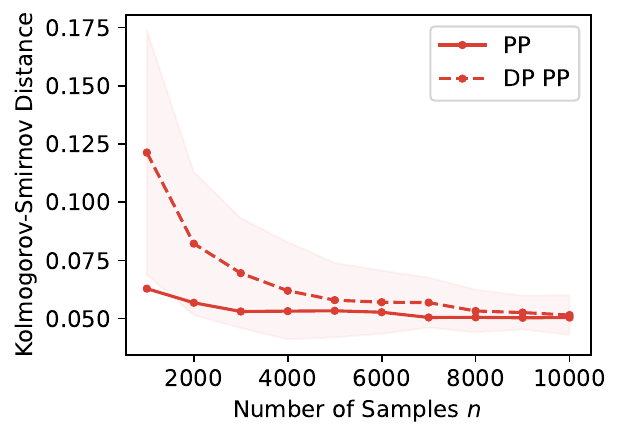}}\hfil
\subfloat[$\mathcal{N}(0,1)$]{\includegraphics[height=3.2cm]{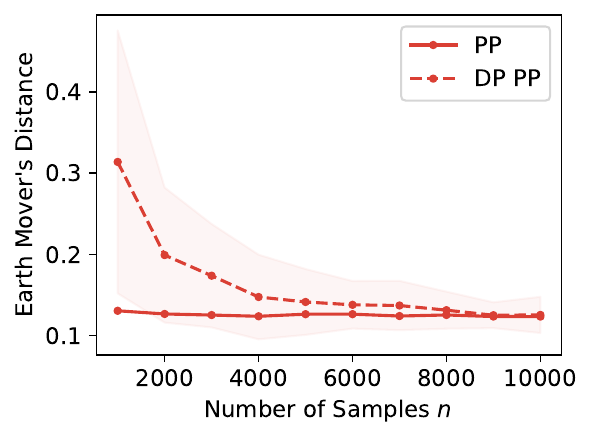}}\hfil
\subfloat[$\mathcal{N}(0,1)$]{\includegraphics[height=3.2cm]{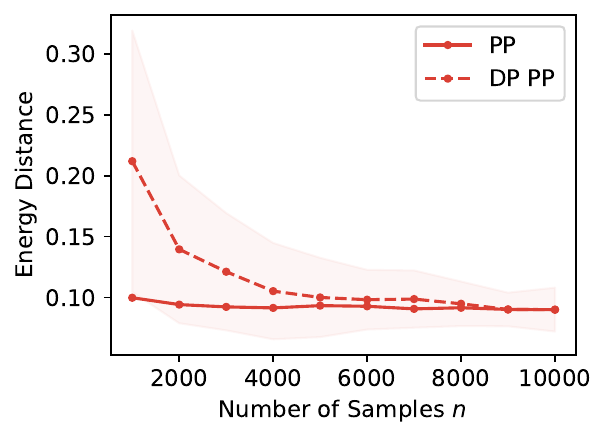}}
\caption{Comparison of distances between \ac{pp}-based approximation and the true CDF under two experimental settings. (a)-(c) Effect of the number of basis functions $m$ with $n=10^4$; (d)-(f) Effect of the sample size $n$ with $m=6$. Experiments were repeated $50$ times with $\epsilon=0.5$ under pure differential privacy.}
\label{fig:pp_para}
\end{figure*}

\begin{figure*}[htb!]
\centering
\subfloat[$\mathcal{N}(0,1)$]{\includegraphics[height=3.2cm]{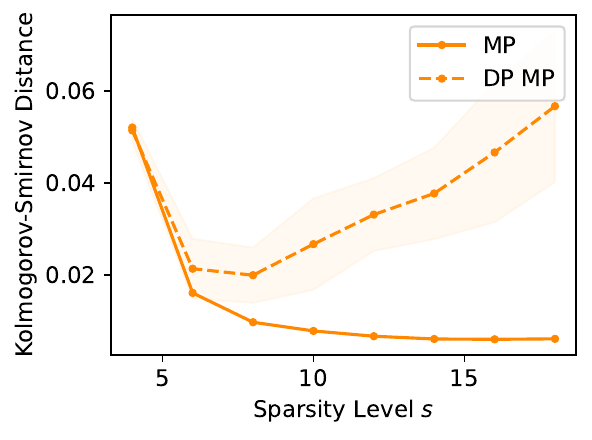}}\hfil
\subfloat[$\mathcal{N}(0,1)$]{\includegraphics[height=3.2cm]{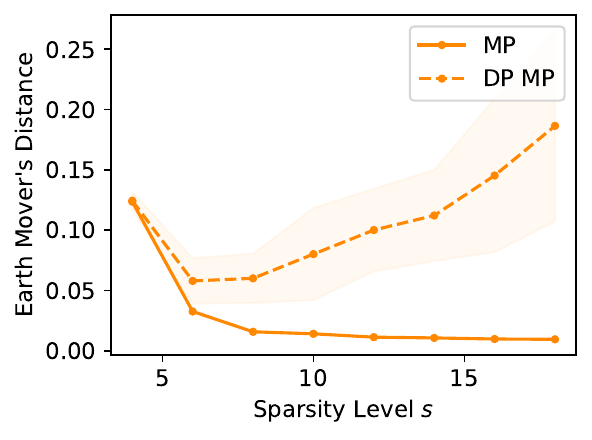}}\hfil
\subfloat[$\mathcal{N}(0,1)$]{\includegraphics[height=3.2cm]{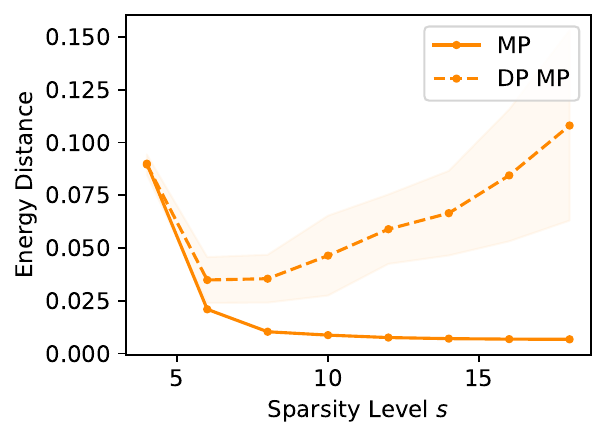}}

\subfloat[$\mathcal{N}(0,1)$]{\includegraphics[height=3.2cm]{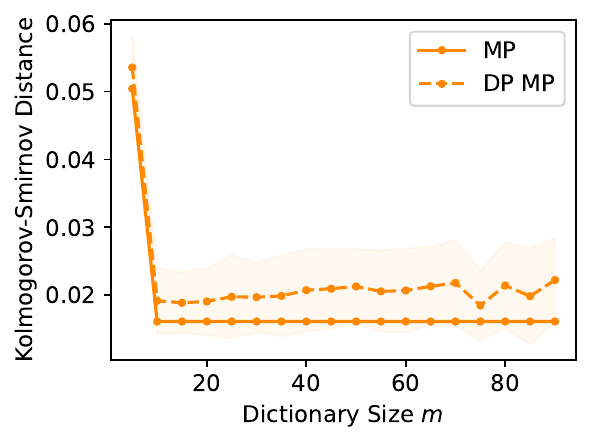}}\hfil
\subfloat[$\mathcal{N}(0,1)$]{\includegraphics[height=3.2cm]{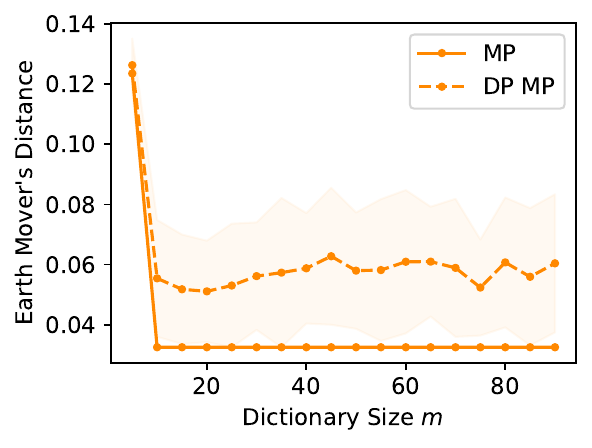}}\hfil
\subfloat[$\mathcal{N}(0,1)$]{\includegraphics[height=3.2cm]{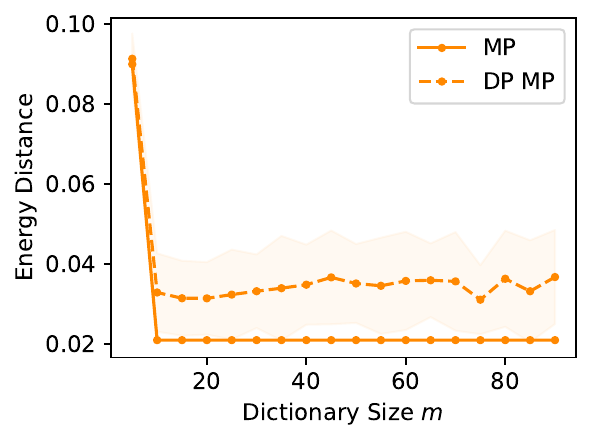}}

\subfloat[$\mathcal{N}(0,1)$]{\includegraphics[height=3.2cm]{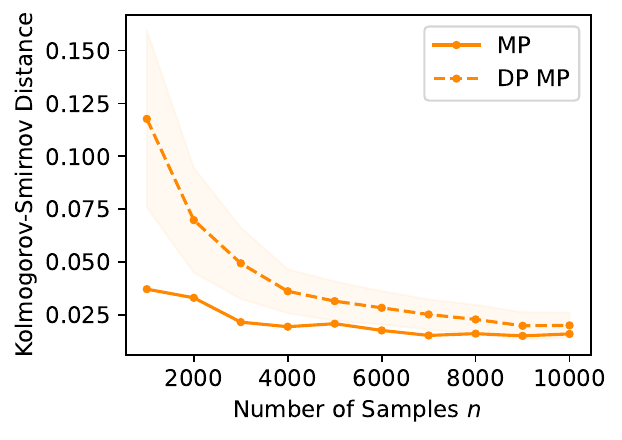}}\hfil
\subfloat[$\mathcal{N}(0,1)$]{\includegraphics[height=3.2cm]{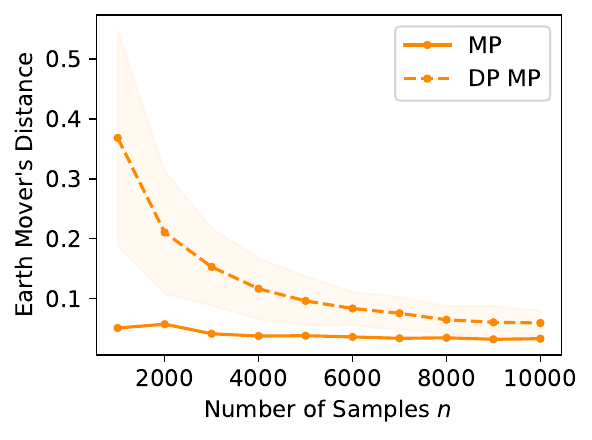}}\hfil
\subfloat[$\mathcal{N}(0,1)$]{\includegraphics[height=3.2cm]{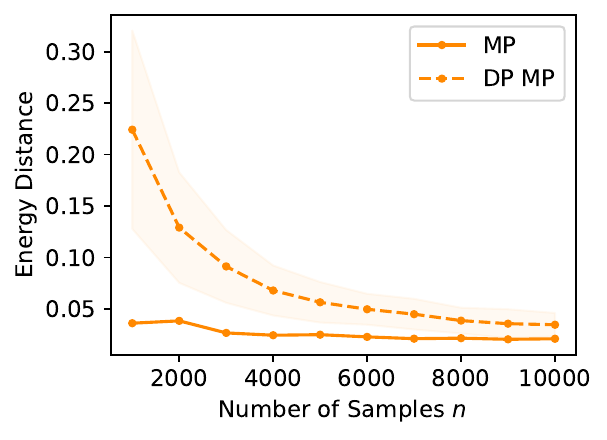}}
\caption{Comparison of distances between \ac{mp}-based approximation and the true CDF under three experimental settings. (a)-(c) Effect of the sparsity level $s$ with $n=10^4$ and a dictionary of $m=40$ Legendre polynomials; (d)-(f) Effect of the dictionary size $m$ with $n=10^4$ and $s=6$; (g)-(i) Effect of the sample size $n$ with a dictionary of $m=40$ Legendre polynomials and $s=6$. Experiments were repeated $50$ times with $\epsilon=0.5$ under pure differential privacy.}
\label{fig:mp_para}
\end{figure*}

\subsubsection{Polynomial Projection}
Figure~\ref{fig:pp_para} (a)-(c) and Figure~\ref{fig:appendix_pp_parameters} (a)-(f) in Appendix~\ref{appendix:figures} illustrate the effect of $m$ on approximation performance. In the non-private setting, increasing $m$ consistently reduces the error between the projected \ac{ecdf} and the true CDF, as the polynomial space becomes more expressive. In the differentially private setting, however, the error first decreases and then increases. This is because a richer polynomial space initially improves approximation quality, but a larger $m$ requires releasing more coefficients, which in turn necessitates adding more noise. Consequently, the discrepancy between the approximated and true CDF tends to grow when $m$ becomes large. Based on experimental results, values of $m$ between 5 and 8 are empirically preferable.

The effect of the sample size $n$ on approximation performance is shown in Figure~\ref{fig:pp_para} (d)-(f) and Figure~\ref{fig:appendix_pp_parameters} (g)-(l) in Appendix~\ref{appendix:figures}. In the non-private setting, the error remains nearly constant as $n$ increases. This reflects the inherent approximation limit imposed by the predefined polynomial space, since even the best projection within this space cannot perfectly match the true CDF. As mentioned previously, increasing $n$ reduces the error in differentially private settings, and the performance gradually approaches that of the non-private setting.

\subsubsection{Matching Pursuit}
The effect of the sparsity level $s$ on the approximation performance of the matching pursuit method is illustrated in Figure~\ref{fig:mp_para} (a)-(c) and Figure~\ref{fig:appendix_mp_parameters} (a)-(f) in Appendix~\ref{appendix:figures}. In the non-private setting, the error decreases as $s$ increases and eventually converges to the approximation limit determined by the expressive power of the dictionary. In the differentially private setting, however, increasing $s$ does not necessarily reduce the error. Similar to the \ac{pp} method, using more basis functions requires injecting additional noise into the released indices and coefficients, which may offset the potential gains in approximation accuracy.
We also study the effect of the dictionary size $m$, as illustrated in Figure~\ref{fig:mp_para} (d)-(f) and Figure~\ref{fig:appendix_mp_parameters} (g)-(l) in Appendix~\ref{appendix:figures}, and find that enlarging $m$ does not necessarily reduce the error in the differentially private setting. 
Consistent with the \ac{pp} method, increasing the sample size $n$ consistently reduces the approximation error in the private setting, as shown in Figure~\ref{fig:mp_para} (g)-(i) and Figure~\ref{fig:appendix_mp_parameters} (m)-(r) in Appendix~\ref{appendix:figures}.

\subsection{Comparison of Methods}
\label{sec:compare_methods}
To evaluate the performance of our method, we conducted experiments on both synthetic and real-world datasets. 
Figure~\ref{fig:comparison} and Figure~\ref{fig:appendix_comparison} in Appendix~\ref{appendix:figures} present comparative examples of our methods alongside existing baselines. The first column shows the \ac{hq} method, the second column shows the TB method, the third column shows the \ac{aq} method, and the last two columns depict our proposed methods: polynomial projection and sparse approximation via matching pursuit. With a small bin number in the HQ method, noticeable discrepancies are observed between the CDF approximation (green solid line) and the true CDF (gray solid line), even without perturbation. The TB method provides a more accurate approximation by using a finer hierarchical partition. The AQ method yields an accurate CDF approximation (blue solid line), but its privacy-preserving counterpart (blue dashed line) tends to focus on regions with the steepest slope, as these regions correspond to greater changes in the quantiles. Our methods accurately fit the true CDF, regardless of whether privacy protection is considered. Notably, the matching pursuit approach, which leverages a larger dictionary and sparse function selection, achieves a more accurate approximation than the polynomial projection method, as indicated by the red and orange solid lines.

\begin{figure*}[ht!]
\centering
\subfloat[$\mathcal{N}(0,1)$]{\includegraphics[height=2.5cm]{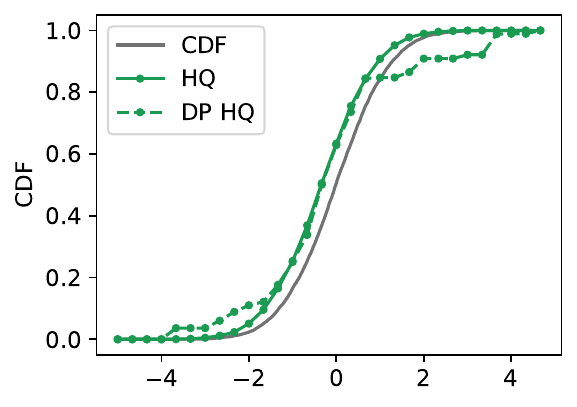}}
\subfloat[$\mathcal{N}(0,1)$]{\includegraphics[height=2.5cm]{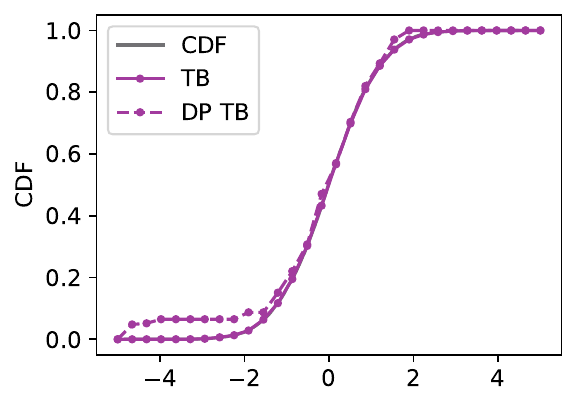}}
\subfloat[$\mathcal{N}(0,1)$]{\includegraphics[height=2.5cm]{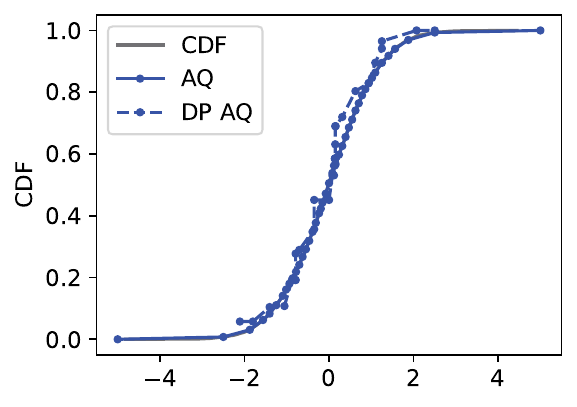}}
\subfloat[$\mathcal{N}(0,1)$]{\includegraphics[height=2.5cm]{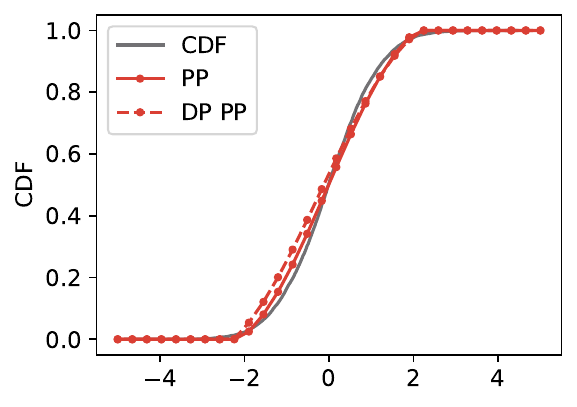}}
\subfloat[$\mathcal{N}(0,1)$]{\includegraphics[height=2.5cm]{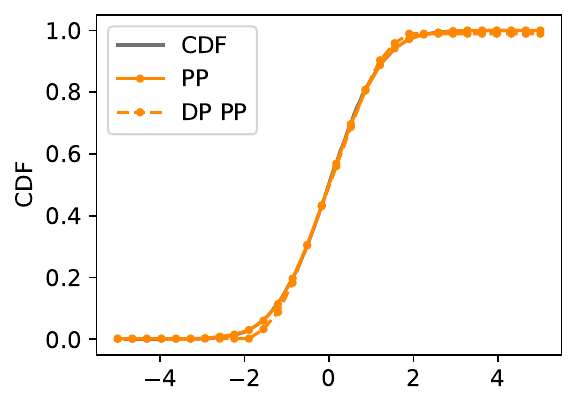}}

\subfloat[Airbnb Data]{\includegraphics[height=2.5cm]{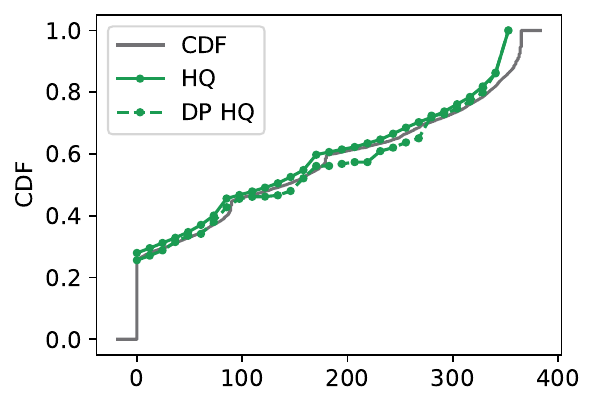}}
\subfloat[Airbnb Data]{\includegraphics[height=2.5cm]{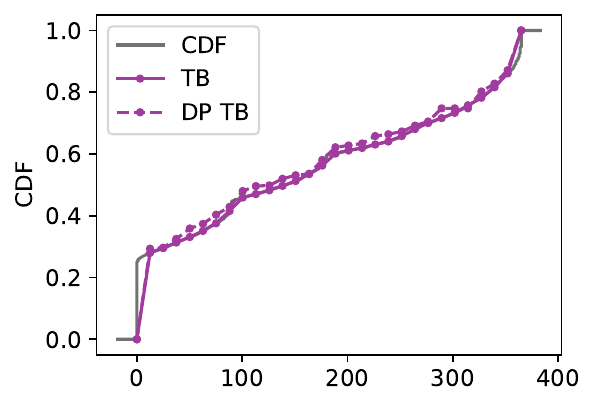}}
\subfloat[Airbnb Data]{\includegraphics[height=2.5cm]{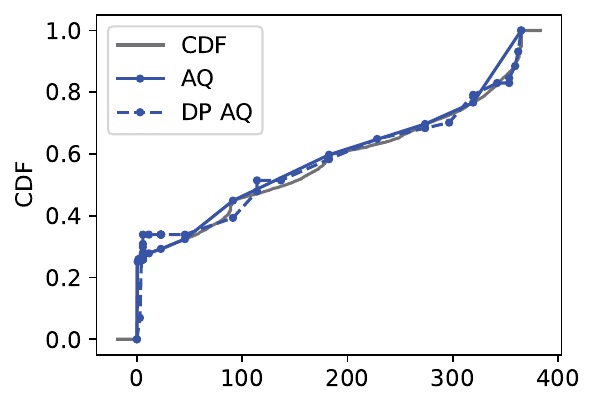}}
\subfloat[Airbnb Data]{\includegraphics[height=2.5cm]{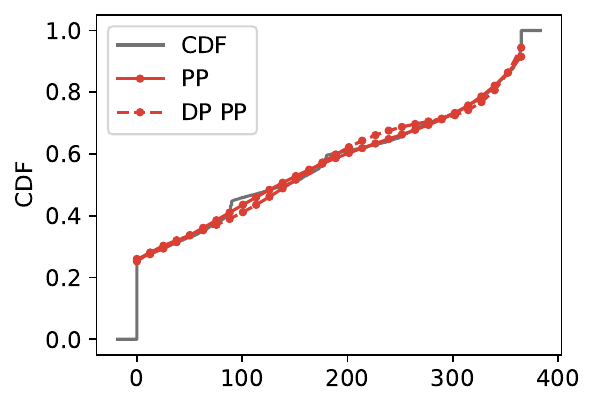}}
\subfloat[Airbnb Data]{\includegraphics[height=2.5cm]{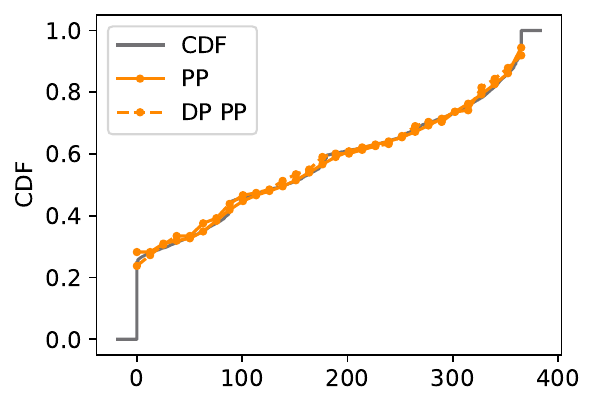}}

\caption{Comparison of \ac{dp} approximation methods for the normal distribution and the \href{https://www.kaggle.com/code/thomaskonstantin/u-s-airbnb-analysis-and-price-prediction/input}{U.S. Airbnb data} under pure differential privacy with $\epsilon=0.1$. (a)-(e) Normal distribution $\mathcal{N}(0,1)$ with $n=10^4$; (f)-(j) U.S. Airbnb Data, where \texttt{availability\_365} records the number of days within a year that a listing is available for booking. The original dataset contains $226030$ records, from which a $10\%$ subsample ($n=22603$) is drawn. For both datasets, \ac{hq} uses $30$ bins, TB uses $1024$ leaves, \ac{aq} runs for $40$ iterations, \ac{pp} employs $6$ basis functions, and \ac{mp} selects $7$ basis functions from a dictionary of $40$ Legendre atoms.}
\label{fig:comparison}
\end{figure*}

To quantitatively compare the performance of different methods, Figure~\ref{fig:distance} and Figure~\ref{fig:appendix_distance} in Appendix~\ref{appendix:figures} present experimental results across various distributions and datasets. For ease of comparison, all experiments are conducted under pure differential privacy with $\delta=0$. In practice, more advanced composition techniques can be applied with a small nonzero $\delta$ to achieve tighter privacy accounting and improve utility. We systematically investigate the hyperparameters of each method using grid search and select representative values for subsequent experiments based on the resulting error between the private CDF and the true CDF. For example, for HQ, we vary the number of bins from $10$ to $100$ in increments of $10$. The relative performance of the baseline methods varies across distributions and evaluation metrics. The \ac{hq} method generally exhibits relatively large errors, while the TB method can perform particularly well. For smooth unimodal distributions, \ac{aq} struggles to capture fine-grained variations, whereas for distributions with pronounced local variations, it achieves better approximation. Overall, our proposed \ac{pp} and \ac{mp} methods demonstrate competitive and consistent performance across varying privacy levels.


\begin{figure*}[ht!]
\centering
\subfloat[$\mathcal{N}(0,1)$]{\includegraphics[height=3.2cm]{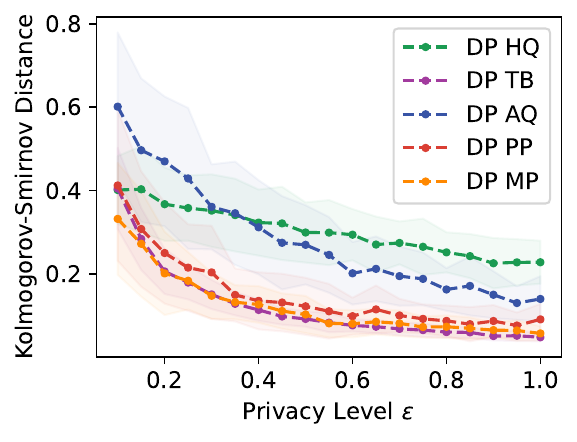}} \hfil
\subfloat[$\mathcal{N}(0,1)$]{\includegraphics[height=3.2cm]{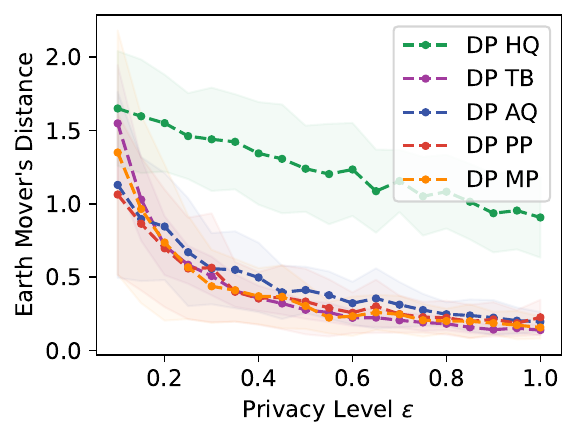}} \hfil
\subfloat[$\mathcal{N}(0,1)$]{\includegraphics[height=3.2cm]{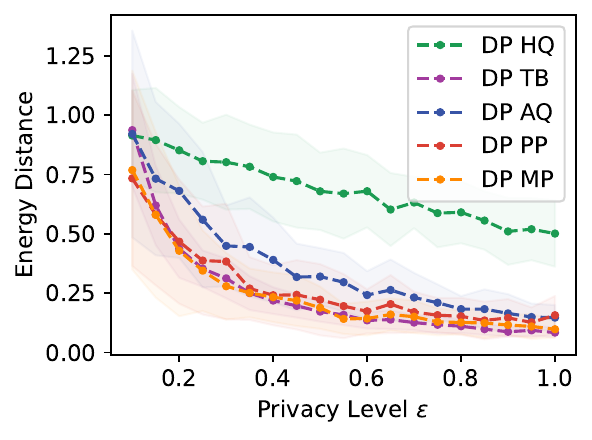}}

\subfloat[Airbnb Data]{\includegraphics[height=3.2cm]{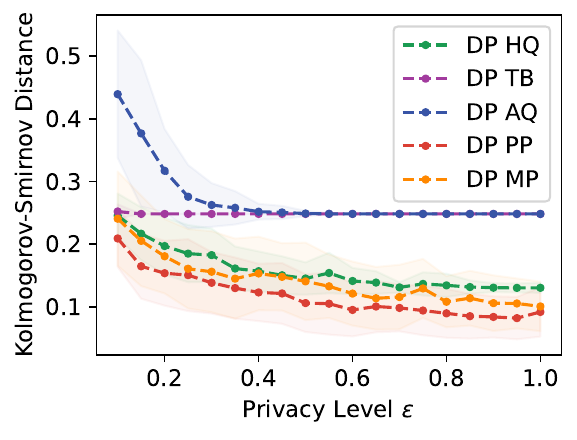}}\hfil
\subfloat[Airbnb Data]{\includegraphics[height=3.2cm]{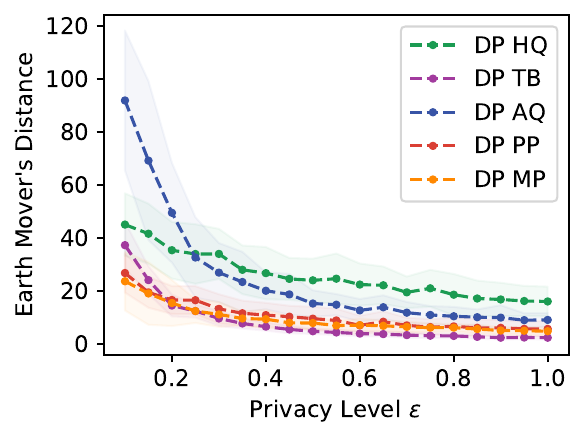}}\hfil
\subfloat[Airbnb Data]{\includegraphics[height=3.2cm]{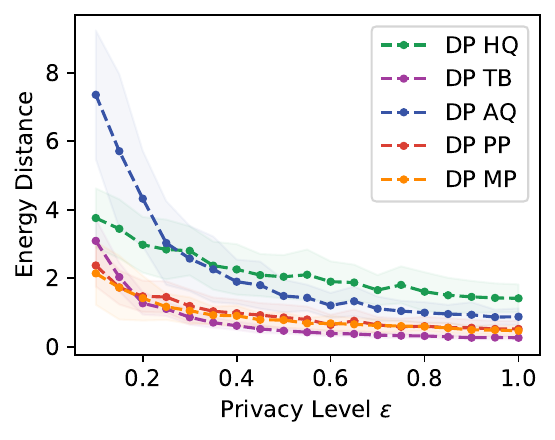}} 

\caption{Comparison of distances between different DP CDF methods and the true CDF under three metrics. (a)-(c) Normal distribution $\mathcal{N}(0,1)$ with $n=10^3$; (d)-(f) U.S. Airbnb Data with $n=2260$. Each experiment was repeated $50$ times under pure differential privacy, where $\epsilon$ denotes the overall privacy budget. For both datasets, HQ uses $40$ bins, TB uses $1024$ leaves, AQ runs for $40$ iterations, PP employs $6$ basis functions, and MP selects $5$ basis functions from a dictionary of $40$ Legendre atoms.}
\label{fig:distance}
\end{figure*}

\begin{figure*}[ht!]
\centering
\subfloat[$\mathcal{N}(0,1)$]{\includegraphics[height=3.2cm]{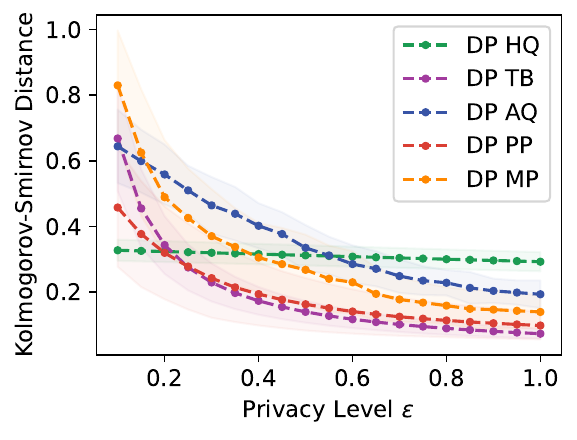}} \hfil
\subfloat[$\mathcal{N}(0,1)$]{\includegraphics[height=3.2cm]{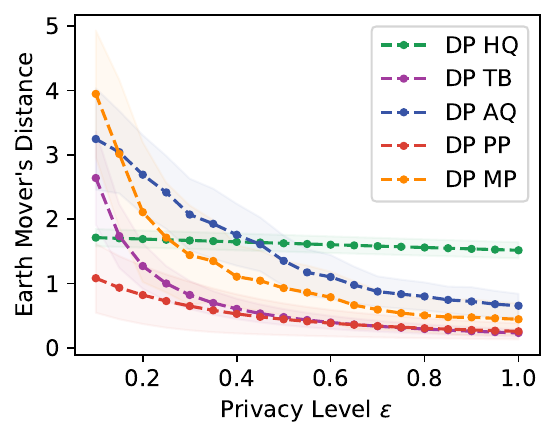}} \hfil
\subfloat[$\mathcal{N}(0,1)$]{\includegraphics[height=3.2cm]{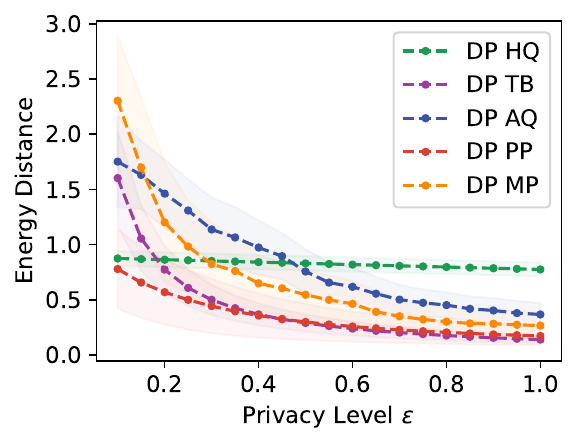}}

\subfloat[Airbnb Data]{\includegraphics[height=3.2cm]{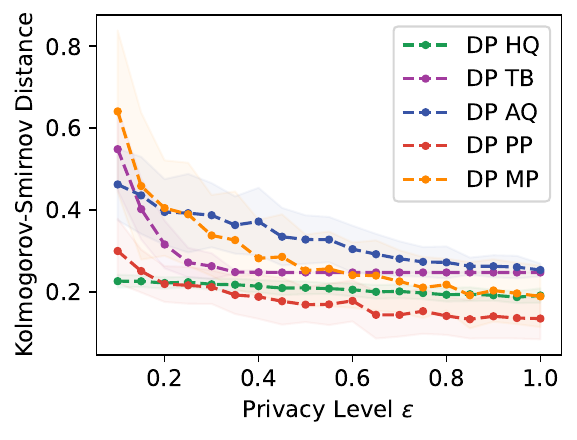}}\hfil
\subfloat[Airbnb Data]{\includegraphics[height=3.2cm]{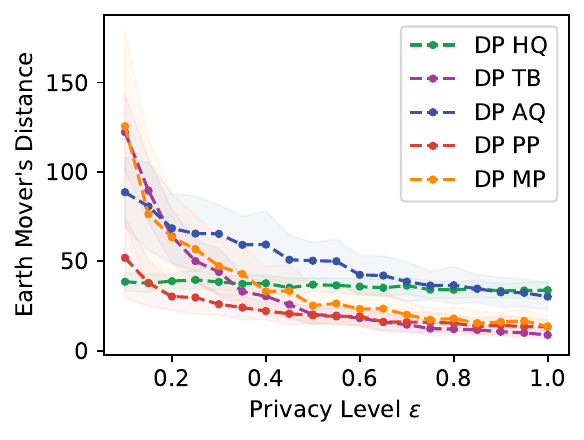}}\hfil
\subfloat[Airbnb Data]{\includegraphics[height=3.2cm]{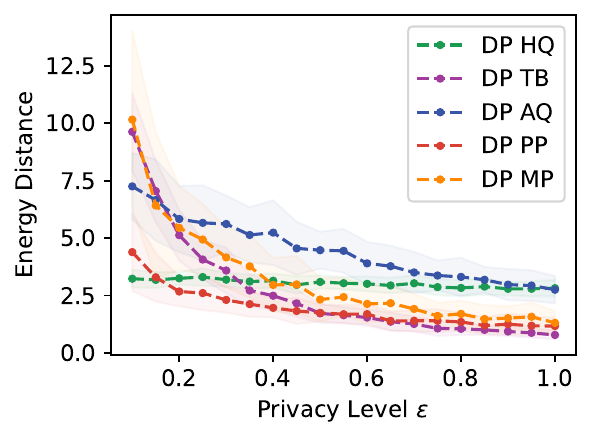}} 

\caption{Comparison of distances between different DP CDF methods and the true CDF under three metrics in a decentralized setting with $10$ sites. (a)-(c) Normal distribution $\mathcal{N}(0,1)$ with $n=200$ samples per site; (d)-(f) U.S. Airbnb Data with $n=226$ samples per site. Each experiment was repeated $50$ times under pure differential privacy, where $\epsilon$ denotes the overall privacy budget. For both datasets, HQ uses $40$ bins, TB uses $1024$ leaves, AQ runs for $40$ iterations, PP employs $6$ basis functions, and MP selects $5$ basis functions from a dictionary of $40$ Legendre atoms.}
\label{fig:decentralized}
\end{figure*}

\begin{figure*}[ht!]
\centering
\subfloat[$\mathcal{N}(0,1)$]{\includegraphics[height=3.2cm]{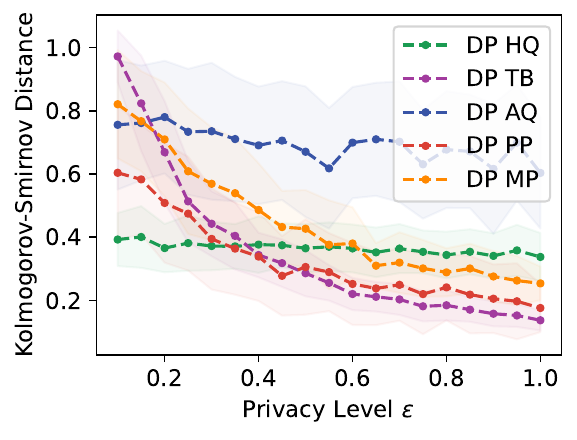}} \hfil
\subfloat[$\mathcal{N}(0,1)$]{\includegraphics[height=3.2cm]{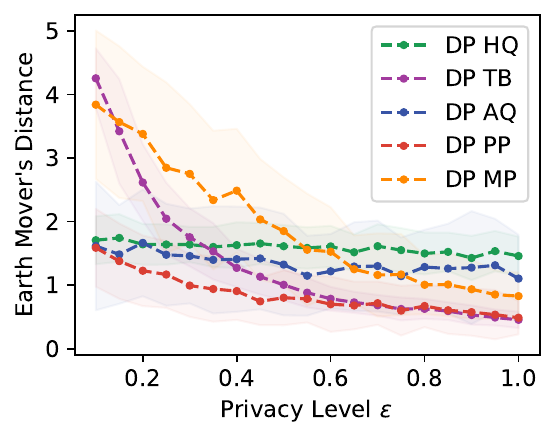}} \hfil
\subfloat[$\mathcal{N}(0,1)$]{\includegraphics[height=3.2cm]{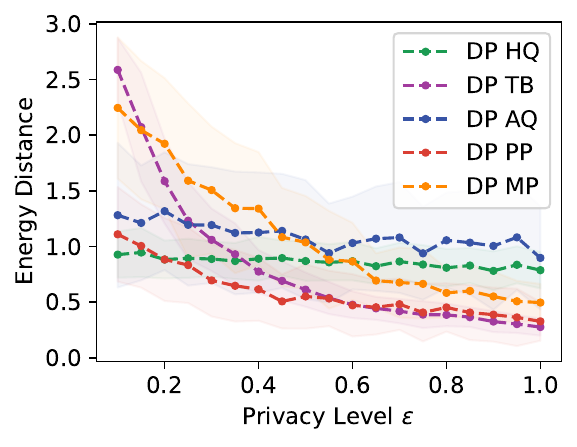}}

\subfloat[Airbnb Data]{\includegraphics[height=3.2cm]{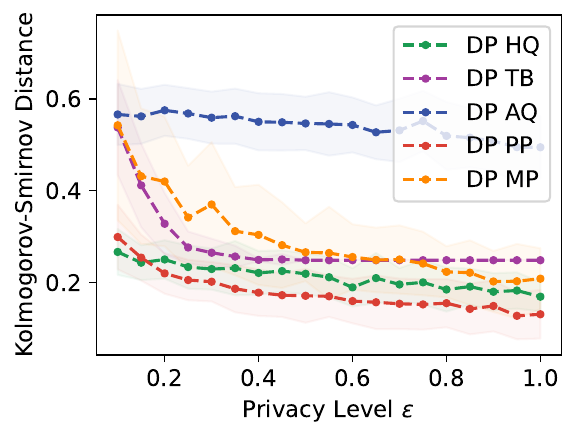}}\hfil
\subfloat[Airbnb Data]{\includegraphics[height=3.2cm]{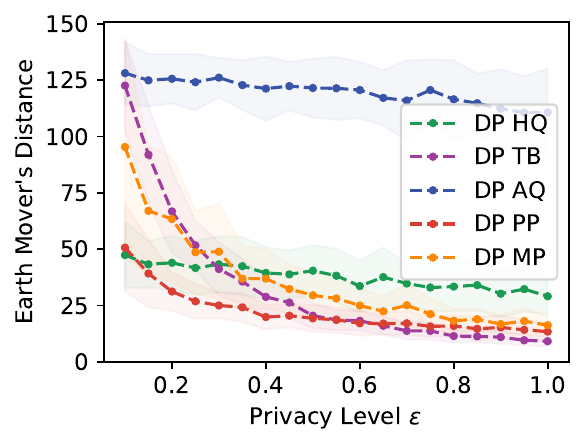}}\hfil
\subfloat[Airbnb Data]{\includegraphics[height=3.2cm]{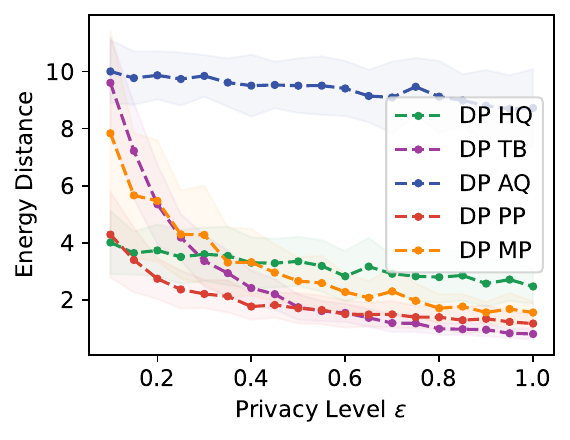}}

\caption{Comparison of distances between different DP CDF methods and the true CDF under three metrics in a newly collected data setting, where the CDF is updated for a total of $10$ rounds. (a)-(c) Normal distribution $\mathcal{N}(0,1)$ with $100$ newly collected samples per round; (d)-(f) U.S. Airbnb Data with $226$ newly collected samples per round. Each experiment was repeated $50$ times under pure differential privacy, where $\epsilon$ denotes the overall privacy budget. For both datasets, HQ uses $40$ bins, TB uses $1024$ leaves, AQ runs for $40$ iterations, PP employs $6$ basis functions, and MP selects $5$ basis functions from a dictionary of $40$ Legendre atoms.}
\label{fig:new_data}
\end{figure*}

\begin{figure*}[ht!]
\centering
\subfloat[]{\includegraphics[height=3.2cm]{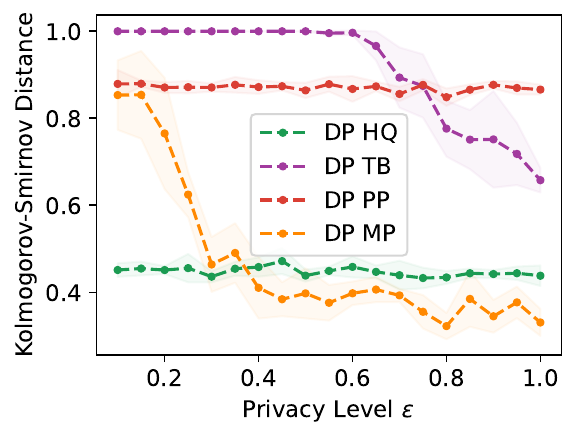}} \hfil
\subfloat[]{\includegraphics[height=3.2cm]{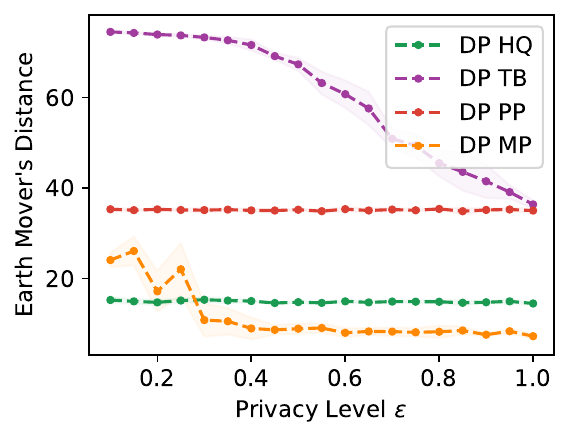}} \hfil
\subfloat[]{\includegraphics[height=3.2cm]{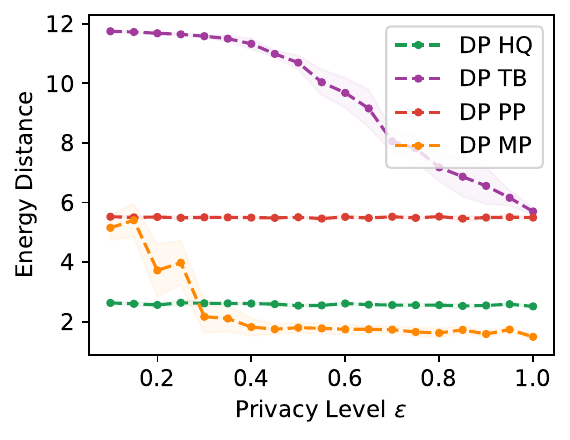}}

\caption{Comparison of distances between different DP CDF methods and the true CDF under three metrics for a two-dimensional standard normal distribution with correlation coefficient $\rho=0.5$. Each experiment was repeated $50$ times with $n=10^3$ under pure differential privacy, where $\epsilon$ denotes the overall privacy budget. HQ uses $40$ bins per dimension, TB uses $512$ leaves per dimension, PP employs $6$ basis functions per dimension, and MP selects $5$ basis functions from a dictionary of $80$ atoms.}
\label{fig:highdim}
\end{figure*}

\begin{algorithm}[htb!]
\caption{Decentralized Differentially Private Legendre Polynomial Projection}
\begin{algorithmic}
\STATE {\textbf{Input:}} Data held by $S$ sites, Legendre polynomial basis $\{P_i\}_{i=0}^{m}$, and privacy parameters $\epsilon$ and $\delta$.
\STATE {\textbf{Output:}} Privacy-preserving coefficients $\{\tilde{c}_i\}_{i=0}^{m}$ and $\dpempcdf$.
\STATE \textbf{Do:}
\begin{enumerate}[label=\arabic*.]
    \item For each site $q \in [1,S]$, let $n_q$ be the site sample size and compute the local moment vector $\boldsymbol{\mu}^{(q)} = [\mu_1^{(q)}, \mu_2^{(q)}, \ldots, \mu_{m+1}^{(q)}]^\top$, where $\mu_j^{(q)} = \frac{1}{n_q} \sum x_k^j$ for $j \in [1,m+1]$.
    \item At each site, add noise to the local moment vector: $\tilde{\boldsymbol{\mu}}^{(q)} = \boldsymbol{\mu}^{(q)} + \mathbf{z}^{(q)}$, where the distribution and scale of $\mathbf{z}^{(q)}$ are determined by the selected privacy parameters and mechanism.
    \item Each site sends $\tilde{\boldsymbol{\mu}}^{(q)}$ to the central server.
    \item The central server computes the weighted aggregate $\tilde{\boldsymbol{\mu}} = \sum_{q=1}^{S} \frac{n_q}{\sum_{r=1}^{S} n_r} \tilde{\boldsymbol{\mu}}^{(q)}$.
    \item Calculate the \ac{dp} coefficients $\tilde{c}_i = \alpha_i \sum_{j=0}^{i} \beta_{i,j} (1-\tilde{\mu}_{j+1})$ for $i \in [0,m]$.
    \item Construct the \ac{dp} \ac{ecdf}: $\dpempcdf(x) = \sum_{i=0}^{m} \tilde{c}_i e_i(x)$.

\end{enumerate}
\STATE {\textbf{End}}
\end{algorithmic}
\label{alg:decentralized_dp_pp}
\end{algorithm}

\begin{algorithm}[htb!]
\caption{Decentralized Differentially Private Approximation via Matching Pursuit}
\begin{algorithmic}
\STATE
\STATE {\textbf{Input:}} Data held by $S$ sites, dictionary $\mathcal{D}=\{\phi_j\}_{j=1}^{m}$, sparsity level $s$, and privacy parameters.
\STATE {\textbf{Output:}} List of privacy-preserving coefficients $\{\tilde{c}_i\}_{i=1}^{s}$ and indices of the corresponding atoms $\{\tilde{I}_i\}_{i=1}^{s}$.
\STATE {\textbf{Initialization:}} $\tilde{F}_0 \leftarrow 0$, $\mathcal{I}\leftarrow\emptyset$.
\STATE {\textbf{For $i=1$ to $s$:}}
\begin{enumerate}[label=\arabic*.]
    \item At each site $q \in [1,S]$, compute the local residual $\tilde{r}_i^{(q)} = F_n^{(q)}-\tilde{F}_{i-1}$.
    \item Each site privately selects an atom from the unselected dictionary: $\tilde{I}_i^{(q)} = \argmax_{j\notin\mathcal{I}} (| \langle \tilde{r}_i^{(q)},\phi_j \rangle | + z_{i,j}^{(q)})$,
    where the distribution and scale of $z_{i,j}^{(q)}$ are determined by the selected privacy parameters and mechanism.
    \item Each site computes a noisy coefficient for the selected atom: $\tilde{c}_i^{(q)} = \langle \tilde{r}_i^{(q)}, \phi_{\tilde{I}_i^{(q)}} \rangle + \xi_i^{(q)}$.
    \item Each site sends $\tilde{I}_i^{(q)}$ and $\tilde{c}_i^{(q)}$ to the central server.
    \item The central server selects the atom receiving the largest number of votes: $\tilde{I}_i = \operatorname{mode} (\tilde{I}_i^{(1)}, \ldots, \tilde{I}_i^{(S)})$.
    \item Let $\mathcal{S}_i = \{q: \tilde{I}_i^{(q)} = \tilde{I}_i \}$.
    The central server aggregates the noisy coefficients reported by the sites in $\mathcal{S}_i$: $\tilde{c}_i = \frac{1}{|\mathcal{S}_i|} \sum_{q\in\mathcal{S}_i} \tilde{c}_i^{(q)}$.
    \item Update the global approximation: $\tilde{F}_i = \tilde{F}_{i-1} + \tilde{c}_i \phi_{\tilde{I}_i}$,
    and update $\mathcal{I} \leftarrow \mathcal{I} \cup \{\tilde{I}_i\}$.
\end{enumerate}
\STATE {\textbf{End}}
\end{algorithmic}
\label{alg:decentralized_dp_mp}
\end{algorithm}

Our methods offer additional benefits in decentralized settings. Suppose $S$ sites collect sensitive data and send privatized information to an untrusted central server for CDF computation. When the \ac{aq} or \ac{mp} method is applied, multiple rounds of communication between the central server and each site are required. Under the same communication constraints, the \ac{mp} method achieves a better approximation than \ac{aq}. In contrast, the \ac{hq}, TB, and \ac{pp} methods require only a single round of communication, in which each site sends its noisy histogram, noisy tree representation, or noisy values of $\mu_i$, $i \in [1,m+1]$, respectively, to the central server. The decentralized implementations of our \ac{pp} and \ac{mp} methods are summarized in Algorithm~\ref{alg:decentralized_dp_pp} and Algorithm~\ref{alg:decentralized_dp_mp}, respectively. Figure~\ref{fig:decentralized} and Figure~\ref{fig:appendix_decentralized} in Appendix~\ref{appendix:figures} demonstrate the performance of different methods in the decentralized setting. Our \ac{pp} and \ac{mp} methods perform particularly well for smooth distributions and achieve competitive performance across the evaluated settings. Considering its competitive approximation performance and simple one-round communication procedure, the \ac{pp} method is a strong candidate for decentralized settings.

Our methods also offer practical benefits when incorporating newly collected data. Specifically, for \ac{pp}, we compute the noisy moment $\frac{1}{n_{\text{new}}} \sum_{k=1}^{n_{\text{new}}} x_k^j$ for the newly collected data and combine it with the previously stored noisy moment $\frac{1}{n_{\text{old}}} \sum_{k=1}^{n_{\text{old}}} x_k^j$ to obtain a noisy estimate of the updated moment $\frac{1}{n_{\text{new}}+n_{\text{old}}} \sum_{k=1}^{n_{\text{old}}+n_{\text{new}}} x_k^j$. Similarly, for \ac{mp}, we compute the residual between the CDF of the newly collected data and the previously stored CDF approximation, approximate this residual using the private matching pursuit procedure, and use the resulting approximation to update the stored CDF approximation. Therefore, neither \ac{pp} nor \ac{mp} requires access to the previously collected raw data. In contrast, the \ac{aq} method requires revisiting the old data to update the CDF approximation, thereby increasing the privacy cost. The \ac{hq} and TB methods can also incorporate newly collected data without revisiting the old data as long as their underlying partitions remain fixed. Figure~\ref{fig:new_data} and Figure~\ref{fig:appendix_new_data} in Appendix~\ref{appendix:figures} demonstrate the performance of different methods when incorporating newly collected data. The TB method performs well in this setting, while \ac{pp} also achieves strong performance and generally outperforms \ac{mp}. It should be noted that the strong performance of TB in these experiments assumes a fixed tree structure. If the tree structure changes, previously collected data may need to be revisited, leading to additional privacy costs and potentially reduced utility. In addition, \ac{pp} offers practical advantages. Reconstructing the DP CDF requires storing only a few values, determined by the number of basis functions $m$, whereas other methods typically need to store many more values depending on the bin count, tree size, iteration number, or the entire dictionary. Furthermore, \ac{pp} directly provides moment information without extra computation.

While \ac{pp} offers advantages in decentralized and newly collected data settings, \ac{mp} is particularly well-suited to high-dimensional CDF approximation. As the dimensionality increases, the number of basis functions required by \ac{pp} grows rapidly, making accurate approximation increasingly challenging. In contrast, \ac{mp} selects only a small number of informative atoms from a larger dictionary, allowing it to exploit sparse representations without using all candidate basis functions. \ac{hq} and TB can also be extended to high-dimensional settings, however, the number of histogram bins or tree nodes grows rapidly with dimensionality, leading to increased estimation and privacy costs. Moreover, the \ac{aq} method considered in this work does not directly extend to multivariate CDF estimation due to the lack of a natural total ordering in multiple dimensions. Figure~\ref{fig:highdim} shows that \ac{mp} performs particularly well as the privacy budget increases.

\begin{figure*}[ht!]
\centering
\subfloat[$\mathcal{N}(0,1)$]{\includegraphics[height=3cm]{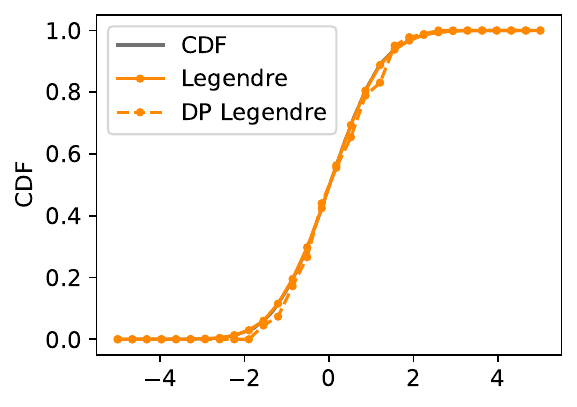}} \hfil
\subfloat[$\mathcal{N}(0,1)$]{\includegraphics[height=3cm]{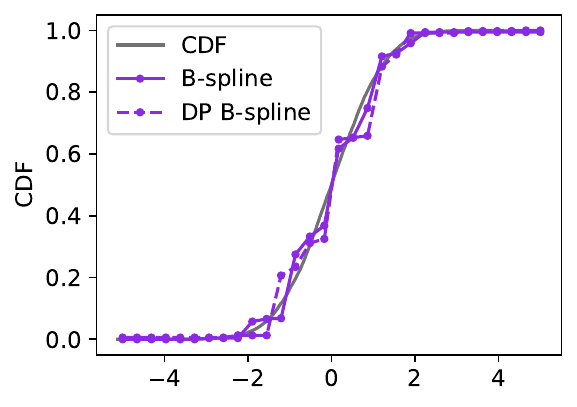}} \hfil
\subfloat[$\mathcal{N}(0,1)$]{\includegraphics[height=3cm]{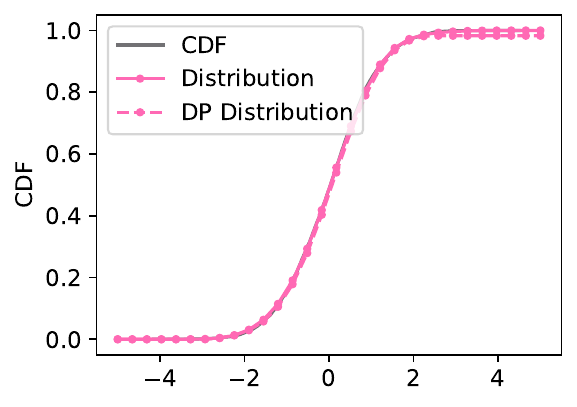}}
\caption{Comparison of CDF reconstruction using different dictionaries with parameters $n=10^4$, $\epsilon=0.5$, $\delta=n^{-3/2}$, and sparsity level $s=30$. (a) Dictionary constructed from $200$ Legendre polynomials; (b) Dictionary constructed from $109$ B-spline functions of degree 0 and 1; (c) Dictionary constructed from $400$ normal CDFs with varying means and variances.}
\label{fig:comparison_dl}
\end{figure*}

\begin{figure*}[ht!]
\centering
\subfloat[$\mathcal{N}(0,1)$]{\includegraphics[height=3.2cm]{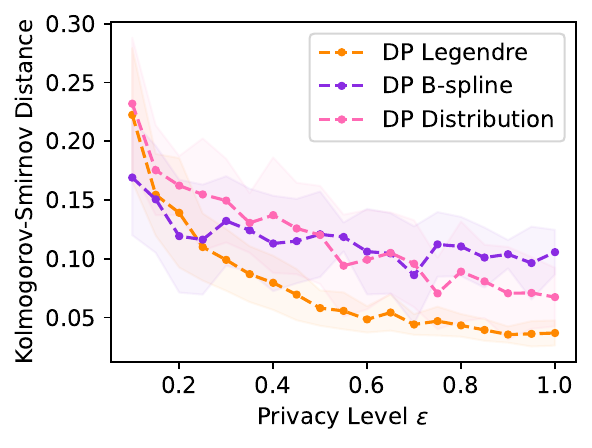}} \hfil
\subfloat[$\mathcal{N}(0,1)$]{\includegraphics[height=3.2cm]{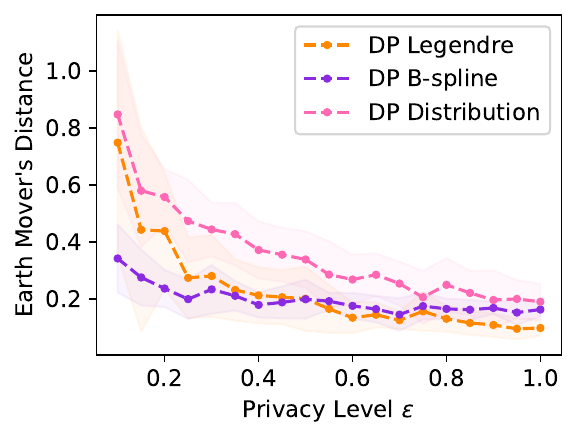}} \hfil
\subfloat[$\mathcal{N}(0,1)$]{\includegraphics[height=3.2cm]{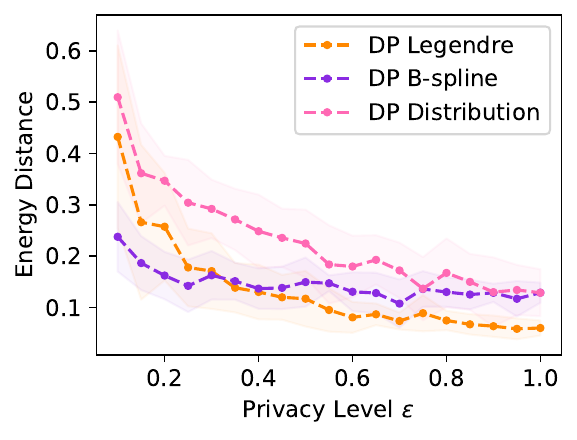}}
\caption{Comparison of the approximation distances between the DP CDF obtained using different dictionaries and the true CDF. Each experiment was repeated 50 times with $n=10^4$, $\delta=n^{-3/2}$, and sparsity level $s=30$.}
\label{fig:distance_dl}
\end{figure*}

\subsection{Exploration of Dictionary Compositions}
\label{sec:dict_compose}
In this section, we explore three representative families of functions for constructing the dictionary: Legendre polynomials, B-splines, and normal distribution CDFs with varying means and variances. These choices illustrate distinct structural properties: Legendre polynomials are orthogonal, B-splines are non-orthogonal, and normal distribution CDFs are non-orthogonal and inherently monotonic. The dictionary is not limited to these three families and may be constructed from arbitrary functions, such as empirical CDFs derived from real-world data.

Figure~\ref{fig:comparison_dl} and Figure~\ref{fig:appendix_comparison_dl} in Appendix~\ref{appendix:figures} present the reconstruction results. These results demonstrate that, as long as the dictionary is sufficiently rich, the CDF can be well approximated. Nevertheless, the approximation quality varies across different dictionaries. B-splines perform particularly well in approximating complex multimodal distributions, as their local support enables flexible adaptation to local variations. In contrast, dictionaries constructed from normal distribution CDFs perform less effectively in such cases. This limitation arises because normal CDFs are inherently smooth and S-shaped, constrained to represent unimodal and symmetric distributions. Even when combined, they lack the flexibility to capture the step-like structures associated with multiple peaks. Moreover, their global support makes local adjustments inefficient, often requiring a large number of atoms to approximate complex patterns.

Figure~\ref{fig:distance_dl} and Figure~\ref{fig:appendix_distance_dl} in Appendix~\ref{appendix:figures} show the approximation distance between the DP CDF and the true CDF. These results further confirm that, for complex distributions, dictionaries constructed from B-splines provide a more effective representation than the other two types.

\section{Future Directions}

The main idea of this work lies in projecting the CDF into a predefined function space and applying differential privacy to the coefficients to protect the estimated CDF. Building on polynomial projection, we extend the framework to obtain sparse approximations from arbitrary function spaces constructed by dictionaries. Several promising directions remain open. Our current framework is primarily developed in the $L^2$ space, and extending the analysis to more general $L^p$ spaces may provide broader insights. For \ac{mp} with orthonormal dictionaries, the current error bound may be loose, motivating the development of a tighter bound. It is also important to further investigate non-orthogonal dictionaries and develop corresponding theoretical guarantees. Although we have explored the extension to multivariate distributions, a more comprehensive analysis remains an important direction for future work. Another direction is to systematically investigate different post-processing techniques and their effects on the utility of the reconstructed CDF. It is also worthwhile to explore theoretical connections with robust statistics~\cite{HuberRonchetti:09robust}, whose modeling principles align with the foundations of differential privacy~\cite{dwork2009differential,SlavkovicM:21robust,AvellaMedina:21robust}, potentially leading to universally valid privacy guarantees.

\section*{Acknowledgments}

The authors would like to express their sincere gratitude to Vince D. Calhoun and Sergey Plis from the TReNDS Center for insightful discussions and constructive suggestions throughout this work. 
The authors also acknowledge the use of ChatGPT for checking for typographical errors, grammar errors, and suggested rephrasing of the manuscript.



\bibliographystyle{IEEEtran}
\bibliography{references}

@inproceedings{tao2025differentially,
  title={Differentially private distribution estimation using functional approximation},
  author={Tao, Ye and Sarwate, Anand D.},
  booktitle={IEEE International Conference on Acoustics, Speech and Signal Processing},
  pages={1--5},
  doi = {10.1109/ICASSP49660.2025.10890461},
  url = {https://doi.org/10.1109/ICASSP49660.2025.10890461},
  year={2025},
}

@inproceedings{DworkMNS:06sensitivity,
	address = {Berlin, Heidelberg},
	author = {Cynthia Dwork and Frank McSherry and Kobbi Nissim and Adam Smith},
	booktitle = {Theory of Cryptography},
	doi = {10.1007/11681878_14},
	editor = {Shai Halevi and Tal Rabin},
	month = {March 4--7},
	pages = {265--284},
	publisher = {Springer},
	series = {Lecture Notes in Computer Science},
	title = {Calibrating Noise to Sensitivity in Private Data Analysis},
	volume = {3876},
	year = {2006},
	url = {http://dx.doi.org/10.1007/11681878_14}}

@inproceedings{dwork2009differential,
	address = {New York, NY, USA},
	author = {Cynthia Dwork and Jing Lei},
	booktitle = {Proceedings of the 41st Annual ACM Symposium on Theory of Computing},
	doi = {10.1145/1536414.1536466},
	pages = {371--380},
	publisher = {ACM},
	title = {Differential Privacy and Robust Statistics},
	year = {2009},
	url = {http://dx.doi.org/10.1145/1536414.1536466}}

@inproceedings{liu2019differential,
author = {Liu, Ao and Xia, Lirong and Duchowski, Andrew and Bailey, Reynold and Holmqvist, Kenneth and Jain, Eakta},
title = {Differential privacy for eye-tracking data},
year = {2019},
isbn = {9781450367097},
publisher = {Association for Computing Machinery},
address = {New York, NY, USA},
url = {https://doi.org/10.1145/3314111.3319823},
doi = {10.1145/3314111.3319823},
booktitle = {Proceedings of the 11th ACM Symposium on Eye Tracking Research \& Applications},
articleno = {28},
numpages = {10},
location = {Denver, Colorado},
series = {ETRA '19}
}

@inproceedings{tao2024privacy,
  title={Privacy-preserving visualization of brain functional network connectivity},
  author={Tao, Ye and Sarwate, Anand D and Panta, Sandeep and Plis, Sergey and Calhoun, Vince D},
  booktitle={IEEE International Symposium on Biomedical Imaging},
  pages={1--5},
  url = {https://doi.org/10.1109/ISBI56570.2024.10635222},
  doi = {10.1109/ISBI56570.2024.10635222},
  year={2024},
}

@article{dvoretzky1956asymptotic,
  title={Asymptotic minimax character of the sample distribution function and of the classical multinomial estimator},
  author={Dvoretzky, Aryeh and Kiefer, Jack and Wolfowitz, Jacob},
  journal={The Annals of Mathematical Statistics},
  volume={27},
  number={3},
  pages={642--669},
  year={1956},
  publisher={JSTOR},
  url={https://www.jstor.org/stable/2237374}
}

@online{mckenna2021estimating,
  title= {Estimating a cumulative distribution function with differential privacy},
  author= {Alex McKenna},
  url={https://medium.com/sarus/estimating-a-cumulative-distribution-function-with-differential-privacy-54433fab45c7}
}

@article{kroll2021density,
  title={On density estimation at a fixed point under local differential privacy},
  author={Kroll, Martin},
  volume = {15},
  pages = {1783--1813},
  year={2021},
  number = {1},
  journal = {Electronic Journal of Statistics},
  publisher = {Institute of Mathematical Statistics and Bernoulli Society},
  url = {https://doi.org/10.1214/21-EJS1830},
  doi = {10.1214/21-EJS1830}
}

@InProceedings{wagner2023fast,
  title = 	 {Fast Private Kernel Density Estimation via Locality Sensitive Quantization},
  author =       {Wagner, Tal and Naamad, Yonatan and Mishra, Nina},
  booktitle = 	 {Proceedings of the 40th International Conference on Machine Learning},
  pages = 	 {35339--35367},
  year = 	 {2023},
  editor = 	 {Krause, Andreas and Brunskill, Emma and Cho, Kyunghyun and Engelhardt, Barbara and Sabato, Sivan and Scarlett, Jonathan},
  volume = 	 {202},
  series = 	 {Proceedings of Machine Learning Research},
  month = 	 {23--29 Jul},
  publisher =    {PMLR},
  url = 	 {https://proceedings.mlr.press/v202/wagner23a.html},
}

@misc{liu2024differentially,
      title={Differentially Private Kernel Density Estimation}, 
      author={Erzhi Liu and Jerry Yao-Chieh Hu and Alex Reneau and Zhao Song and Han Liu},
      year={2025},
      eprint={2409.01688},
      archivePrefix={arXiv},
      primaryClass={cs.DS},
      url={https://arxiv.org/abs/2409.01688}, 
}

@article{cooley1965algorithm,
  title={{An algorithm for the machine calculation of complex Fourier series}},
  author={Cooley, James W and Tukey, John W},
  journal={Mathematics of Computation},
  volume={19},
  month = 4,
  number={90},
  pages={297--301},
  year={1965},
  publisher={JSTOR},
  url = {https://doi.org/10.2307/2003354}
}

@book{oppenheim1999discrete,
  title={Discrete-Time Signal Processing (2nd Edition)},
  author={Alan V Oppenhein and Ronald W Schafer and John R Buck},
  year={1999},
  address = {Upper Saddle River, NJ, USA},
  publisher={Prentice Hall}
}

@book{tolstov2012fourier,
  title={Fourier Series},
  author={Tolstov, Georgi P},
  year={1962},
  address = {Englewood Cliffs, NJ, USA},
  publisher={Prentice Hall}
}

@book{luenberger1997optimization,
  title={Optimization by Vector Space Methods},
  author={Luenberger, David G},
  year={1969},
  address={New York},
  publisher={John Wiley \& Sons}
}

@book{kantorovich2014functional,
  title={Functional Analysis},
  author={Kantorovich, Leonid Vitalevich and Akilov, Gleb Pavlovich},
  year={1982},
  address = {Elmsford, NY, USA},
  publisher={Pergamon Press}
}

@book{yosida2012functional,
  title={Functional Analysis},
  author={Yosida, K{\^o}saku},
  edition = {6th},
  series = {Classics in Mathematics},
  volume={123},
  year={1980},
  address = {Berlin},
  publisher={Springer},
  doi = {10.1007/978-3-642-61859-8},
  url = {https://doi.org/10.1007/978-3-642-61859-8}
}

@inproceedings{alda2017bernstein,
	author = {Francesco Ald{\`{a}} and Benjamin I. P. Rubinstein},
	bibsource = {dblp computer science bibliography, https://dblp.org},
	booktitle = {Proceedings of the 31st {AAAI} Conference on Artificial Intelligence},
	editor = {Satinder Singh and Shaul Markovitch},
	pages = {1705--1711},
	publisher = {{AAAI} Press},
	title = {{The Bernstein mechanism: Function release under differential privacy}},
	year = {2017},
	url = {https://doi.org/10.1609/aaai.v31i1.10884 }}

@article{zhang2012functional,
	Author = {Zhang, J. and Zhang, Z. and Xiao, X. and Yang, Y. and Winslett, M.},
	Doi = {10.14778/2350229.2350253},
	Issn = {2150-8097},
	Issue_Date = {July 2012},
	Journal = {Proc. VLDB Endow.},
	Month = jul,
	Number = {11},
	Numpages = {12},
	Pages = {1364--1375},
	Title = {Functional mechanism: Regression analysis under differential privacy},
	Volume = {5},
	Year = {2012},
	url= {http://dx.doi.org/10.14778/2350229.2350253}}

@article{DworkRoth,
  title={The algorithmic foundations of differential privacy},
  author={Dwork, Cynthia and Roth, Aaron and others},
  journal={Foundations and Trends{\textregistered} in Theoretical Computer Science},
  volume={9},
  number={3--4},
  pages={211--407},
  year={2014},
  publisher={Now Publishers, Inc.},
  doi = {10.1561/0400000042},
  url = {https://doi.org/10.1561/0400000042}
}

@article{dwork2014algorithmic,
	address = {Hanover, MA, USA},
	author = {Dwork, Cynthia and Roth, Aaron},
	doi = {10.1561/0400000042},
	issue_date = {August 2014},
	journal = {Foundations and Trends in Theoretical Computer Science},
	month = 8,
	number = {3--4},
	numpages = {197},
	pages = {211--407},
	publisher = {Now Publishers Inc.},
	title = {The Algorithmic Foundations of Differential Privacy},
	volume = {9},
	year = {2014},
	url = {https://doi.org/10.1561/0400000042}}

@inproceedings{mcsherry2007mechanism,
	author = {Frank McSherry and Kunal Talwar},
	booktitle = {48th Annual IEEE Symposium on Foundations of Computer Science},
	doi = {10.1109/FOCS.2007.41},
	month = {10},
	pages = {94--103},
	title = {Mechanism Design via Differential Privacy},
	year = {2007},
	url = {http://dx.doi.org/10.1109/FOCS.2007.41}}

@article{geng2016optimal,
author = {Geng, Quan and Viswanath, Pramod},
title = {The Optimal Noise-Adding Mechanism in Differential Privacy},
year = {2016},
issue_date = {Feb. 2016},
publisher = {IEEE Press},
volume = {62},
number = {2},
issn = {0018-9448},
url = {https://doi.org/10.1109/TIT.2015.2504967},
doi = {10.1109/TIT.2015.2504967},
journal = {IEEE Transactions on Information Theory},
month = feb,
pages = {925–951},
numpages = {27}
}

@article{geng2016optimal2,
author = {Geng, Quan and Viswanath, Pramod},
title = {Optimal Noise Adding Mechanisms for Approximate Differential Privacy},
year = {2016},
issue_date = {Feb. 2016},
publisher = {IEEE Press},
volume = {62},
number = {2},
issn = {0018-9448},
url = {https://doi.org/10.1109/TIT.2015.2504972},
doi = {10.1109/TIT.2015.2504972},
journal = {IEEE Transactions on Information Theory},
month = feb,
pages = {952–969},
numpages = {18}
}

@article{liu2018generalized,
  title={Generalized {Gaussian} mechanism for differential privacy},
  author={Liu, Fang},
  journal={IEEE Transactions on Knowledge and Data Engineering},
  volume={31},
  number={4},
  pages={747--756},
  year={2018},
  doi = {10.1109/TKDE.2018.2845388},
  url = {https://doi.org/10.1109/TKDE.2018.2845388}
}

@inproceedings{alghamdi2022cactus,
  title={Cactus mechanisms: Optimal differential privacy mechanisms in the large-composition regime},
  author={Alghamdi, Wael and Asoodeh, Shahab and Calmon, Flavio P and Kosut, Oliver and Sankar, Lalitha and Wei, Fei},
  booktitle={IEEE International Symposium on Information Theory},
  pages={1838--1843},
  year={2022},
}

@article{geng2015staircase,
	author = {Geng, Quan and Kairouz, Peter and Oh, Sewoong and Viswanath, Pramod},
	doi = {10.1109/JSTSP.2015.2425831},
	journal = {IEEE Journal of Selected Topics in Signal Processing},
	month = {10},
	number = {7},
	pages = {1176-1184},
	title = {The Staircase Mechanism in Differential Privacy},
	volume = {9},
	year = {2015},
	url = {https://doi.org/10.1109/JSTSP.2015.2425831}}

@article{kairouz2014extremal,
	author = {Kairouz, Peter and Oh, Sewoong and Viswanath, Pramod},
	journal = {Journal of Machine Learning Research},
	month = {1},
	number = {17},
	pages = {1--51},
	title = {Extremal Mechanisms for Local Differential Privacy},
	volume = {17},
	year = {2016},
	url = {https://www.jmlr.org/papers/v17/15-135.html}}

@InProceedings{balle2018improving,
  title = 	 {Improving the {G}aussian Mechanism for Differential Privacy: Analytical Calibration and Optimal Denoising},
  author =       {Balle, Borja and Wang, Yu-Xiang},
  booktitle = 	 {Proceedings of the 35th International Conference on Machine Learning},
  pages = 	 {394--403},
  year = 	 {2018},
  editor = 	 {Dy, Jennifer and Krause, Andreas},
  volume = 	 {80},
  series = 	 {Proceedings of Machine Learning Research},
  month = 	 {10--15 Jul},
  publisher =    {PMLR},
  url = 	 {https://proceedings.mlr.press/v80/balle18a.html}
}

@article{kairouz2015composition,
	author = {Kairouz, Peter and Oh, Sewoong and Viswanath, Pramod},
	doi = {10.1109/TIT.2017.2685505},
	journal = {IEEE Transactions on Information Theory},
	month = {6},
	number = {6},
	pages = {4037-4049},
	title = {The Composition Theorem for Differential Privacy},
	volume = {63},
	year = {2017},
	url = {https://doi.org/10.1109/TIT.2017.2685505}}

@inproceedings{mironov2017renyi,
	author = {Mironov, Ilya},
	booktitle = {IEEE 30th Computer Security Foundations Symposium},
	doi = {10.1109/CSF.2017.11},
	month = {8},
	pages = {263--275},
	title = {R\'{e}nyi Differential Privacy},
	year = {2017},
	url = {https://doi.org/10.1109/CSF.2017.11}}

@article{dong2019gaussian,
	author = {Jinshuo Dong and Aaron Roth and Weijie Su},
	doi = {doi.org/10.1111/rssb.12454},
	journal = {Journal of the Royal Statistical Society: Series B},
	month = {2},
	number = {1},
	pages = {3--37},
	title = {Gaussian Differential Privacy},
	volume = {84},
	year = {2021},
	url = {https://doi.org/doi.org/10.1111/rssb.12454}}

@inproceedings{gopi2021numerical,
	author = {Gopi, Sivakanth and Lee, Yin Tat and Wutschitz, Lukas},
	booktitle = {Advances in Neural Information Processing Systems},
	editor = {M. Ranzato and A. Beygelzimer and Y. Dauphin and P.S. Liang and J. Wortman Vaughan},
	pages = {11631--11642},
	publisher = {Curran Associates, Inc.},
	title = {Numerical Composition of Differential Privacy},
	url = {https://proceedings.neurips.cc/paper_files/paper/2021/file/6097d8f3714205740f30debe1166744e-Paper.pdf},
	volume = {34},
	year = {2021}}

@inproceedings{murtagh2015complexity,
	address = {Berlin, Heidelberg},
	author = {Murtagh, Jack and Vadhan, Salil},
	booktitle = {Theory of Cryptography},
	doi = {10.1007/978-3-662-49096-9_7},
	editor = {Kushilevitz, Eyal and Malkin, Tal},
	isbn = {978-3-662-49096-9},
	pages = {157--175},
	publisher = {Springer Berlin Heidelberg},
	title = {The Complexity of Computing the Optimal Composition of Differential Privacy},
	url = {https://doi.org/10.1007/978-3-662-49096-9_7},
	year = {2016}
    }

@article{beaudry2011intuitive,
  author = {Normand J. Beaudry and Renato Renner},
  title = {An intuitive proof of the data processing inequality},
  journal = {Quantum Information \& Computation},
  volume = {12},
  number = {5-6},
  pages = {432--441},
  year = {2012},
  url = {https://doi.org/10.26421/QIC12.5-6-4},
  doi = {10.26421/QIC12.5-6-4},
}

@book{boyd2001chebyshev,
  title={Chebyshev and Fourier Spectral Methods, Second Edition (Revised)},
  author={Boyd, John P},
  year={2001},
  address = {Mineola, NY, USA},
  publisher={Dover}
}

@book{wainwright2019high,
  title={High-Dimensional Statistics: A Non-Asymptotic Viewpoint},
  author={Wainwright, Martin J},
  year={2019},
  address = {Cambridge},
  publisher={Cambridge University Press}
}

@article{lavine1995nonparametric,
  title={A nonparametric {Bayes} method for isotonic regression},
  author={Lavine, Michael and Mockus, A},
  journal={Journal of Statistical Planning and Inference},
  volume={46},
  number={2},
  pages={235--248},
  year={1995},
  publisher={Elsevier},
  url = {https://doi.org/10.1016/0378-3758(94)00106-6},
  doi = {10.1016/0378-3758(94)00106-6}
}

@article{mallat1993matching,
  title={{Matching pursuits with time-frequency dictionaries}},
  author={Mallat, St{\'e}phane G and Zhang, Zhifeng},
  journal={IEEE Transactions on Signal Processing},
  volume={41},
  number={12},
  pages={3397--3415},
  year={1993},
  publisher={IEEE},
  url = {https://doi.org/10.1109/78.258082},
  doi = {10.1109/78.258082}
}

@article{ding2021permute,
  title={{The permute-and-flip mechanism is identical to report-noisy-max with exponential noise}},
  author={Ding, Zeyu and Kifer, Daniel and Steinke, Thomas and Wang, Yuxin and Xiao, Yingtai and Zhang, Danfeng and others},
  journal={arXiv preprint arXiv:2105.07260},
  year={2021}
}

@article{lilliefors1967kolmogorov,
  title={{On the {K}olmogorov-{S}mirnov test for normality with mean and variance unknown}},
  author={Lilliefors, Hubert W},
  journal={Journal of the American statistical Association},
  volume={62},
  number={318},
  pages={399--402},
  year={1967},
  publisher={Taylor \& Francis},
  url = {https://doi.org/10.1080/01621459.1967.10482916},
  doi = {10.1080/01621459.1967.10482916}
}

@inproceedings{rubner1998metric,
  title={A metric for distributions with applications to image databases},
  author={Rubner, Yossi and Tomasi, Carlo and Guibas, Leonidas J},
  booktitle={IEEE 6th International Conference on Computer Vision},
  pages={59--66},
  year={1998}
}

@article{rizzo2016energy,
  title={Energy distance},
  author={Rizzo, Maria L and Sz{\'e}kely, G{\'a}bor J},
  journal={Wiley Interdisciplinary Reviews: Computational Statistics},
  volume={8},
  number={1},
  pages={27--38},
  year={2016},
  publisher={Wiley Online Library}
}

@book{HuberRonchetti:09robust,
	address = {New York, New York, USA},
	author = {Peter J. Huber and Elvezio M. Ronchetti},
	doi = {10.1002/9780470434697},
	publisher = {John Wiley \& Sons, Inc},
	series = {Wiley Series in Probability and Statistics},
	title = {Robust Statistics, Second Edition},
	year = {2009},
	url = {https://doi.org/10.1002/9780470434697}}

@techreport{SlavkovicM:21robust,
	author = {Aleksandra Slavkovi\'{c} and Roberto Molinari},
	copyright = {Creative Commons Attribution 4.0 International},
	doi = {10.48550/ARXIV.2108.08266},
	institution = {ArXiV},
	month = {8},
	number = {arXiv:2108.08266 [cs.CR]},
	publisher = {arXiv},
	title = {Perturbed {M}-{E}stimation: A Further Investigation of Robust Statistics for Differential Privacy},
	year = {2021},
	url = {https://arxiv.org/abs/2108.08266}}

@article{AvellaMedina:21robust,
	author = {Marco Avella-Medina},
	doi = {10.1080/01621459.2019.1700130},
	journal = {Journal of the American Statistical Association},
	number = {534},
	pages = {969--983},
	publisher = {Taylor & Francis},
	title = {Privacy-Preserving Parametric Inference: A Case for Robust Statistics},
	volume = {116},
	year = {2021},
	url = {https://doi.org/10.1080/01621459.2019.1700130}}

@book{carslaw1921introduction,
  title={Introduction to the Theory of Fourier's Series and Integrals},
  author={Carslaw, Horatio Scott},
  volume={1},
  year={1921},
  address = {London},
  publisher={Macmillan and Company}
}

@book{prautzsch2002bezier,
  title={B{\'e}zier and B-spline Techniques},
  author={Prautzsch, Hartmut and Boehm, Wolfgang and Paluszny, Marco},
  year={2002},
  url = {https://doi.org/10.1007/978-3-662-04919-8},
  address = {Berlin, Heidelberg},
  publisher={Springer}
}

@article{QardajiYL:13hierarchical,
author = {Qardaji, Wahbeh and Yang, Weining and Li, Ninghui},
title = {Understanding hierarchical methods for differentially private histograms},
year = {2013},
issue_date = {September 2013},
publisher = {VLDB Endowment},
volume = {6},
number = {14},
issn = {2150-8097},
url = {https://doi.org/10.14778/2556549.2556576},
doi = {10.14778/2556549.2556576},
journal = {Proc. VLDB Endow.},
month = sep,
pages = {1954–1965},
numpages = {12}
}

@INPROCEEDINGS{CormodePSSY:12spatial,
  author={Cormode, Graham and Procopiuc, Cecilia and Srivastava, Divesh and Shen, Entong and Yu, Ting},
  booktitle={2012 IEEE 28th International Conference on Data Engineering}, 
  title={Differentially Private Spatial Decompositions}, 
  year={2012},
  volume={},
  number={},
  pages={20-31},
  doi={10.1109/ICDE.2012.16},
  url = {https://10.1109/ICDE.2012.16}
  }

@InProceedings{CormodeB:22sample,
  title = 	 { Sample-and-threshold differential privacy: Histograms and applications },
  author =       {Bharadwaj, Akash and Cormode, Graham},
  booktitle = 	 {Proceedings of The 25th International Conference on Artificial Intelligence and Statistics},
  pages = 	 {1420--1431},
  year = 	 {2022},
  editor = 	 {Camps-Valls, Gustau and Ruiz, Francisco J. R. and Valera, Isabel},
  volume = 	 {151},
  series = 	 {Proceedings of Machine Learning Research},
  month = 	 {28--30 Mar},
  publisher =    {PMLR},
  url = 	 {https://proceedings.mlr.press/v151/cormode22a.html}
}

@inproceedings{KorolovaKMN:09clicks,
author = {Korolova, Aleksandra and Kenthapadi, Krishnaram and Mishra, Nina and Ntoulas, Alexandros},
title = {Releasing search queries and clicks privately},
year = {2009},
isbn = {9781605584874},
publisher = {Association for Computing Machinery},
address = {New York, NY, USA},
url = {https://doi.org/10.1145/1526709.1526733},
doi = {10.1145/1526709.1526733},
booktitle = {Proceedings of the 18th International Conference on World Wide Web},
pages = {171–180},
numpages = {10},
location = {Madrid, Spain},
series = {WWW '09}
}

@inproceedings{BunNS:16concepts,
author = {Bun, Mark and Nissim, Kobbi and Stemmer, Uri},
title = {Simultaneous Private Learning of Multiple Concepts},
year = {2016},
isbn = {9781450340571},
publisher = {Association for Computing Machinery},
address = {New York, NY, USA},
url = {https://doi.org/10.1145/2840728.2840747},
doi = {10.1145/2840728.2840747},
booktitle = {Proceedings of the 2016 ACM Conference on Innovations in Theoretical Computer Science},
pages = {369–380},
numpages = {12},
location = {Cambridge, Massachusetts, USA},
series = {ITCS '16}
}

@ARTICLE{XiaoWG:11wavelet,
  author={Xiao, Xiaokui and Wang, Guozhang and Gehrke, Johannes},
  journal={IEEE Transactions on Knowledge and Data Engineering}, 
  title={Differential Privacy via Wavelet Transforms}, 
  year={2011},
  volume={23},
  number={8},
  pages={1200-1214},
  doi={10.1109/TKDE.2010.247},
  url = {https://doi.org/10.1109/TKDE.2010.247/}}

@inproceedings{BassilyS:15succinct,
author = {Bassily, Raef and Smith, Adam},
title = {Local, Private, Efficient Protocols for Succinct Histograms},
year = {2015},
isbn = {9781450335362},
publisher = {Association for Computing Machinery},
address = {New York, NY, USA},
url = {https://doi.org/10.1145/2746539.2746632},
doi = {10.1145/2746539.2746632},
booktitle = {Proceedings of the Forty-Seventh Annual ACM Symposium on Theory of Computing},
pages = {127–135},
numpages = {9},
location = {Portland, Oregon, USA},
series = {STOC '15}
}

@InProceedings{AcharyaSZ:19hadamard,
  title = 	 {Hadamard Response: Estimating Distributions Privately, Efficiently, and with Little Communication},
  author =       {Acharya, Jayadev and Sun, Ziteng and Zhang, Huanyu},
  booktitle = 	 {Proceedings of the Twenty-Second International Conference on Artificial Intelligence and Statistics},
  pages = 	 {1120--1129},
  year = 	 {2019},
  editor = 	 {Chaudhuri, Kamalika and Sugiyama, Masashi},
  volume = 	 {89},
  series = 	 {Proceedings of Machine Learning Research},
  month = 	 {16--18 Apr},
  publisher =    {PMLR},
  url = 	 {https://proceedings.mlr.press/v89/acharya19a.html}
}

@inproceedings {WangBLJ:17local,
author = {Tianhao Wang and Jeremiah Blocki and Ninghui Li and Somesh Jha},
title = {Locally Differentially Private Protocols for Frequency Estimation},
booktitle = {26th USENIX Security Symposium (USENIX Security 17)},
year = {2017},
isbn = {978-1-931971-40-9},
address = {Vancouver, BC},
pages = {729--745},
url = {https://www.usenix.org/conference/usenixsecurity17/technical-sessions/presentation/wang-tianhao},
publisher = {USENIX Association},
month = aug
}

@article{XuZXYYW:13boundaries,
	author = {Xu, Jia and Zhang, Zhenjie and Xiao, Xiaokui and Yang, Yin and Yu, Ge and Winslett, Marianne},
	date = {2013/12/01},
	doi = {10.1007/s00778-013-0309-y},
	id = {Xu2013},
	isbn = {0949-877X},
	journal = {The VLDB Journal},
	number = {6},
	pages = {797--822},
	title = {Differentially private histogram publication},
	volume = {22},
	year = {2013},
	url = {https://doi.org/10.1007/s00778-013-0309-y}}

@inproceedings{BCDKMT07,
	address = {New York, NY, USA},
	author = {Boaz Barak and Kamalika Chaudhuri and Cynthia Dwork and Satyen Kale and Frank McSherry and Kunal Talwar},
	booktitle = {Proceedings of the Twenty-Sixth ACM SIGMOD-SIGACT-SIGART Symposium on Principles of Database Systems (PODS '07)},
	doi = {10.1145/1265530.1265569},
	pages = {273-282},
	publisher = {ACM},
	title = {Privacy, accuracy, and consistency too: a holistic solution to contingency table release},
	url = {http://dx.doi.org/10.1145/1265530.1265569},
	year = {2007}
    }

@article{BalcerV:19finite,
	author = {Victor Balcer and Salil Vadhan},
	doi = {10.29012/jpc.679},
	journal = {Journal of Privacy and Confidentiality},
	month = 9,
	number = {2},
	title = {Differential Privacy on Finite Computers},
	volume = {9},
	year = {2019},
	url = {https://doi.org/10.29012/jpc.679}}

@article{rameshwar2026optimal,
  title={Optimal Tree-Based Mechanisms for Differentially Private Approximate {CDFs}},
  author={Rameshwar, V Arvind and Tandon, Anshoo and Sharma, Abhay},
  journal={IEEE Transactions on Information Theory},
  year={2026},
  publisher={IEEE},
  url = {https://doi.org/10.1109/TIT.2026.3704677},
  doi = {10.1109/TIT.2026.3704677}
}

@book{devore1993constructive,
  title={Constructive Approximation},
  author={DeVore, Ronald A and Lorentz, George G},
  series ={Grundlehren der mathematischen Wissenschaften},
  volume = {303},
  year={1993},
  address = {Berlin},
  publisher={Springer}
}

@article{canuto1982approximation,
  title={Approximation results for orthogonal polynomials in {Sobolev} spaces},
  author={Canuto, Claudio and Quarteroni, Alfio},
  journal={Mathematics of Computation},
  volume={38},
  number={157},
  pages={67--86},
  year={1982},
  url={https://doi.org/10.1090/S0025-5718-1982-0637287-3},
  doi = {10.1090/S0025-5718-1982-0637287-3}
}

@article{verfurth1999note,
  title={A note on polynomial approximation in {Sobolev} spaces},
  author={Verf{\"u}rth, R{\"u}diger},
  journal={ESAIM: Mathematical Modelling and Numerical Analysis},
  volume={33},
  number={4},
  pages={715--719},
  year={1999},
  publisher={EDP Sciences},
  doi = {10.1051/m2an:1999159},
  url = {https://doi.org/10.1051/m2an:1999159}
}

\newpage

\clearpage
\appendices
\section{Alternative Functional Families}
\label{appendix:alternative_family}

\begin{figure*}[!h]
\centering
\subfloat[Fourier-8]{\includegraphics[height=3.2cm]{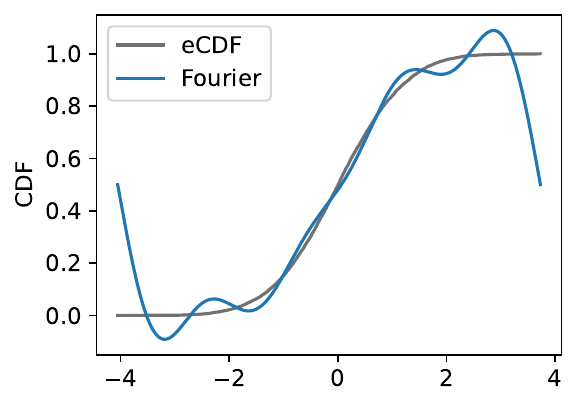}}
\subfloat[Fourier-16]{\includegraphics[height=3.2cm]{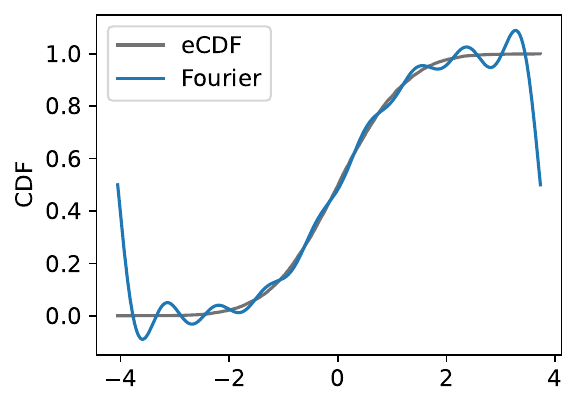}}
\subfloat[Fourier-24]{\includegraphics[height=3.2cm]{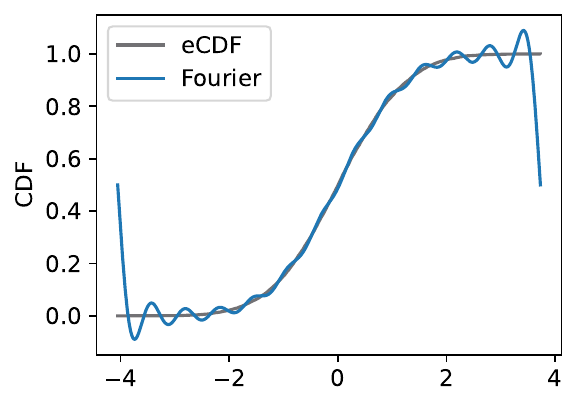}}
\subfloat[Fourier-32]{\includegraphics[height=3.2cm]{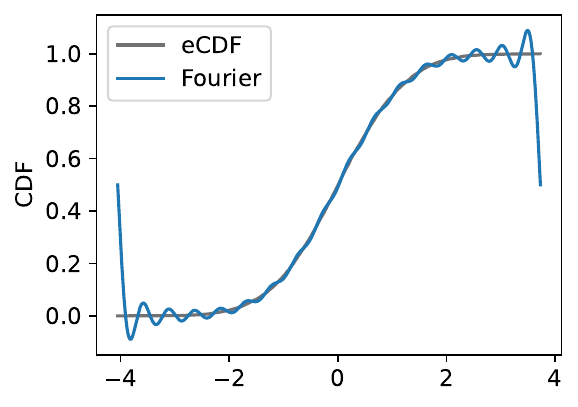}}\\

\subfloat[B-spline(0)-8]{\includegraphics[height=3.2cm]{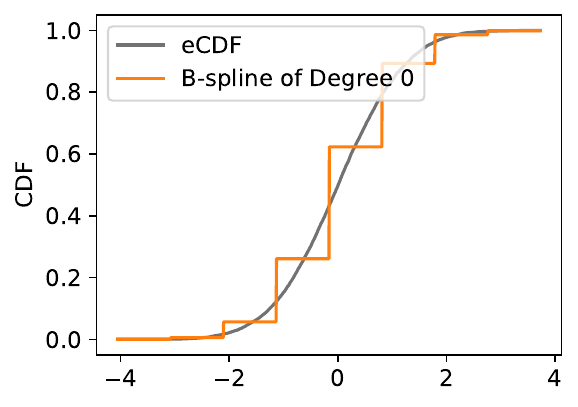}}
\subfloat[B-spline(0)-16]{\includegraphics[height=3.2cm]{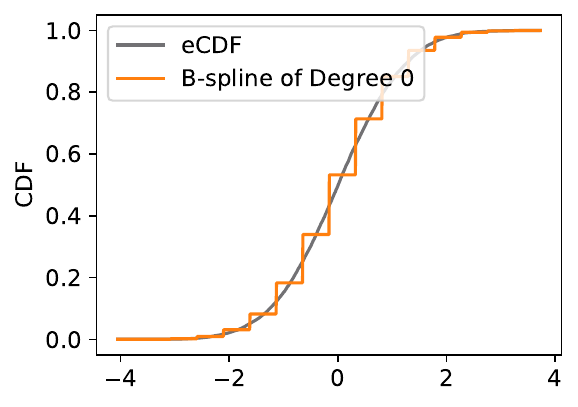}}
\subfloat[B-spline(0)-24]{\includegraphics[height=3.2cm]{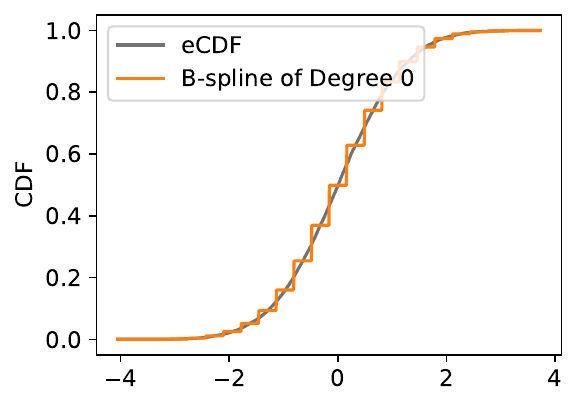}}
\subfloat[B-spline(0)-32]{\includegraphics[height=3.2cm]{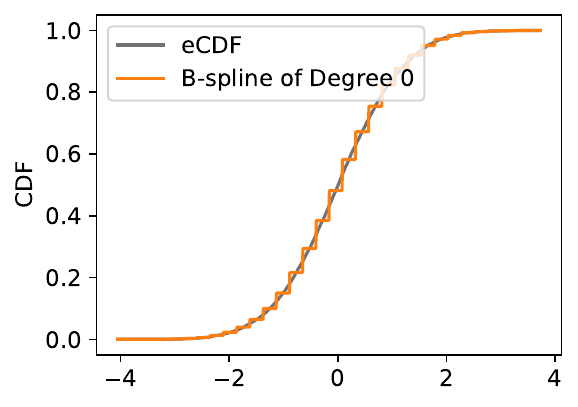}}\\

\subfloat[B-spline(1)-8]{\includegraphics[height=3.2cm]{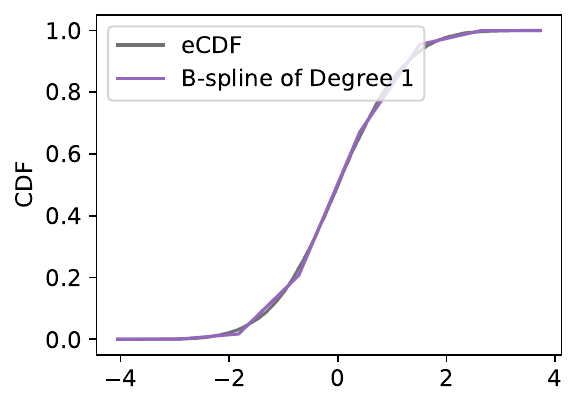}}
\subfloat[B-spline(1)-16]{\includegraphics[height=3.2cm]{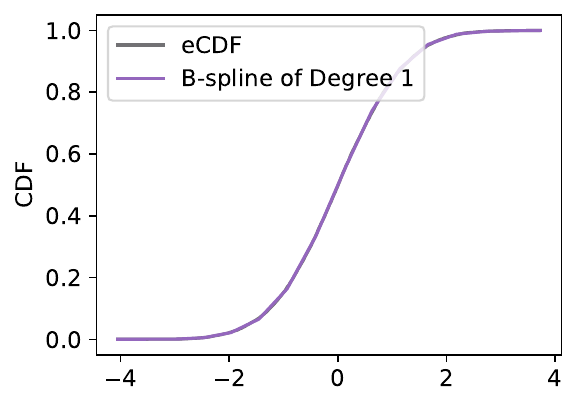}}
\subfloat[B-spline(1)-24]{\includegraphics[height=3.2cm]{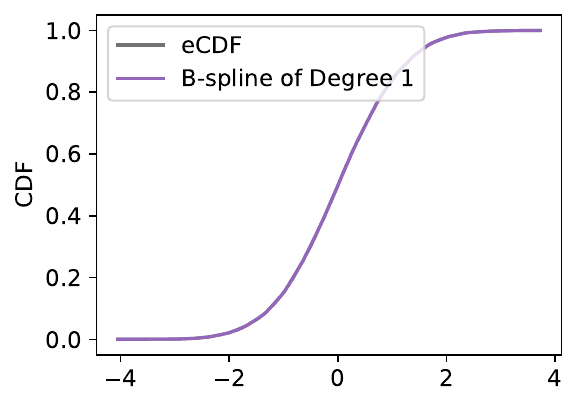}}
\subfloat[B-spline(1)-32]{\includegraphics[height=3.2cm]{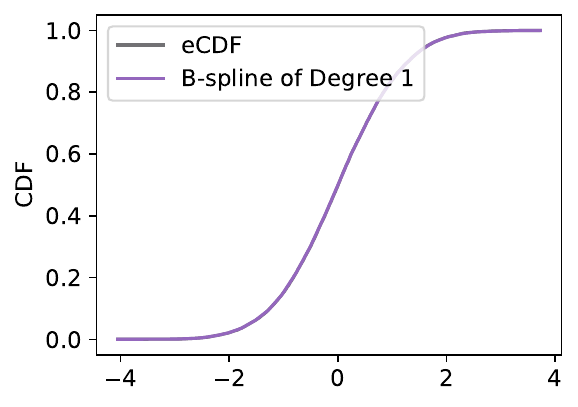}}\\

\subfloat[Legendre-8]{\includegraphics[height=3.2cm]{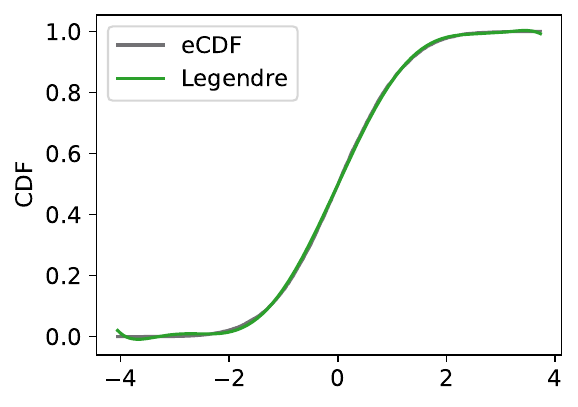}}
\subfloat[Legendre-16]{\includegraphics[height=3.2cm]{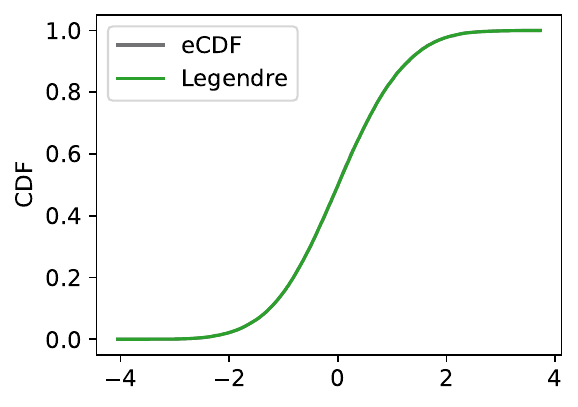}}
\subfloat[Legendre-24]{\includegraphics[height=3.2cm]{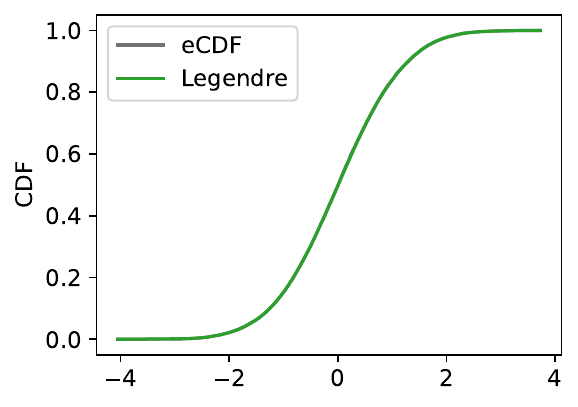}}
\subfloat[Legendre-32]{\includegraphics[height=3.2cm]{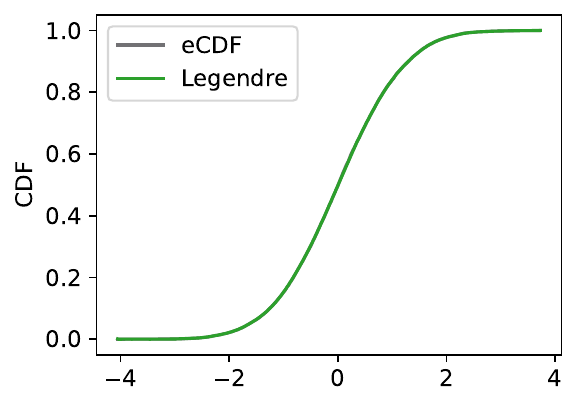}}

\caption{Approximation of the \ac{ecdf} of the normal distribution using predefined functional bases. (a)-(d) present Fourier approximations with 8, 16, 24, and 32 basis functions. While the accuracy in the central region improves with more bases, overshoot and ringing effects remain at the domain boundaries, highlighting the limitations of Fourier bases for discontinuous, non-periodic functions and making them unsuitable for \ac{ecdf} approximation. (e)-(h) B-spline of degree 0; (i)-(l) B-spline of degree 1; (m)-(p) Legendre polynomial approximations, each using the same number of basis functions.}
\label{fig:compare_base}
\end{figure*}

In our approach, we selected a polynomial family, using Legendre polynomials as an example, to serve as the basis for approximating the \ac{ecdf}. Polynomial families are orthogonal and defined on a bounded interval, providing a natural framework for smooth, global approximation. Although an \ac{ecdf} is a step function, polynomial approximation produces a smooth curve that captures the overall distribution shape while removing abrupt jumps.

We also explored other function families. Fourier-based approaches decompose a function into sines and cosines, which are inherently periodic. Applying a Fourier series to a discontinuous, bounded step function like an \ac{ecdf} often induces the Gibbs phenomenon~\cite{carslaw1921introduction}, resulting in overshoot and ringing artifacts near discontinuities. Such effects result in less natural and visually distorted reconstructions (see Figure~\ref{fig:compare_base} (a)-(d)).

B-splines~\cite{prautzsch2002bezier} provide another flexible option. Since an \ac{ecdf} can be viewed as a superposition of several step functions, a B-spline of degree 0, also known as a rectangular or step function, forms a natural and suitable basis for approximation. Figure~\ref{fig:compare_base} (e)-(h) illustrates that fitting an \ac{ecdf} with these basis functions produces a very rigid, piecewise constant curve that cannot capture any smooth variations in the data. It only roughly approximates the stepwise shape of the \ac{ecdf} and does not provide a smooth estimate. Nevertheless, increasing the number of basis functions improves the approximation, producing a piecewise constant curve that more closely follows the \ac{ecdf} steps. Higher-degree B-splines, such as degree 1, also known as piecewise linear functions, are continuous. They provide a smoother approximation than degree 0 but are not as smooth as higher-degree B-splines, such as degree 3. Figure~\ref{fig:compare_base} (i)-(l) shows that degree 1 approximations tend to exhibit sharp corners at the knots. Increasing the number of basis functions further improves the approximation; for example, in Figure~\ref{fig:compare_base} (l), these sharp corners become visually imperceptible.

Through these exploratory experiments, we found that polynomial families offer smoother, high-quality approximations. Due to their orthogonality, they simplify computation and theoretical analysis within our privacy-preserving framework. Nonetheless, the B-spline experiments inspired the idea of constructing a dictionary that includes atoms from multiple bases, leveraging the unique properties of each to better approximate \ac{ecdf}s in this extended work.

\section{Proofs}
\label{appendix:proofs}

\ecdfapprox*
\begin{proof}
Because $\mathcal{P}$ is spanned by the Legendre polynomials $\{P_0, P_1, \ldots, P_m\}$ and these polynomials are orthogonal to each other, the orthonormal basis of $\mathcal{P}$ is given by
\begin{align*}
e_i(x) = \sqrt{\frac{2i+1}{2}} P_i(x) = \sqrt{\frac{2i+1}{2}} 2^i \sum_{j=0}^i x^j \binom{i}{j} \binom{\frac{i+j-1}{2}}{i}.
\end{align*}
Using the Gram-Schmidt procedure, we get
\begin{align}
\label{eq:nondp_approximation_1}
\optempcdf(x) = \sum_{i=0}^m \langle \empcdf, e_i \rangle e_i(x).
\end{align}
It is ensured that $\optempcdf(x)$ is the optimal approximation because for all $i' \in[0,m]$, we have
\begin{align*}
\langle \empcdf - \optempcdf, e_{i'} \rangle &= \left\langle \empcdf -\sum_{i=0}^m \langle \empcdf, e_i \rangle e_i, e_{i'} \right\rangle 
\\ &= \langle \empcdf, e_{i'} \rangle - \sum_{i=0}^m \langle \empcdf, e_i \rangle \cdot \langle e_i, e_{i'} \rangle 
\\ &= 0.
\end{align*}
To solve~\eqref{eq:nondp_approximation_1}, we need to find the coefficients $\langle \empcdf, e_i \rangle$ which are given by
\begin{align}
&\langle \empcdf, e_i \rangle \nonumber
\\ = &\int_{-1}^1 \empcdf(x) e_i(x) \, dx \nonumber
\\ = &2^i \sqrt{\frac{2i+1}{2}} \int_{-1}^1 \empcdf(x) \sum_{j=0}^i x^j \binom{i}{j} \binom{\frac{i+j-1}{2}}{i} \, dx \nonumber
\\ = &2^i \sqrt{\frac{2i+1}{2}} \sum_{j=0}^i \left(\binom{i}{j} \binom{\frac{i+j-1}{2}}{i} \int_{-1}^1 \empcdf(x) x^j \, dx \right). \label{eq:nondp_approximation_2}
\end{align}
For any nonnegative integer $j$, we have
\begin{align}
\int_{-1}^1 \empcdf(x) x^j \, dx &= \int_{x_1}^{x_2} \frac{x^j}{n} \, dx + \int_{x_2}^{x_3} \frac{2x^j}{n} \, dx + \cdots \nonumber
\\ &\quad + \int_{x_{n-1}}^{x_n} \frac{(n-1)x^j}{n} \, dx + \int_{x_n}^1 \frac{n x^j}{n} \, dx \nonumber
\\ &= \frac{x_2^{j+1}-x_1^{j+1}}{n(j+1)} +\frac{2(x_3^{j+1}-x_2^{j+1})}{n(j+1)} + \cdots \nonumber
\\ &\quad + \frac{(n-1)(x_n^{j+1}-x_{n-1}^{j+1})}{n(j+1)} + \frac{n(1-x_n^{j+1})}{n(j+1)} \nonumber
\\ &= \frac{1}{j+1} \left(1-\frac{1}{n}\sum_{k=1}^n x_k^{j+1} \right). \label{eq:nondp_approximation_3}
\end{align}
Substituting~\eqref{eq:nondp_approximation_3} into~\eqref{eq:nondp_approximation_2}, we obtain
\begin{align*}
\optempcdf(x) = \sum_{i=0}^m \langle \empcdf, e_i \rangle e_i =  \sum_{i=0}^m \alpha_i \sum_{j=0}^i \left(\beta_{i,j} \left(1-\mu_{j+1}\right)\right) e_i(x), 
\end{align*}
where $\alpha_i = 2^i \sqrt{\frac{2i+1}{2}}$, $\beta_{i,j} = \frac{1}{j+1} \binom{i}{j} \binom{\frac{i+j-1}{2}}{i}$, and $\mu_{j+1} = \frac{1}{n}\sum_{k=1}^n x_k^{j+1}$ represents mean of the $(j+1)$-th moment of the data.
\end{proof}

\begin{theorem}[Optimal Approximation of eCDF in High Dimensions]
\label{theorem:nondp_approximation_multi}
Let the dataset consist of $n$ i.i.d. samples $\mathbf{x}_1, \mathbf{x}_2, \dots, \mathbf{x}_n \in [-1, 1]^d$, where each sample is a $d$-dimensional vector $\mathbf{x}_k = (x_{k,1}, x_{k,2}, \dots, x_{k,d})$. The multi-dimensional eCDF $F_n(\mathbf{x}) = \frac{1}{n} \sum_{k=1}^n \prod_{j=1}^d \mathbb{I}(x_{k,j} \le x_j)$ can be projected onto the $d$-dimensional tensor-product Legendre polynomial space spanned by $\{e_{\mathbf{i}}(\mathbf{x}) = \prod_{j=1}^d e_{i_j}(x_j)\}$. The optimal approximation $\hat{F}_n(\mathbf{x})$ that minimizes the $L^2([-1,1]^d)$ distance to $F_n(\mathbf{x})$ is given by:
\begin{equation*}
\hat{F}_n(\mathbf{x}) = \sum_{\mathbf{i}} c_{\mathbf{i}} e_{\mathbf{i}}(\mathbf{x}),
\end{equation*}
where $\mathbf{i} = (i_1, i_2, \dots, i_d)$ is the multi-index of the polynomial degrees, and the high-dimensional projection coefficients $c_{\mathbf{i}}$ can be explicitly computed as:
\begin{equation}
c_{\mathbf{i}} = \left( \prod_{j=1}^d \alpha_{i_j} \right) \sum_{\mathbf{0} \le \mathbf{l} \le \mathbf{i}} \left( \prod_{j=1}^d \beta_{i_j, l_j} \right) \hspace{-1mm} \left[ \frac{1}{n} \sum_{k=1}^n \prod_{j=1}^d (1 - x_{k,j}^{l_j+1}) \right],
\label{eq:highdim:coeff}
\end{equation}
where $\alpha_{i_j} = 2^{i_j} \sqrt{\frac{2i_j+1}{2}}$, and $\beta_{i_j, l_j} = \frac{1}{l_j+1} \binom{i_j}{l_j} \binom{\frac{i_j+l_j-1}{2}}{i_j}$.
\end{theorem}
\begin{proof}
By the projection theorem in Hilbert spaces, the optimal coefficients are given by the inner product $c_{\mathbf{i}} = \langle F_n, e_{\mathbf{i}} \rangle$. Expanding the inner product over the $d$-dimensional domain $[-1,1]^d$, we have:
\begin{align*}
c_{\mathbf{i}} &= \int_{[-1,1]^d} F_n(\mathbf{x}) e_{\mathbf{i}}(\mathbf{x}) d\mathbf{x} \\
&= \int_{-1}^1 \dots \int_{-1}^1 \left[ \frac{1}{n} \sum_{k=1}^n \prod_{j=1}^d \mathbb{I}(x_{k,j} \le x_j) \right] 
\\
&\hspace{4cm} \prod_{j=1}^d e_{i_j}(x_j) dx_1 \dots dx_d.
\end{align*}
By Fubini's Theorem, we can interchange the summation over the samples and the multi-dimensional integration, which allows the joint integral to decouple into a product of 1D integrals for each dimension $j$:
\begin{align*}
c_{\mathbf{i}} &= \frac{1}{n} \sum_{k=1}^n \prod_{j=1}^d \left[ \int_{-1}^1 \mathbb{I}(x_{k,j} \le x_j) e_{i_j}(x_j) dx_j \right].
\end{align*}
Substituting the monomial expansion of the 1D orthonormal basis $e_{i_j}(x_j) = \alpha_{i_j} \sum_{l_j=0}^{i_j} (l_j + 1) \beta_{i_j, l_j} x_j^{l_j}$ into the decoupled integral yields:
\begin{align*}
&\int_{-1}^1 \mathbb{I}(x_{k,j} \le x_j) e_{i_j}(x_j) dx_j 
\\ =& \alpha_{i_j} \sum_{l_j=0}^{i_j} (l_j + 1) \beta_{i_j, l_j} \int_{x_{k,j}}^1 x_j^{l_j} dx_j \\
=& \alpha_{i_j} \sum_{l_j=0}^{i_j} \beta_{i_j, l_j} (1 - x_{k,j}^{l_j+1}).
\end{align*}
Recombining the product over all dimensions $j = 1, \dots, d$, the coefficient $c_{\mathbf{i}}$ becomes:
\begin{align*}
c_{\mathbf{i}} &= \frac{1}{n} \sum_{k=1}^n \prod_{j=1}^d \left[ \alpha_{i_j} \sum_{l_j=0}^{i_j} \beta_{i_j, l_j} (1 - x_{k,j}^{l_j+1}) \right].
\end{align*}
By collecting the product constants $\alpha_{i_j}$ and expanding the product of the inner finite sums into a sum over multi-indices $\mathbf{l} = (l_1, \dots, l_d)$, we obtain \eqref{eq:highdim:coeff}, which completes the proof.
\end{proof}

\lemmafour*
\begin{proof}
The approximation errors can be expressed as follows:
\begin{align*}
\|f - \hat{f}\|_2^2 &= \left\langle f - \sum_{i=1}^s \langle f, \phi_i \rangle \phi_i, \, f - \sum_{i=1}^s \langle f, \phi_i \rangle \phi_i \right\rangle 
\\ &= \|f\|_2^2 - \sum_{i=1}^s |\langle f, \phi_i \rangle|^2, \\
\|f - \hat{f}^s\|_2^2 &= \left\langle f - \sum_{i=1}^s \langle f, \phi_{I_i} \rangle \phi_{I_i}, \, f - \sum_{i=1}^s \langle f, \phi_{I_i} \rangle \phi_{I_i} \right\rangle 
\\ &= \|f\|_2^2 - \sum_{i=1}^s |\langle f, \phi_{I_i} \rangle|^2.
\end{align*}
Since $\{\phi_{I_i}\}_{i=1}^s$ corresponds to the $s$ basis functions in $\mathcal{D}_m$ with the largest absolute coefficients:
\[ 
\sum_{i=1}^s |\langle f, \phi_{I_i} \rangle|^2 \geq \sum_{i=1}^s |\langle f, \phi_i \rangle|^2.
\]
This implies $\|f - \hat{f}^s\|_2^2 \leq \|f - \hat{f}\|_2^2$, with equality if and only if $\{\phi_{I_i}\}_{i=1}^s = \{\phi_i\}_{i=1}^s$.
\end{proof}

\lemmafive*
\begin{proof}
We prove by induction on the sparsity level $s$. When $s=1$, \ac{mp} selects the basis function 
\[
\phi_{I_1} = \argmax_i |\langle f, \phi_i \rangle|,
\]
which is exactly the basis function with the largest absolute inner product. Hence, the claim holds. Suppose the \ac{mp} algorithm selects the top $s-1$ basis functions with the largest absolute inner products. We now prove the claim also holds for sparsity level $s$. At step $s$, the residual is
\[
r_{s-1} = f - \sum_{i=1}^{s-1} \langle f, \phi_{I_i} \rangle \phi_{I_i}.
\]
Since the $\{\phi_i\}$ form an orthonormal basis, we have
\[
\langle r_{s-1}, \phi_i \rangle = \langle f, \phi_i \rangle - \sum_{i=1}^{s-1} \langle f, \phi_{I_i} \rangle \langle \phi_{I_i}, \phi_i \rangle.
\]
This simplifies to:
\[
\langle r_{s-1}, \phi_i \rangle =
\begin{cases}
0, & \text{if } i \in \{I_1, \ldots, I_{s-1}\}, \\
\langle f, \phi_i \rangle, & \text{otherwise}.
\end{cases}
\]
Hence, \ac{mp} selects the next basis function with the largest $|\langle f, \phi_i \rangle|$ not yet chosen, as claimed.
\end{proof}

\senip*
\begin{proof}
Let $D = \{x_k\}_{k=1}^n$ and $D' = \{x_k'\}_{k=1}^n$ be two neighboring datasets differing in at most one data point. The sensitivity is defined as the maximum difference in the inner product over all such neighboring pairs:
\begin{align*}
&\Delta_{\text{\ac{mp}}} =\max_{D, D'} \left| \langle \tilde{r}_{i, D}, \phi \rangle - \langle \tilde{r}_{i, D'}, \phi \rangle \right|
\\ &=\max_{D, D'} \left|  \left\langle F_D-\sum_{j=1}^{i-1} \tilde{c}_j \phi_{\tilde{I}_j}, \phi  \right\rangle - \left\langle F_{D'}-\sum_{j=1}^{i-1} \tilde{c}_j \phi_{\tilde{I}_j}, \phi \right\rangle \right|
\\ &=\max_{D, D'} \left| \langle F_D, \phi \rangle - \langle F_{D'}, \phi \rangle \right|.
\end{align*}
Note that at a fixed iteration $i$, the coefficients $\tilde{c}_j$ and basis functions $\phi_{\tilde{I}_j}$ for $j < i$ are determined in earlier steps and remain constant when evaluating the sensitivity at step $i$. Therefore, the difference between the two inner products reduces to the difference between the \ac{ecdf}, as the residual terms from previous steps cancel out. We can further expand this as:
\begin{align*}
&\Delta_{\text{\ac{mp}}} =\max_{D, D'} \left| \int F_D(x) \phi(x) \, dx - \int F_{D'}(x) \phi(x) \, dx \right|
\\ &=\frac{1}{n} \max_{D, D'} \left| \int \left( \sum_{k=1}^n \mathbb{I}(x_k \leq x)-\sum_{k=1}^n \mathbb{I}(x_k' \leq x)\right) \phi(x) \, dx \right|
\\ & \leq \frac{1}{n} \int \left| \phi(x) \right| \, dx,
\end{align*}
where the last inequality follows from the fact that the two datasets differ in at most one data point, so the integrand differs by at most $\left|\phi(x)\right|$ at each $x$.
\end{proof}

\senaip*
\begin{proof}
By Lemma~\ref{lemma:sen_ip} and the reverse triangle inequality, we have
\begin{align*}
\Delta_{\text{A\ac{mp}}} &=\max_{D, D'} \left| |\langle \tilde{r}_{i, D}, \phi \rangle| - |\langle \tilde{r}_{i, D'}, \phi \rangle| \right|
\\ &\leq \max_{D, D'} \left| \langle \tilde{r}_{i, D}, \phi \rangle - \langle \tilde{r}_{i, D'}, \phi \rangle \right|
\\ &\leq \frac{1}{n} \int \left| \phi(x) \right| \, dx.
\end{align*}
\end{proof}


\begin{lemma}[Tail bound for Erlang distribution]
\label{lemma:erlang_bound}
Let $Z \sim \mathrm{Erlang}(s, \lambda)$, i.e., $Z$ is the sum of $s$ independent exponential random variables with rate $\lambda$.  
For any threshold $T > s/\lambda$, the tail probability can be bounded as
\[
\mathbb{P}(Z > T) \le \exp\Big(s - \lambda T \Big) \left(\frac{\lambda T}{s}\right)^s,
\]
which decreases exponentially as $\lambda$ increases, up to a polynomial factor $(\lambda T/s)^s$.
\end{lemma}

\begin{proof}
Using the Chernoff bound and the moment generating function of $Z$, we have for any $t \in (0,\lambda)$,
\[
\mathbb{P}(Z > T) \le e^{-t T} \mathbb{E}[e^{t Z}] = e^{-t T} (1 - t/\lambda)^{-s}.
\]
To find the optimal $t$ that minimizes the right-hand side, we let $f(t) = e^{-t T} (1 - t/\lambda)^{-s}$ and take the logarithm:
\[
\ln f(t) = -t T - s \ln(1 - t/\lambda),
\]
and differentiate with respect to $t$:
\[
\frac{d}{dt} \ln f(t) = -T + \frac{s}{\lambda - t}.
\]
Setting the derivative to zero gives
\[
t^* = \lambda \Big(1 - \frac{s}{\lambda T}\Big), \quad T > s/\lambda.
\]
Substituting $t^*$ into the bound yields
\[
\mathbb{P}(Z > T) \le \exp(s - \lambda T) \left(\frac{\lambda T}{s}\right)^s.
\]
Moreover, the dependence on $\lambda$ is monotone decreasing in the regime of interest. Indeed, write
\[
f(\lambda)=\exp(s-\lambda T)\Big(\frac{\lambda T}{s}\Big)^s.
\]
Hence
\[
\frac{d}{d\lambda}\ln f(\lambda)=-T+\frac{s}{\lambda}.
\]
For $\lambda>s/T$, we have $-T+s/\lambda<0$, so $\ln f(\lambda)$ is strictly decreasing in $\lambda$.  Therefore $f(\lambda)$ decreases as $\lambda$ increases in the feasible region. In particular, for large $\lambda$ the linear term $-\lambda T$ dominates the logarithm, so
\[
\ln f(\lambda) = -\lambda T + O(\ln(\lambda T)),
\]
and thus $f(\lambda)$ decays exponentially in $\lambda$ up to a polynomial factor.
\end{proof}

\begin{lemma}[Sensitivity of $\boldsymbol{\mu}$]
\label{lemma:sensitivity}
Let datasets $\{x_k\}_{k=1}^n$ and $\{x_k'\}_{k=1}^n$ differ in at most one element. Then, the $\ell_1$ sensitivity and $\ell_2$ sensitivity of $\boldsymbol{\mu}$ are
\[
\Delta_1 = \frac{3m + 4}{2n}, \quad \Delta_2=\sqrt{\frac{5m+8}{2 n^2}}.
\]
\end{lemma}
\begin{proof}
Assume that $\{x_k\}_{k=1}^n$ and $\{x_k'\}_{k=1}^n$ differ only in the $k$-th element.
\begin{align*}
\boldsymbol{\mu} - \boldsymbol{\mu}' &= [\mu_1 - \mu_1', \mu_2 -\mu_2',\ldots, \mu_{m+1}-\mu_{m+1}']^\top
\\ &= \left[\frac{x_k - x_k'}{n}, \frac{x_k^2 - x_k'^2}{n}, \ldots, \frac{x_k^{m+1} - x_k'^{m+1}}{n} \right]^\top.
\end{align*}
Given that $x_k, x_k' \in [-1,1]$, we proceed by analyzing the $\ell_1$ and $\ell_2$ sensitivity of $\boldsymbol{\mu}$ separately.

\noindent \textbf{$\ell_1$ sensitivity:} If $m$ is even, then
\begin{align*}
\|\boldsymbol{\mu} - \boldsymbol{\mu}'\|_1 &= \left|\frac{x_k - x_k'}{n}\right| + \cdots + \left|\frac{x_k^{m+1} - x_k'^{m+1}}{n}\right|
\\ &\leq \frac{2}{n}+\frac{1}{n} + \cdots + \frac{2}{n}
\\ &\leq \frac{3m+4}{2 n}.
\end{align*}
Similarity, if $m$ is odd, $\|\boldsymbol{\mu} - \boldsymbol{\mu}'\|_1 \leq \frac{3m+3}{2 n}$. Thus, the $\ell_1$ sensitivity of $\boldsymbol{\mu}$ is $\frac{3m+4}{2 n}$.

\noindent \textbf{$\ell_2$ sensitivity:} If $m$ is even, then
\begin{align*}
\|\boldsymbol{\mu} - \boldsymbol{\mu}'\|_2 &= \sqrt{\left(\frac{x_k - x_k'}{n}\right)^2+ \cdots + \left(\frac{x_k^{m+1} - x_k'^{m+1}}{n}\right)^2}
\\ &\leq \sqrt{\left(\frac{2}{n}\right)^2+\left(\frac{1}{n}\right)^2 + \cdots + \left(\frac{2}{n}\right)^2}
\\ &\leq \sqrt{\frac{5m+8}{2 n^2}}.
\end{align*}
Similarity, if $m$ is odd, $\|\boldsymbol{\mu} - \boldsymbol{\mu}'\|_2 \leq \sqrt{\frac{5m+5}{2 n^2}}$. Thus, the $\ell_2$ sensitivity of $\boldsymbol{\mu}$ is $\sqrt{\frac{5m+8}{2 n^2}}$.
\end{proof}

\begin{lemma}[Sensitivity of High-Dimensional Moments]
\label{lemma:sensitivity_multi}
Let $\mathbf{m} = (m_1, \dots, m_d)$ be the vector of maximum polynomial degrees for each dimension $j \in \{1, \dots, d\}$. Let $D = \{\mathbf{x}_k\}_{k=1}^n$ and $D' = \{\mathbf{x}_k'\}_{k=1}^n$ be two neighboring datasets differing in at most one element via replacement. Let $\mathbf{M} = \{\mathbf{M}_{\mathbf{l}}\}_{\mathbf{0} \le \mathbf{l} \le \mathbf{m}}$ be the high-dimensional moment tensor where each entry corresponds to the multi-index $\mathbf{l} = (l_1, \dots, l_d)$ and is defined as:
\begin{equation*}
\mathbf{M}_{\mathbf{l}} = \frac{1}{n} \sum_{k=1}^n \prod_{j=1}^d (1 - x_{k,j}^{l_j+1}).
\end{equation*}
Then, the global $\ell_1$ sensitivity $\Delta_1$ and $\ell_2$ sensitivity $\Delta_2$ of the high-dimensional moment tensor $\mathbf{M}$ are bounded by:
\begin{equation*}
\Delta_1 = \frac{1}{n} \prod_{j=1}^d C_{1, j},
\end{equation*}
\begin{equation*}
\Delta_2 = \frac{1}{n} \sqrt{\prod_{j=1}^d C_{2, j}},
\end{equation*}
where for each dimension $j$, the sensitivity constants are given by:
\begin{equation*}
\begin{aligned}
C_{1,j} &=
\begin{cases}
\frac{3m_j+4}{2}, & m_j \text{ even}\\
\frac{3m_j+3}{2}, & m_j \text{ odd}
\end{cases},
&
C_{2,j} &=
\begin{cases}
\frac{5m_j+8}{2}, & m_j \text{ even}\\
\frac{5m_j+5}{2}, & m_j \text{ odd}
\end{cases}.
\end{aligned}
\end{equation*}
\end{lemma}
\begin{proof}
Assume that the two datasets $D$ and $D'$ differ only in the $k$-th element, where $\mathbf{x}_k, \mathbf{x}_k' \in [-1, 1]^d$. The difference between the moment entries $\mathbf{M}_{\mathbf{l}}$ and $\mathbf{M}_{\mathbf{l}}'$ is contributed solely by the $k$-th sample:
\begin{align*}
&\mathbf{M}_{\mathbf{l}} - \mathbf{M}_{\mathbf{l}}' 
\\ =& \left[ \frac{1}{n} \sum_{k=1}^n \prod_{j=1}^d (1 - x_{k,j}^{l_j+1}) \right] - \left[ \frac{1}{n} \sum_{k=1}^n \prod_{j=1}^d (1 - (x_{k,j}')^{l_j+1}) \right] \\
=& \frac{1}{n} \left[ \prod_{j=1}^d (1 - x_{k,j}^{l_j+1}) - \prod_{j=1}^d (1 - (x_{k,j}')^{l_j+1}) \right].
\end{align*}
Given that $x_{k,j}, x_{k,j}' \in [-1, 1]$, the maximum change for each individual dimension's component $(1 - x_{k,j}^{l_j+1})$ depends on whether the exponent $l_j+1$ is odd or even:
\begin{itemize}
    \item If $l_j$ is even, then $x_{k,j}^{l_j+1} \in [-1, 1]$, which implies $(1 - x_{k,j}^{l_j+1}) \in [0, 2]$. The maximum absolute difference between two such terms is $|2 - 0| = 2$.
    \item If $l_j$ is odd, then $x_{k,j}^{l_j+1} \in [0, 1]$, which implies $(1 - x_{k,j}^{l_j+1}) \in [0, 1]$. The maximum absolute difference between two such terms is $|1 - 0| = 1$.
\end{itemize}
Thus, the worst-case maximum absolute value for a single tensor entry variation is bounded by:
\begin{equation*}
|\mathbf{M}_{\mathbf{l}} - \mathbf{M}_{\mathbf{l}}'| \le \frac{1}{n} \prod_{j=1}^d 2^{\mathbb{I}(l_j \text{ is even})},
\end{equation*}
where $\mathbb{I}(\cdot)$ is the indicator function. We now proceed to analyze the $\ell_1$ and $\ell_2$ sensitivities separately.

\noindent \textbf{$\ell_1$ sensitivity:} The global $\ell_1$ sensitivity is computed by summing the maximum absolute differences over all multi-indices $\mathbf{0} \le \mathbf{l} \le \mathbf{m}$:
\begin{align*}
\|\mathbf{M} - \mathbf{M}'\|_1 &= \sum_{\mathbf{0} \le \mathbf{l} \le \mathbf{m}} |\mathbf{M}_{\mathbf{l}} - \mathbf{M}_{\mathbf{l}}'| 
\\ &\le \frac{1}{n} \sum_{l_1=0}^{m_1} \dots \sum_{l_d=0}^{m_d} \prod_{j=1}^d 2^{\mathbb{I}(l_j \text{ is even})} \\
&= \frac{1}{n} \prod_{j=1}^d \left( \sum_{l_j=0}^{m_j} 2^{\mathbb{I}(l_j \text{ is even})} \right).
\end{align*}
For each dimension $j$, within the sequence $l_j \in [0, m_j]$, there are $\lfloor m_j/2 \rfloor + 1$ even numbers and $\lceil m_j/2 \rceil$ odd numbers. Summing these values gives:
\begin{equation*}
\begin{aligned}
\sum_{l_j=0}^{m_j} 2^{\mathbb{I}(l_j \text{ is even})}
&= 2\left(\lfloor m_j/2 \rfloor + 1\right)
   + \left\lceil m_j/2 \right\rceil \\
&=
\begin{cases}
\frac{3m_j+4}{2}, & \text{if } m_j \text{ is even},\\
\frac{3m_j+3}{2}, & \text{if } m_j \text{ is odd}.
\end{cases}
\end{aligned}
\end{equation*}
Thus, the $\ell_1$ sensitivity is $\Delta_1 = \frac{1}{n} \prod_{j=1}^d C_{1, j}$.

\noindent \textbf{$\ell_2$ sensitivity:} Similarly, the global $\ell_2$ sensitivity is derived by summing the squared entry variations over the multi-index space:
\begin{align*}
\|\mathbf{M} - \mathbf{M}'\|_2 &= \sqrt{\sum_{\mathbf{0} \le \mathbf{l} \le \mathbf{m}} |\mathbf{M}_{\mathbf{l}} - \mathbf{M}_{\mathbf{l}}'|^2} 
\\ &\le \frac{1}{n} \sqrt{\sum_{l_1=0}^{m_1} \dots \sum_{l_d=0}^{m_d} \prod_{j=1}^d 4^{\mathbb{I}(l_j \text{ is even})}} \\
&= \frac{1}{n} \sqrt{\prod_{j=1}^d \left( \sum_{l_j=0}^{m_j} 4^{\mathbb{I}(l_j \text{ is even})} \right)}.
\end{align*}
Similarly, we have:
\begin{equation*}
\begin{aligned}
\sum_{l_j=0}^{m_j} 4^{\mathbb{I}(l_j \text{ is even})}
&= 4\left(\lfloor m_j/2 \rfloor + 1\right)
   + \left\lceil m_j/2 \right\rceil \\
&=
\begin{cases}
\frac{5m_j+8}{2}, & \text{if } m_j \text{ is even},\\
\frac{5m_j+5}{2}, & \text{if } m_j \text{ is odd}.
\end{cases}
\end{aligned}
\end{equation*}
Taking the square root completes the derivation, yielding $\Delta_2 = \frac{1}{n} \sqrt{\prod_{j=1}^d C_{2, j}}$.
\end{proof}

\begin{lemma}[Bound on $\|\hat{F}_n-\tilde{F}_n\|_{\infty}$]\label{lemma:privacy_error_bound}
For the Legendre polynomial-based method, the privacy error is bounded as
\[
\|\hat{F}_n-\tilde{F}_n\|_{\infty} \leq \frac{(m+1)^2}{2} \max_{i \in [1,m+1]} |z_i|.
\]
\end{lemma}
\begin{proof}
The bound for $\|\hat{F}_n-\tilde{F}_n\|_{\infty}$ is
\begin{align}
&\|\hat{F}_n-\tilde{F}_n\|_{\infty} \nonumber \\ = &\left\|\sum_{i=0}^m \alpha_i \sum_{j=0}^i (\beta_{i,j} \cdot z_{j+1}) e_i(x) \right\|_{\infty} \nonumber 
\\ = &\left\|\sum_{i=0}^m 2^{i-1} (2i+1) \sum_{j=0}^i (\beta_{i,j} \cdot z_{j+1}) P_i(x) \right\|_{\infty} \nonumber
\\ \leq &\left\| \max_{i \in [1,m+1]} |z_{i}| \cdot \sum_{i=0}^m 2^{i-1} (2i+1) \sum_{j=0}^i \binom{i}{j}\binom{\frac{i+j-1}{2}}{i}  \right\|_{\infty} \nonumber
\\ = &\left\| \max_{i \in [1,m+1]} |z_i| \cdot \sum_{i=0}^m \left(i+\frac{1}{2} \right)\right\|_{\infty} \nonumber
\\ = &\frac{(m+1)^2}{2} \max_{i \in [1,m+1]} |z_i|. \nonumber
\end{align}
The inequality stems from the property of Legendre polynomials, where $|P_i(x)| \leq 1$ for all $x \in [-1,1]$. The equality following the inequality is due to the fact that $P_i(1) = 2^i \sum_{j=0}^i \binom{i}{j} \binom{\frac{i+j-1}{2} }{i}=1$.
\end{proof}


\iso*
\begin{proof}
Since the function $F$ is monotone non-decreasing and bounded on $[0,1]$, we have $F \in \mathcal{C}$. By Lemma~\ref{lemma:2} and Lemma~\ref{lemma:3}, the constraint set $\mathcal{C}$ is a closed convex subset of $L^2([-1,1])$. By definition, $\tilde{F}_n^{\mathrm{iso}}$ is the projection of $\tilde{F}_n$ onto $\mathcal{C}$ in the $L^2$ norm. From the theory of Hilbert spaces (see Theorem 1 in Chapter 3.12 of~\cite{luenberger1997optimization}), the projection $\tilde{F}_n^{\mathrm{iso}}$ satisfies the best approximation property, which implies that
\[
\|\tilde{F}_n -\tilde{F}_n^{\mathrm{iso}}\|_2 \leq \|\tilde{F}_n -  F\|_2,
\]
and 
\[
\langle \tilde{F}_n - \tilde{F}_n^{\mathrm{iso}}, F - \tilde{F}_n^{\mathrm{iso}}\rangle \leq 0.
\]
Thus, we have:
\begin{align*}
&\|\tilde{F}_n - F\|_2^2 
\\ = &\|\tilde{F}_n- \tilde{F}_n^{\mathrm{iso}} + \tilde{F}_n^{\mathrm{iso}} - F\|_2^2 
\\ = &\|\tilde{F}_n- \tilde{F}_n^{\mathrm{iso}}\|_2^2 + \|\tilde{F}_n^{\mathrm{iso}} - F\|_2^2 + 2 \langle \tilde{F}_n- \tilde{F}_n^{\mathrm{iso}}, \tilde{F}_n^{\mathrm{iso}} - F \rangle.
\end{align*}
Since $\langle \tilde{F}_n- \tilde{F}_n^{\mathrm{iso}}, \tilde{F}_n^{\mathrm{iso}} - F \rangle \geq 0$ and $\|\tilde{F}_n- \tilde{F}_n^{\mathrm{iso}}\|_2^2 \geq 0$, we obtain:
\[
\|\tilde{F}_n - F\|_2^2 \geq \|\tilde{F}_n^{\mathrm{iso}} - F\|_2^2.
\]
Taking the square root of both sides gives:
\[
\|\tilde{F}_n^{\mathrm{iso}} - F\|_2 \leq \|\tilde{F}_n - F\|_2.
\]
\end{proof}

\begin{lemma} 
\label{lemma:2}
$\mathcal{C}_1$ is a closed convex subset of $L^2([-1,1])$.
\end{lemma}
\begin{proof}
Let $f_1, f_2 \in \mathcal{C}_1$ and $\lambda \in [0,1]$. Define $f = \lambda f_1 + (1 - \lambda) f_2$. For almost all pairs $x_1 \leq x_2$, we have
\begin{align*}
f(x_1) &= \lambda f_1(x_1) + (1 - \lambda) f_2(x_1) 
\\ &\leq \lambda f_1(x_2) + (1 - \lambda) f_2(x_2) 
\\ &= f(x_2).
\end{align*}
Thus, $f \in \mathcal{C}_1$, and $\mathcal{C}_1$ is convex.

\noindent Let $\{f_n\} \subset \mathcal{C}_1$ be a sequence converging to $f$ in the $L^2([-1,1])$ norm. Since convergence in $L^2$ implies the existence of a subsequence $\{f_{n_k}\}$ that converges to $f$ almost everywhere, we may assume $f_{n_k}(x) \to f(x)$ for almost every $x \in [-1,1]$. Since each $f_{n_k} \in \mathcal{C}_1$, there exists a representative of $f_{n_k}$ that is monotone non-decreasing on $[-1,1]$. It is known that the pointwise a.e. limit of a sequence of monotone non-decreasing functions is also monotone non-decreasing on a set of full measure. Hence, $f(x_1) \leq f(x_2)$ holds for almost every pair $x_1 \leq x_2$. This implies that $f$ is almost everywhere equal to a monotone non-decreasing function. Thus, $f \in \mathcal{C}_1$, and $\mathcal{C}_1$ is closed.
\end{proof}

\begin{lemma}
\label{lemma:3}
$\mathcal{C}_2$ is a closed convex subset of $L^2([-1,1])$.
\end{lemma}
\begin{proof}
Let $f_1, f_2 \in \mathcal{C}_2$ and $\lambda \in [0,1]$. Define $f = \lambda f_1 + (1 - \lambda) f_2$. For any $x \in [-1,1]$, we have
\[
f(x) = \lambda f_1(x) + (1 - \lambda) f_2(x) \in [0,1],
\]
since both $f_1(x)$ and $f_2(x)$ take values in the interval $[0,1]$, and the convex combination of two values in $[0,1]$ also lies in $[0,1]$. Therefore, $f \in \mathcal{C}_2$, and $\mathcal{C}_2$ is convex.

\noindent Let $\{f_n\} \subset \mathcal{C}_2$ be a sequence converging to $f$ in the $L^2([-1,1])$ norm. Since convergence in $L^2$ implies the existence of a subsequence $\{f_{n_k}\}$ such that $f_{n_k}(x) \to f(x)$ almost everywhere for $x \in [-1,1]$, and since each $f_{n_k}$ satisfies $0 \leq f_{n_k}(x) \leq 1$, we conclude that the pointwise limit of the subsequence $\{f_{n_k}\}$ satisfies $0 \leq f(x) \leq 1$ for almost every $\ x \in [-1,1].$ Thus, $f \in \mathcal{C}_2$, and $\mathcal{C}_2$ is closed.
\end{proof}

\begin{theorem}[The Classical Projection Theorem~\cite{luenberger1997optimization}]
\label{theorem:projection}
Let $\mathcal{H}$ be a Hilbert space and $\mathcal{P}$ a closed subspace of $\mathcal{H}$. Corresponding to any vector $F \in \mathcal{H}$, there is a unique vector $\hat{F} \in \mathcal{P}$ such that $\|F - \hat{F} \| \leq \|F-f\|$ for all $f \in \mathcal{P}$. Furthermore, a necessary and sufficient condition that $\hat{F} \in \mathcal{P}$ be the unique minimizing vector is that $F - \hat{F}$ be orthogonal to $\mathcal{P}$.
\end{theorem}

\section{Additional Details on Experimental Results}
\label{appendix:figures}

\subsection{Effect of Parameters}
\label{appendix:effect_para}

\begin{figure*}[b]
\centering
\subfloat[Lognormal$(0, 0.5)$]{\includegraphics[height=2.8cm]{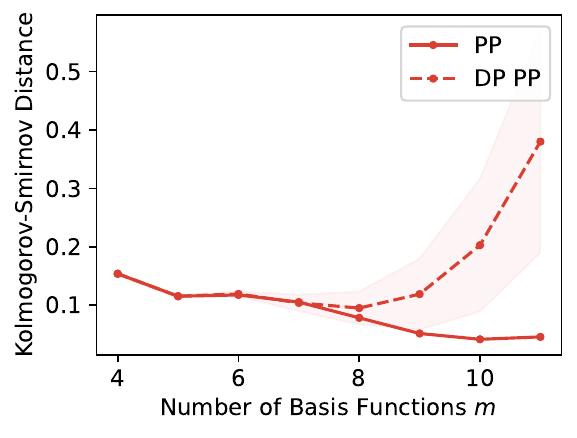}}\hfil
\subfloat[Lognormal$(0, 0.5)$]{\includegraphics[height=2.8cm]{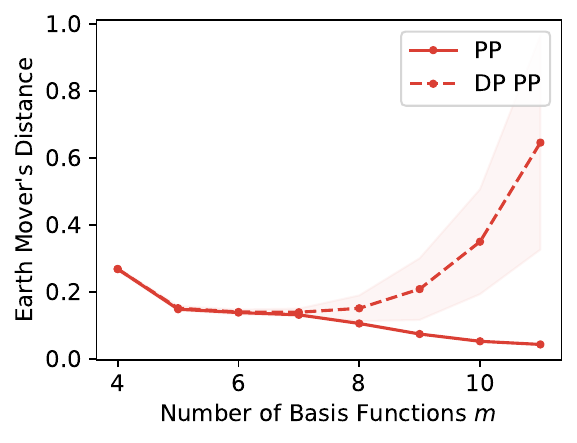}}\hfil
\subfloat[Lognormal$(0, 0.5)$]{\includegraphics[height=2.8cm]{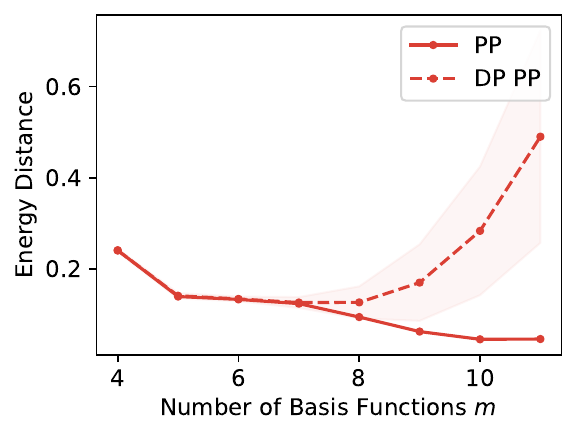}} 

\subfloat[Beta$(10, 2)$]{\includegraphics[height=2.8cm]{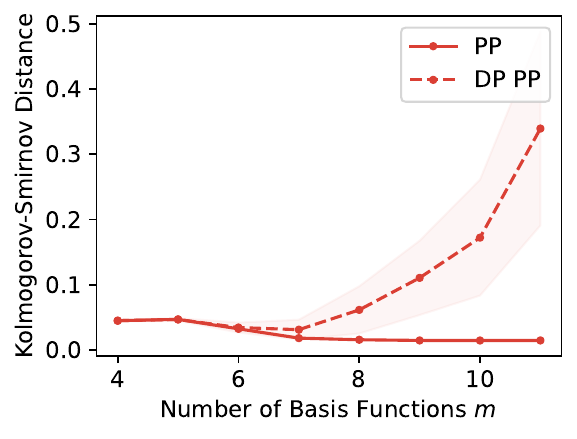}}\hfil
\subfloat[Beta$(10, 2)$]{\includegraphics[height=2.8cm]{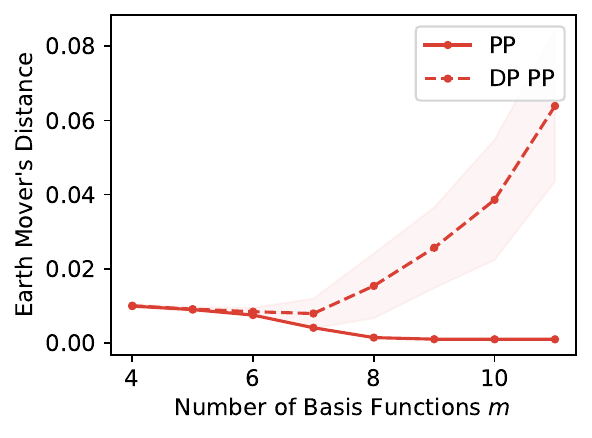}}\hfil
\subfloat[Beta$(10, 2)$]{\includegraphics[height=2.8cm]{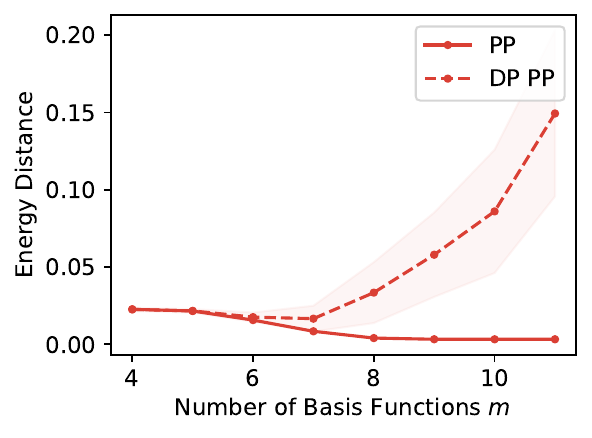}}

\subfloat[Lognormal$(0, 0.5)$]{\includegraphics[height=2.8cm]{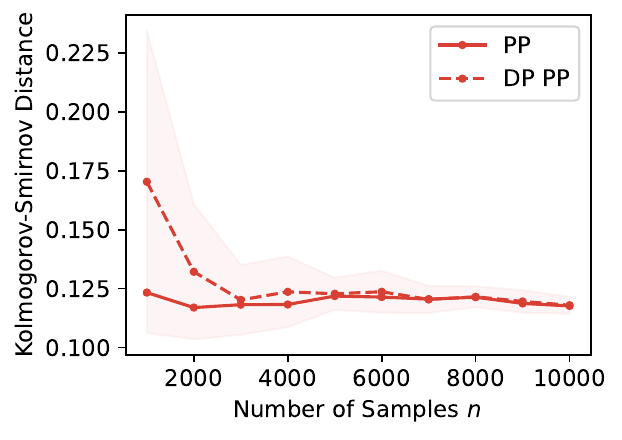}}\hfil
\subfloat[Lognormal$(0, 0.5)$]{\includegraphics[height=2.8cm]{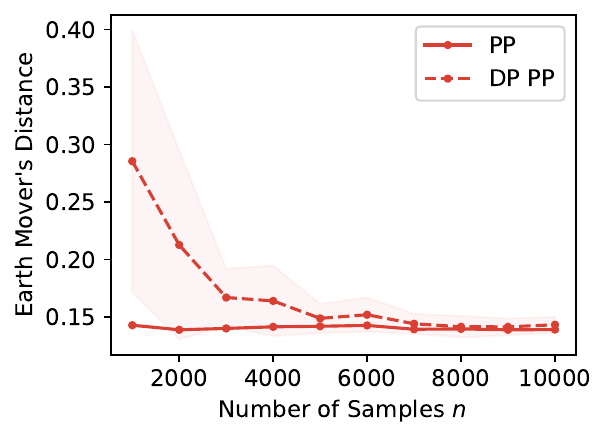}}\hfil
\subfloat[Lognormal$(0, 0.5)$]{\includegraphics[height=2.8cm]{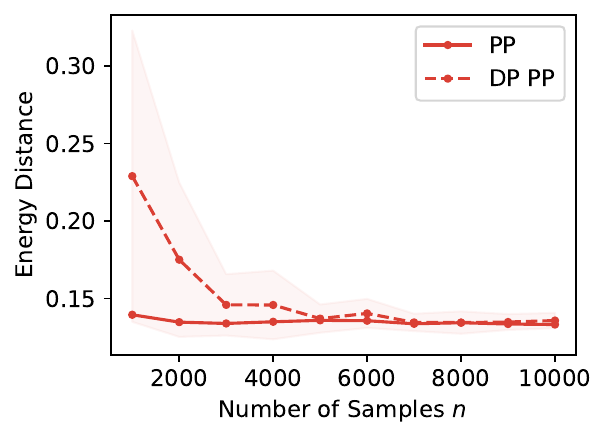}} 

\subfloat[Beta$(10, 2)$]{\includegraphics[height=2.8cm]{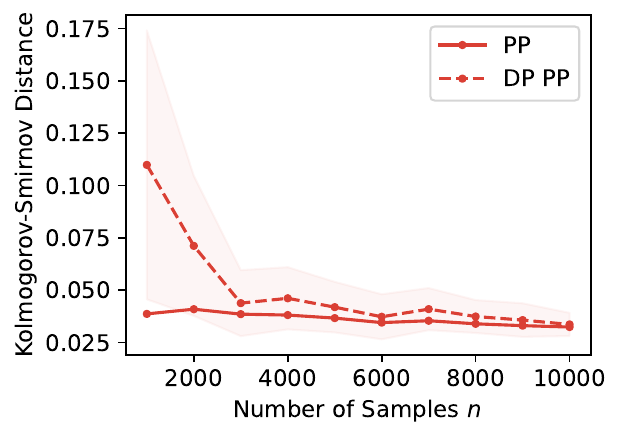}}\hfil
\subfloat[Beta$(10, 2)$]{\includegraphics[height=2.8cm]{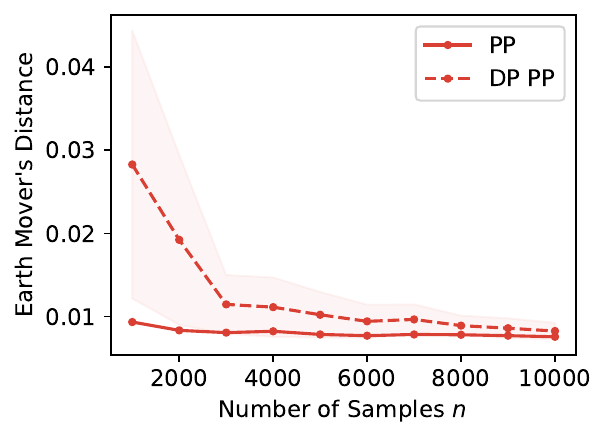}}\hfil
\subfloat[Beta$(10, 2)$]{\includegraphics[height=2.8cm]{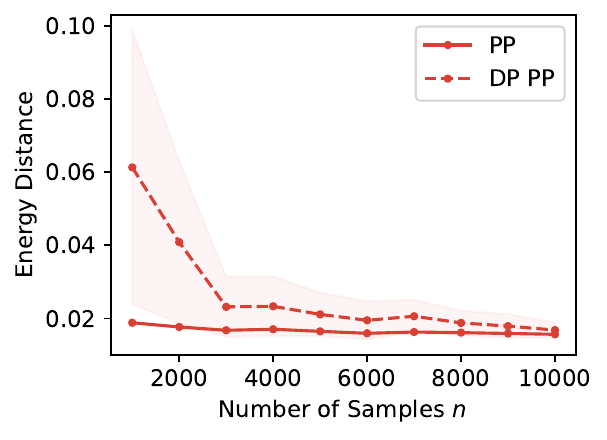}}

\caption{Comparison of distances between \ac{pp}-based approximations and the true CDF under two experimental settings across various distributions. (a)-(f) Effect of the number of basis functions $m$ with $n=10^4$; (g)-(l) Effect of the sample size $n$ with $m=6$. Experiments were repeated $50$ times with $\epsilon=0.5$ under pure differential privacy.}
\label{fig:appendix_pp_parameters}
\end{figure*}

\begin{figure*}[ht!]
\centering
\subfloat[Lognormal$(0, 0.5)$]{\includegraphics[height=2.8cm]{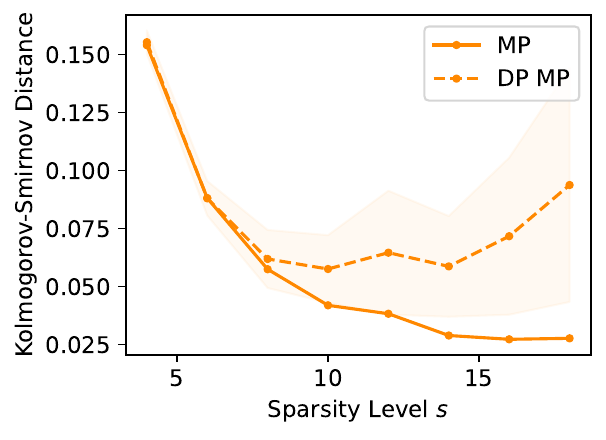}}\hfil
\subfloat[Lognormal$(0, 0.5)$]{\includegraphics[height=2.8cm]{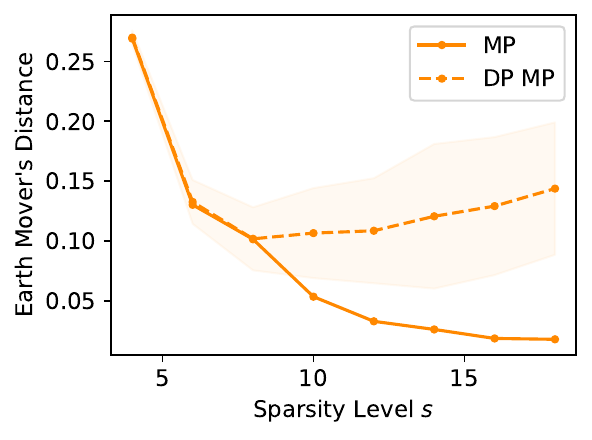}}\hfil
\subfloat[Lognormal$(0, 0.5)$]{\includegraphics[height=2.8cm]{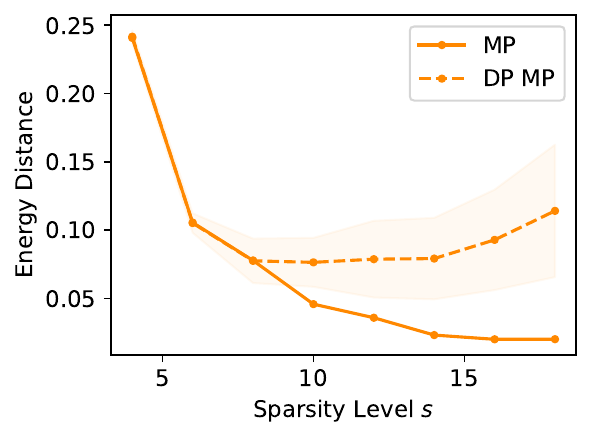}} 

\subfloat[Beta$(10, 2)$]{\includegraphics[height=2.8cm]{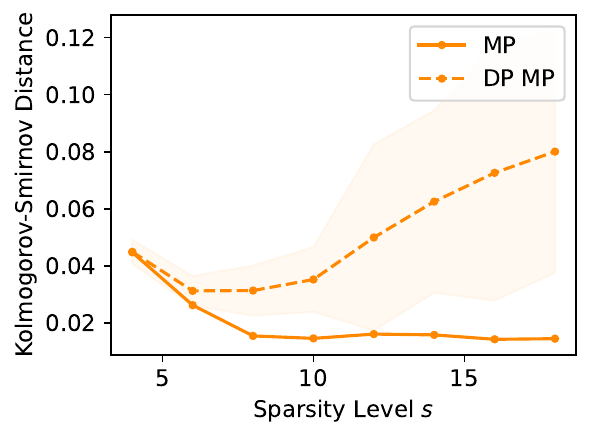}}\hfil
\subfloat[Beta$(10, 2)$]{\includegraphics[height=2.8cm]{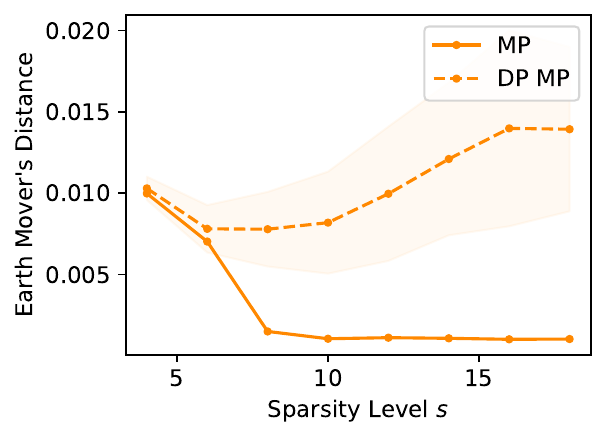}}\hfil
\subfloat[Beta$(10, 2)$]{\includegraphics[height=2.8cm]{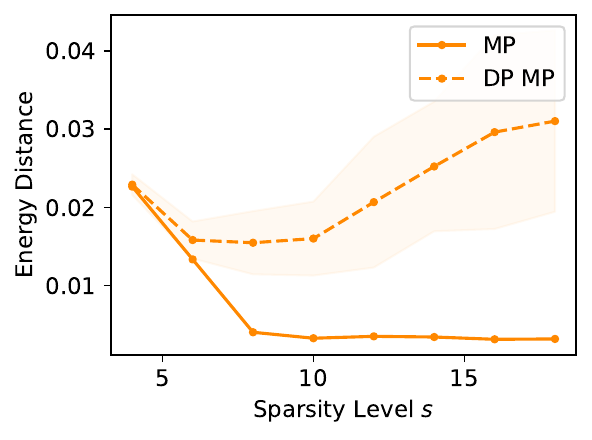}}

\subfloat[Lognormal$(0, 0.5)$]{\includegraphics[height=2.8cm]{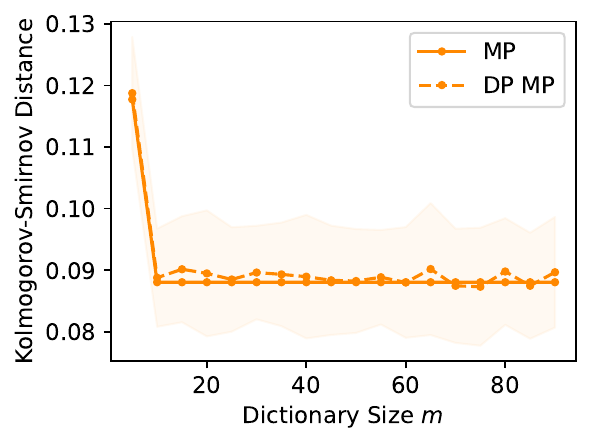}}\hfil
\subfloat[Lognormal$(0, 0.5)$]{\includegraphics[height=2.8cm]{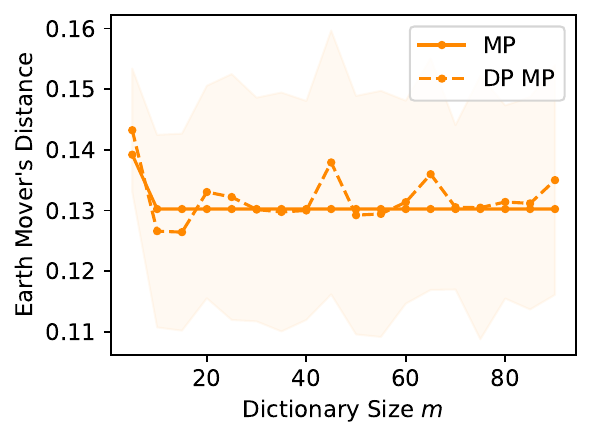}}\hfil
\subfloat[Lognormal$(0, 0.5)$]{\includegraphics[height=2.8cm]{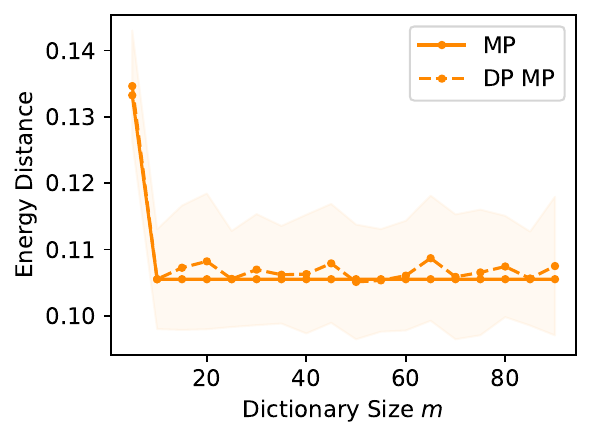}} 

\subfloat[Beta$(10, 2)$]{\includegraphics[height=2.8cm]{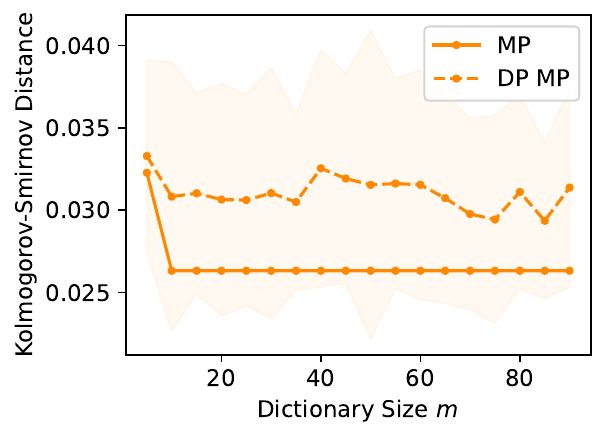}}\hfil
\subfloat[Beta$(10, 2)$]{\includegraphics[height=2.8cm]{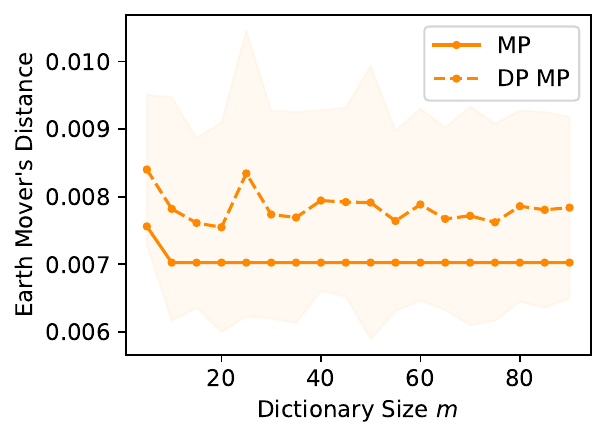}}\hfil
\subfloat[Beta$(10, 2)$]{\includegraphics[height=2.8cm]{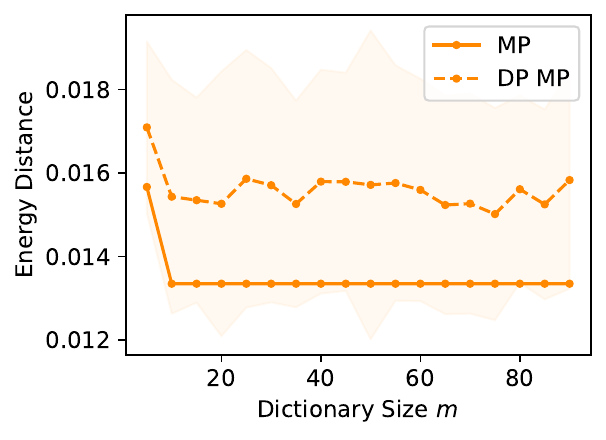}}

\subfloat[Lognormal$(0, 0.5)$]{\includegraphics[height=2.8cm]{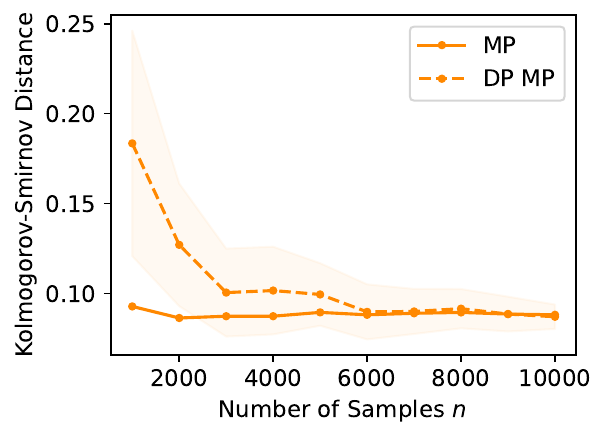}}\hfil
\subfloat[Lognormal$(0, 0.5)$]{\includegraphics[height=2.8cm]{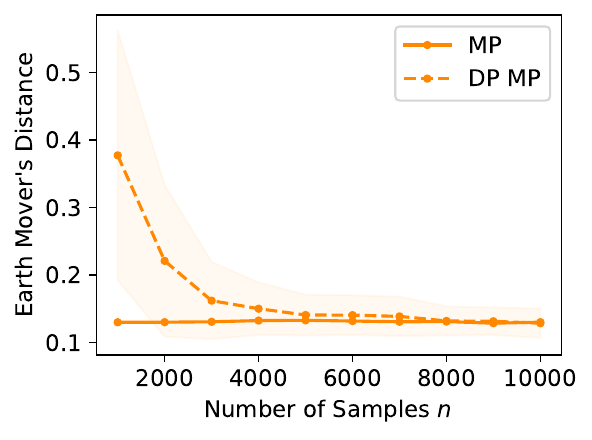}}\hfil
\subfloat[Lognormal$(0, 0.5)$]{\includegraphics[height=2.8cm]{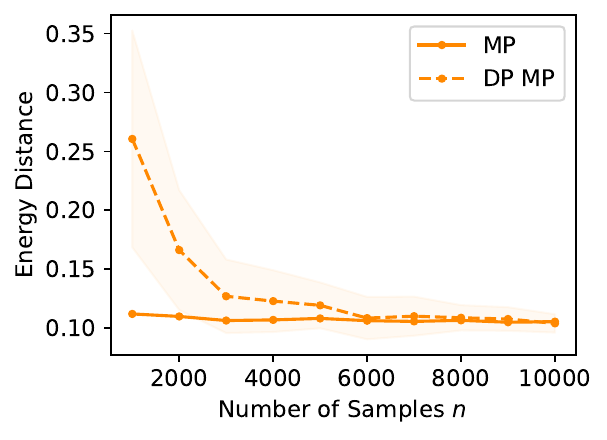}} 

\subfloat[Beta$(10, 2)$]{\includegraphics[height=2.8cm]{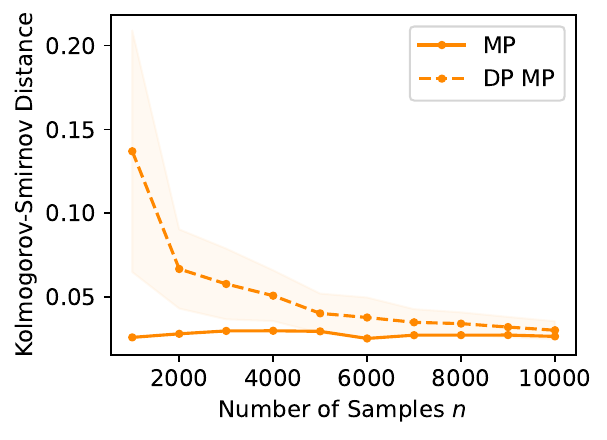}}\hfil
\subfloat[Beta$(10, 2)$]{\includegraphics[height=2.8cm]{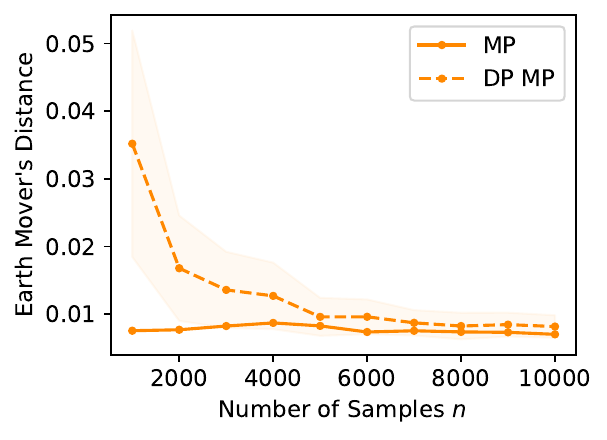}}\hfil
\subfloat[Beta$(10, 2)$]{\includegraphics[height=2.8cm]{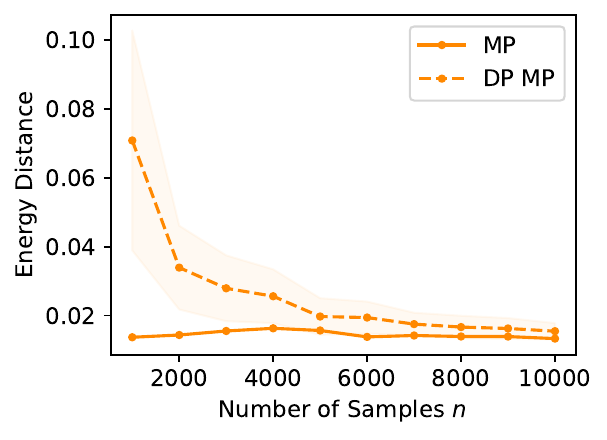}}

\caption{Comparison of distances between \ac{mp}-based approximations and the true CDF under three experimental settings across various distributions. (a)-(f) Effect of the sparsity level $s$ with $n=10^4$ and a dictionary of $m=40$ Legendre polynomials; (g)-(l) Effect of the dictionary size $m$ with $n=10^4$ and $s=6$; (m)-(r) Effect of the sample size $n$ with a dictionary of $m=40$ Legendre polynomials and $s=6$. Experiments were repeated $50$ times with $\epsilon=0.5$ under pure differential privacy.}
\label{fig:appendix_mp_parameters}
\end{figure*}

Figure~\ref{fig:appendix_pp_parameters} shows the effects of $m$ and $n$ on the approximation performance of the \ac{pp} method. Figure~\ref{fig:appendix_mp_parameters} shows the effects of sparsity level $s$, dictionary size $m$, and sample size $n$ on the approximation performance of the \ac{mp} method. These results are consistent with the trends discussed in the main text: in non-private settings, increasing $s$ or $m$ improves approximation accuracy up to the inherent limits of the chosen function space, while in differentially private settings, the gains from larger $m$ or $s$ can be offset by the additional noise required. Increasing the sample size $n$ consistently reduces the approximation error under privacy constraints.

\subsection{Comparison of Methods}
\label{appendix:compare_methods}
\begin{figure*}[t]
\centering
\subfloat[Lognormal$(0, 0.5)$]{\includegraphics[height=2.5cm]{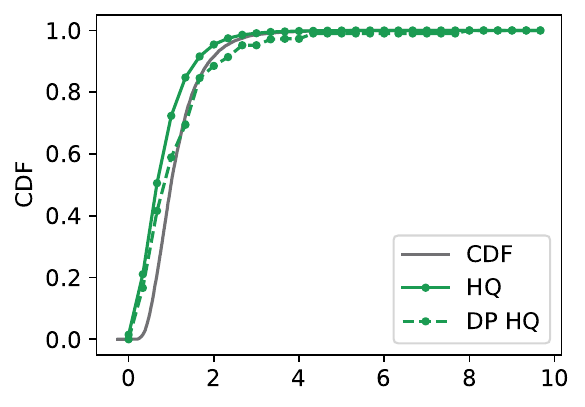}}
\subfloat[Lognormal$(0, 0.5)$]{\includegraphics[height=2.5cm]{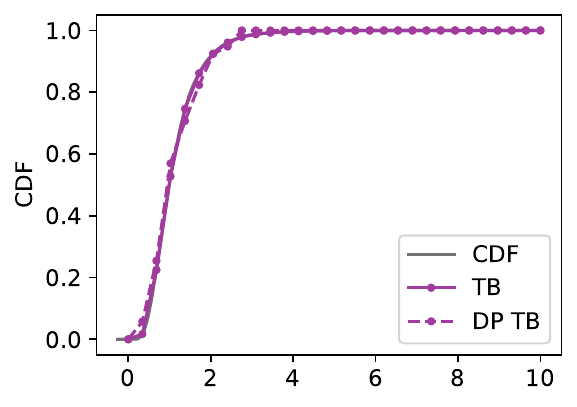}}
\subfloat[Lognormal$(0, 0.5)$]{\includegraphics[height=2.5cm]{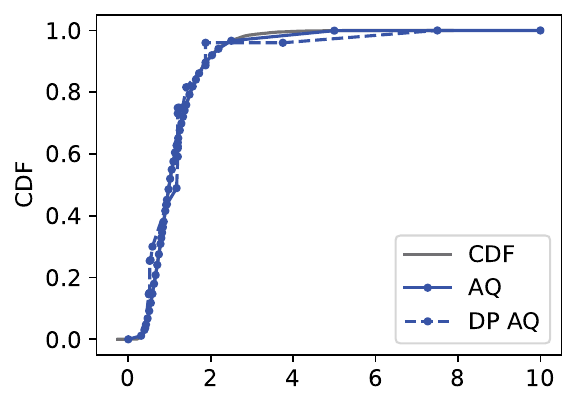}}
\subfloat[Lognormal$(0, 0.5)$]{\includegraphics[height=2.5cm]{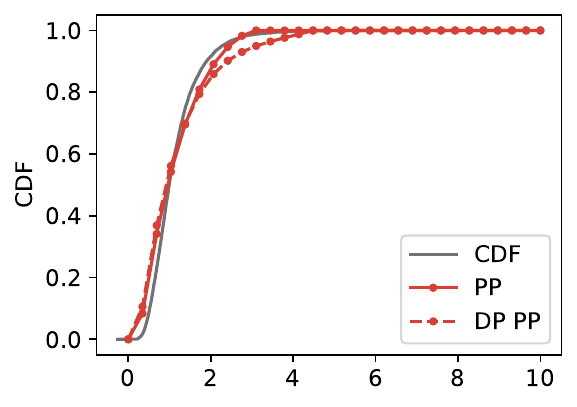}}
\subfloat[Lognormal$(0, 0.5)$]{\includegraphics[height=2.5cm]{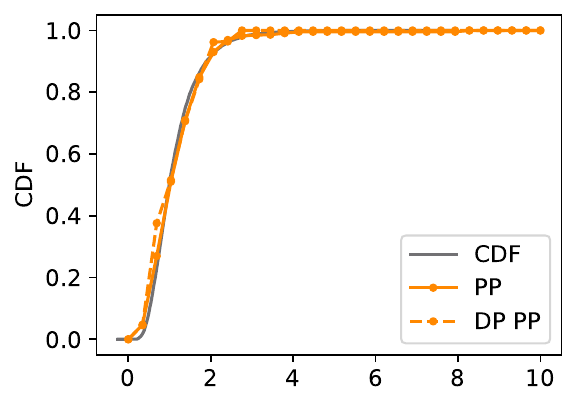}}

\subfloat[Beta$(10, 2)$]{\includegraphics[height=2.5cm]{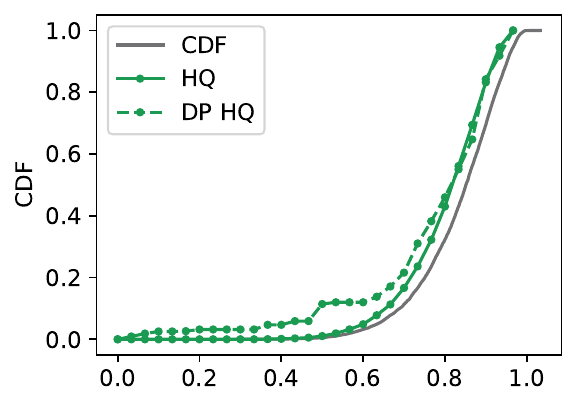}}
\subfloat[Beta$(10, 2)$]{\includegraphics[height=2.5cm]{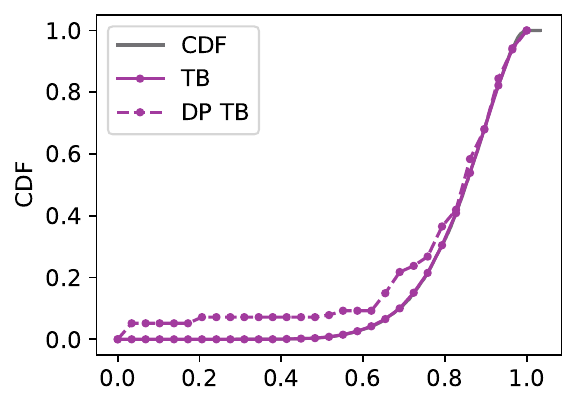}}
\subfloat[Beta$(10, 2)$]{\includegraphics[height=2.5cm]{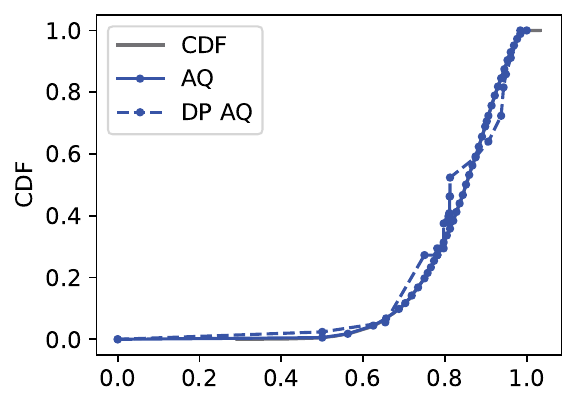}}
\subfloat[Beta$(10, 2)$]{\includegraphics[height=2.5cm]{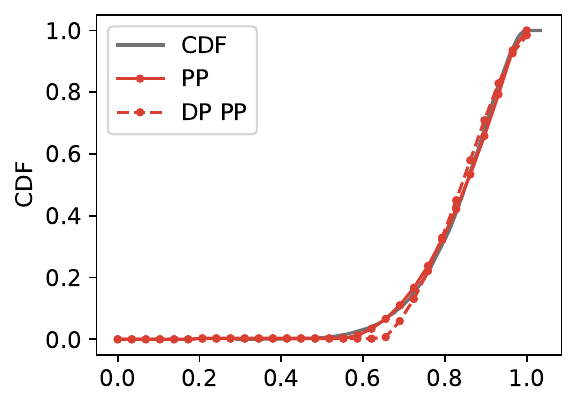}}
\subfloat[Beta$(10, 2)$]{\includegraphics[height=2.5cm]{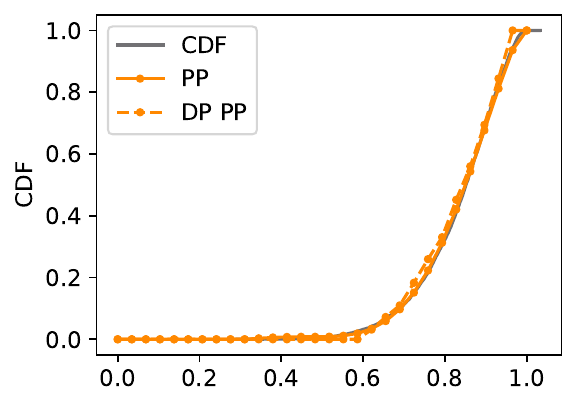}}

\subfloat[Lyft Data]{\includegraphics[height=2.5cm]{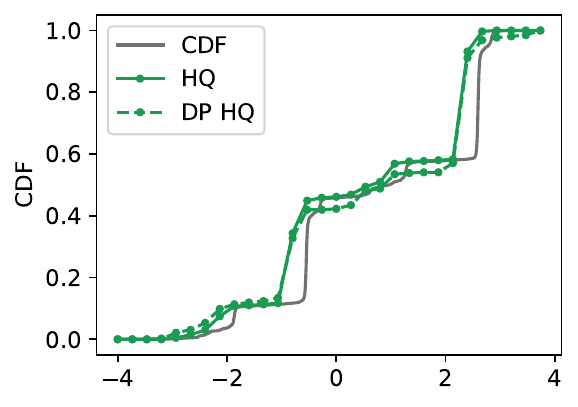}}
\subfloat[Lyft Data]{\includegraphics[height=2.5cm]{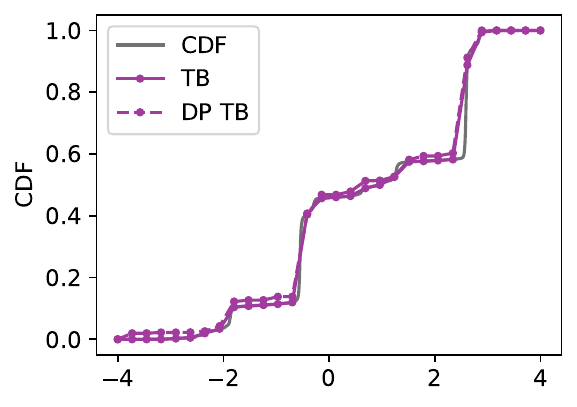}}
\subfloat[Lyft Data]{\includegraphics[height=2.5cm]{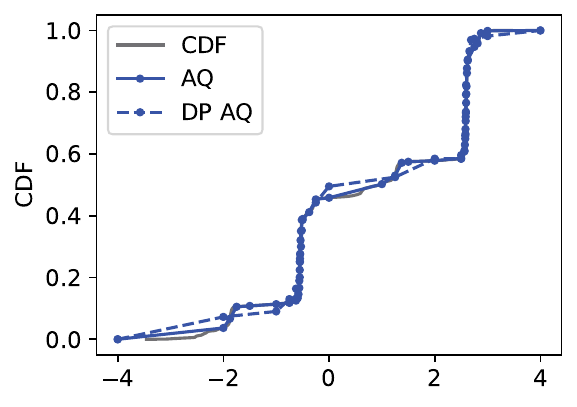}}
\subfloat[Lyft Data]{\includegraphics[height=2.5cm]{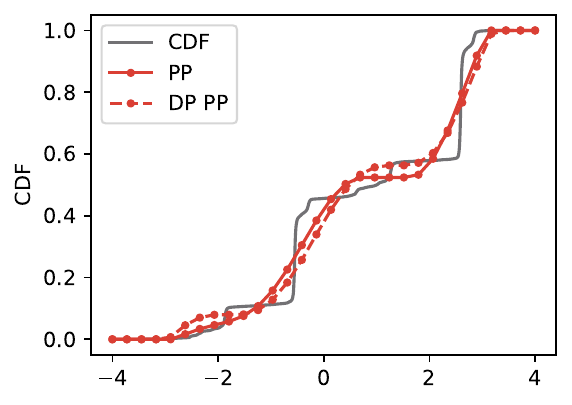}}
\subfloat[Lyft Data]{\includegraphics[height=2.5cm]{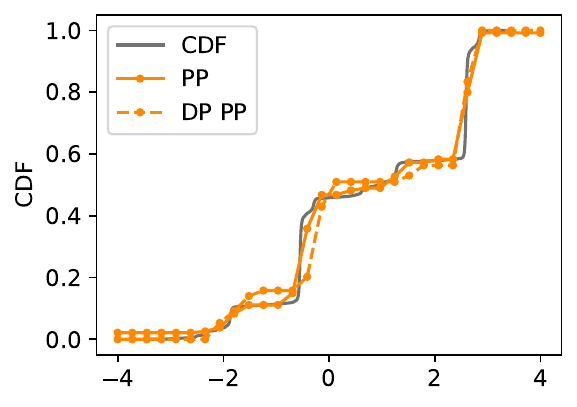}}

\caption{Comparison of \ac{dp} approximation methods across synthetic distributions and a real-world dataset under pure differential privacy with $\epsilon=0.1$. (a)-(e) Lognormal$(0,0.5)$ distribution with $n=10^4$; (f)-(j) Beta$(10,2)$ distribution with $n=10^4$; and (k)-(o) \href{https://www.kaggle.com/code/gaborfodor/eda-3d-object-detection-challenge/notebook}{Lyft 3D Object Detection Data}, where \texttt{yaw} represents the rotation angle of a 3D bounding box around the vertical axis, indicating the orientation of the object in the horizontal plane. The original dataset contains $638179$ records, from which a $5\%$ subsample ($n=31909$) is drawn. For all datasets, \ac{hq} uses $30$ bins, TB uses $1024$ leaves, and \ac{aq} runs for $40$ iterations. For (d)(e)(i)(j), \ac{pp} employs $6$ basis functions and \ac{mp} selects $7$ basis functions from a dictionary of $40$ Legendre atoms. For (n)(o), \ac{pp} employs $10$ basis functions and \ac{mp} selects $20$ basis functions from a dictionary of $40$ Legendre atoms.}
\label{fig:appendix_comparison}
\end{figure*}

\begin{figure*}[ht!]
\centering
\subfloat[Lognormal$(0, 0.5)$]{\includegraphics[height=2.8cm]{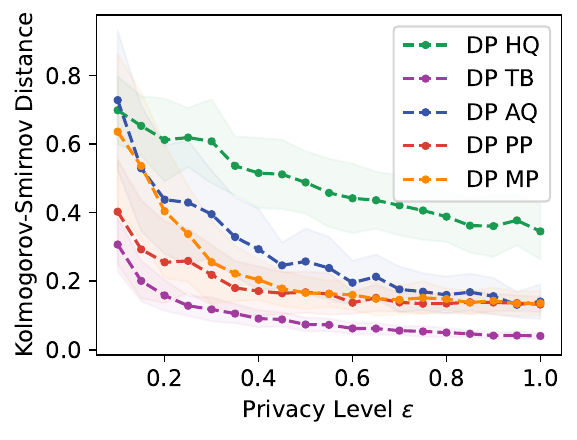}}\hfil
\subfloat[Lognormal$(0, 0.5)$]{\includegraphics[height=2.8cm]{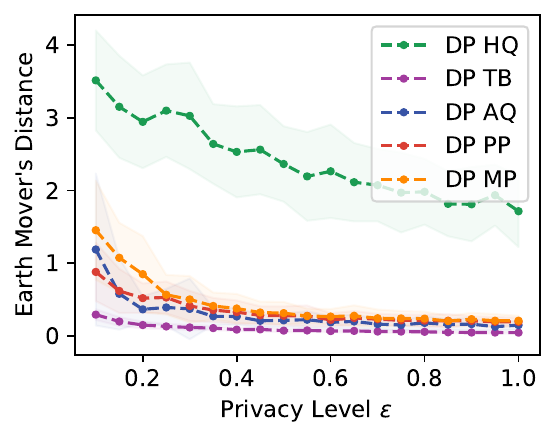}}\hfil
\subfloat[Lognormal$(0, 0.5)$]{\includegraphics[height=2.8cm]{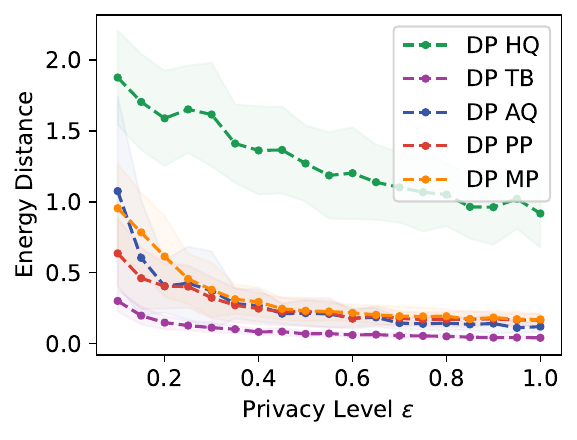}} 

\subfloat[Beta$(10, 2)$]{\includegraphics[height=2.8cm]{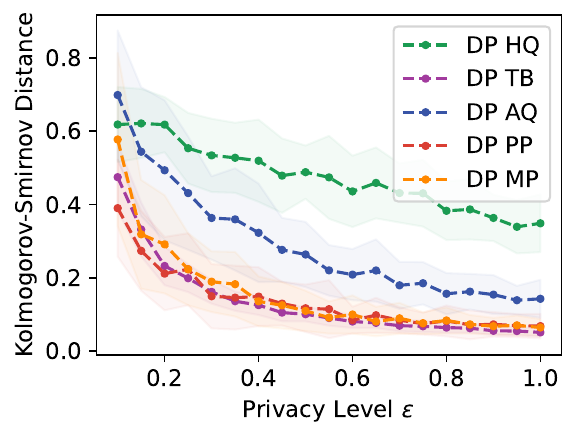}}\hfil
\subfloat[Beta$(10, 2)$]{\includegraphics[height=2.8cm]{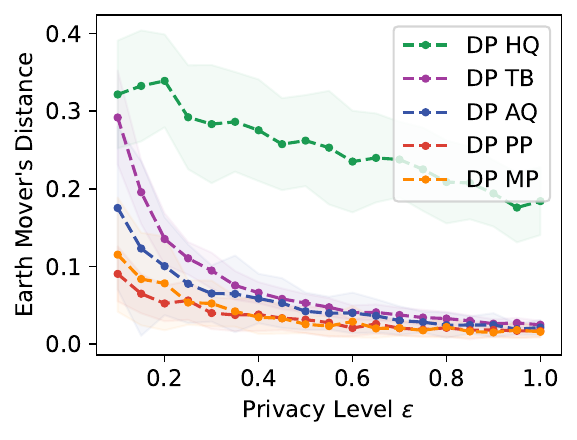}}\hfil
\subfloat[Beta$(10, 2)$]{\includegraphics[height=2.8cm]{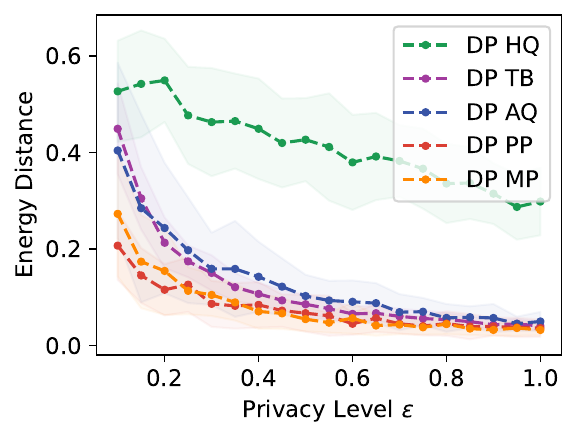}}

\subfloat[Lyft Data]{\includegraphics[height=2.8cm]{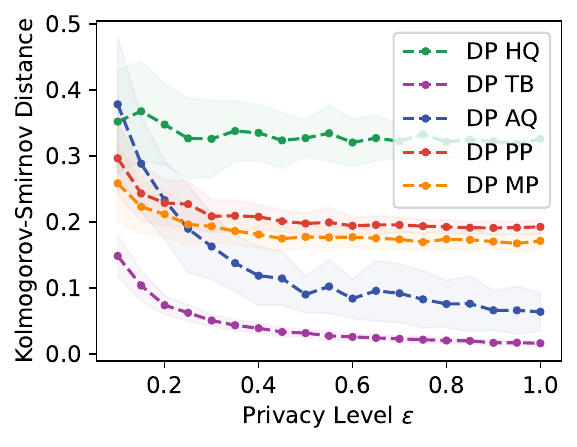}}\hfil
\subfloat[Lyft Data]{\includegraphics[height=2.8cm]{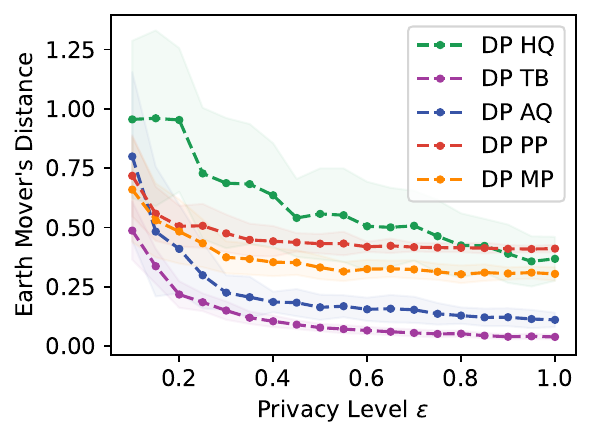}}\hfil
\subfloat[Lyft Data]{\includegraphics[height=2.8cm]{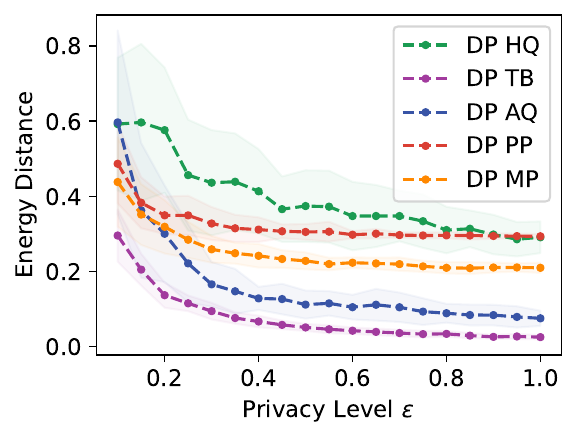}}

\caption{Comparison of distances between different DP CDF methods and the true CDF under three metrics across synthetic distributions and a real-world dataset. (a)-(c) Lognormal$(0,0.5)$ distribution with $n=10^3$; (d)-(f) Beta$(10,2)$ distribution with $n=10^3$; and (g)-(i) \href{https://www.kaggle.com/code/gaborfodor/eda-3d-object-detection-challenge/notebook}{Lyft 3D Object Detection Data}, where a subsample of $n=3191$ is drawn from the original dataset. Each experiment was repeated $50$ times under pure differential privacy, where $\epsilon$ denotes the overall privacy budget. For all datasets, HQ uses $40$ bins, TB uses $1024$ leaves, AQ runs for $40$ iterations, and PP employs $6$ basis functions. For (a)-(f), MP selects $5$ basis functions from a dictionary of $40$ Legendre atoms, while for (g)-(i), MP selects $6$ basis functions from the same dictionary.}
\label{fig:appendix_distance}
\end{figure*}

\begin{figure*}[ht!]
\centering
\subfloat[Lognormal$(0, 0.5)$]{\includegraphics[height=2.8cm]{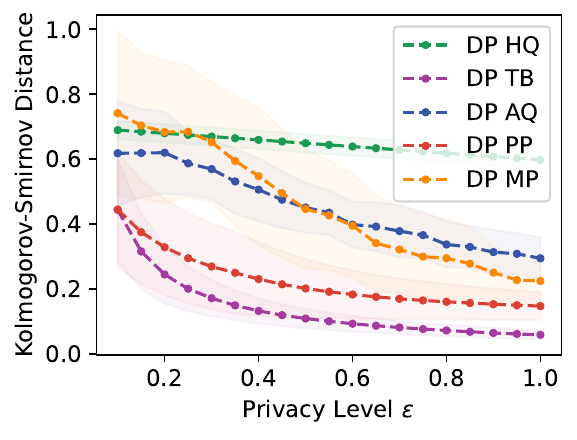}}\hfil
\subfloat[Lognormal$(0, 0.5)$]{\includegraphics[height=2.8cm]{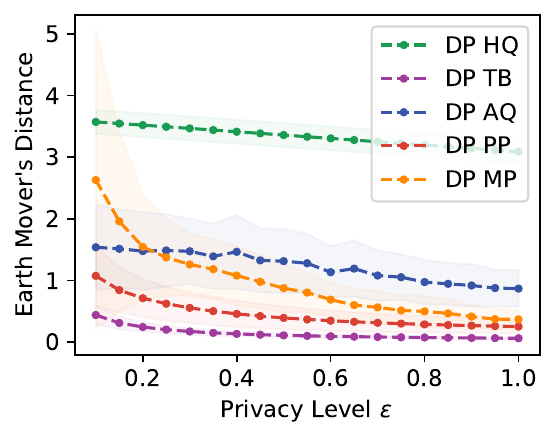}}\hfil
\subfloat[Lognormal$(0, 0.5)$]{\includegraphics[height=2.8cm]{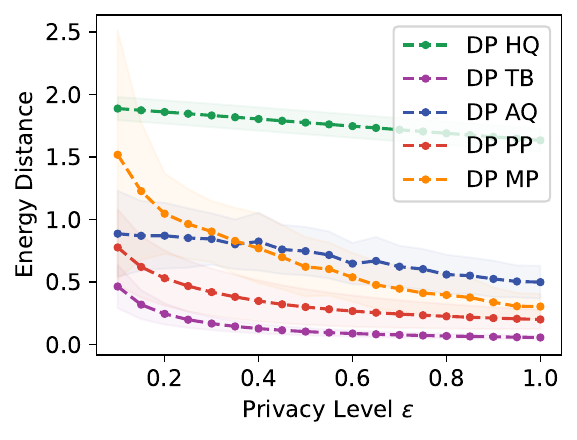}} 

\subfloat[Beta$(10, 2)$]{\includegraphics[height=2.8cm]{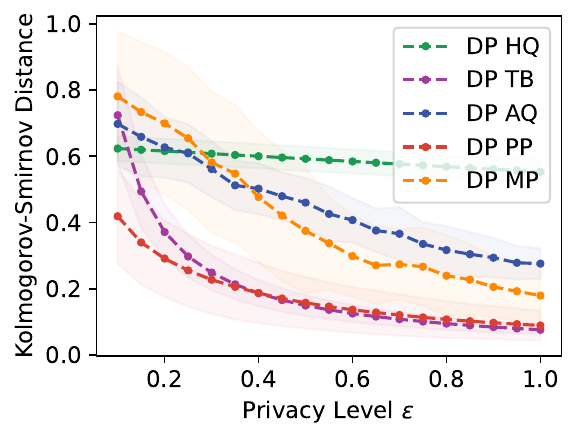}}\hfil
\subfloat[Beta$(10, 2)$]{\includegraphics[height=2.8cm]{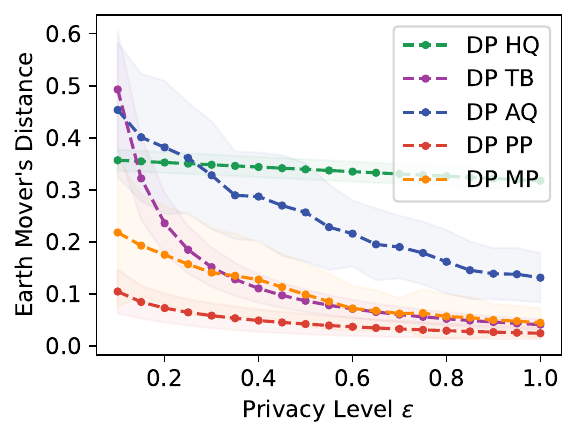}}\hfil
\subfloat[Beta$(10, 2)$]{\includegraphics[height=2.8cm]{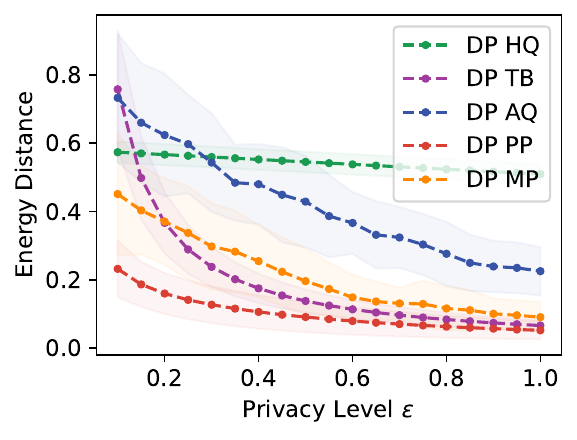}}

\subfloat[Lyft Data]{\includegraphics[height=2.8cm]{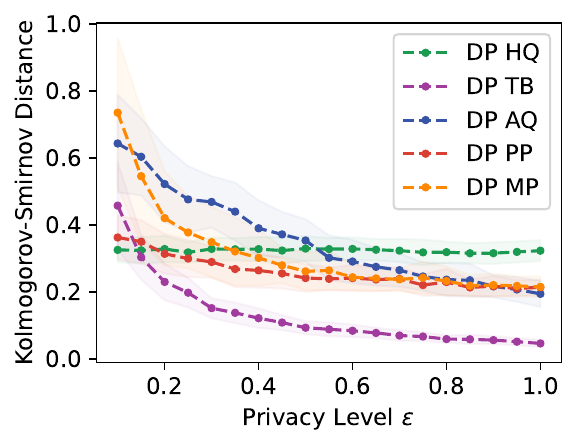}}\hfil
\subfloat[Lyft Data]{\includegraphics[height=2.8cm]{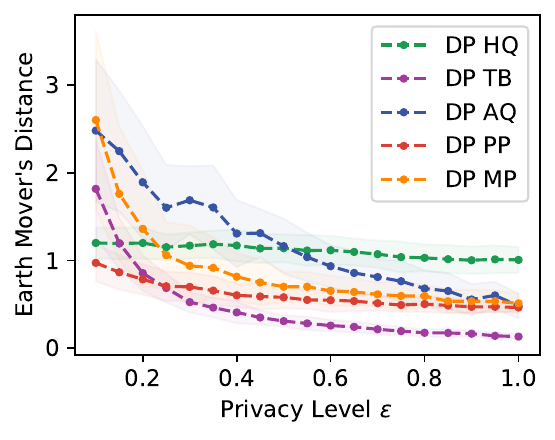}}\hfil
\subfloat[Lyft Data]{\includegraphics[height=2.8cm]{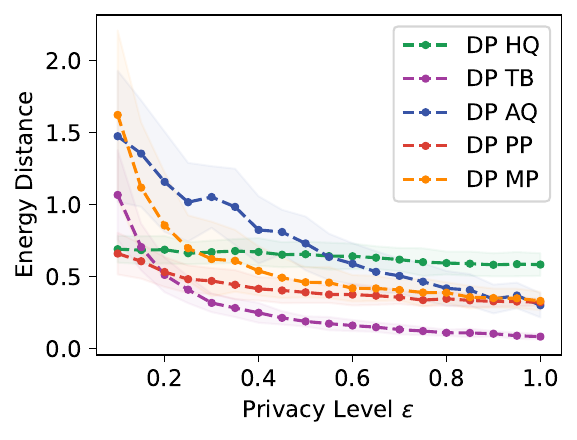}}

\caption{Comparison of distances between different DP CDF methods and the true CDF under three metrics in a decentralized setting with $10$ sites across synthetic distributions and a real-world dataset. (a)-(c) Lognormal$(0,0.5)$ distribution with $n=200$ samples per site; (d)-(f) Beta$(10,2)$ distribution with $n=200$ samples per site; and (g)-(i) \href{https://www.kaggle.com/code/gaborfodor/eda-3d-object-detection-challenge/notebook}{Lyft 3D Object Detection Data} with $n=320$ samples per site. Each experiment was repeated $50$ times under pure differential privacy, where $\epsilon$ denotes the overall privacy budget. For all datasets, HQ uses $40$ bins, TB uses $1024$ leaves, AQ runs for $40$ iterations, and PP employs $6$ basis functions. For (a)-(f), MP selects $5$ basis functions from a dictionary of $40$ Legendre atoms, while for (g)-(i), MP selects $6$ basis functions from the same dictionary.}
\label{fig:appendix_decentralized}
\end{figure*}

\begin{figure*}[ht!]
\centering
\subfloat[Lognormal$(0, 0.5)$]{\includegraphics[height=2.8cm]{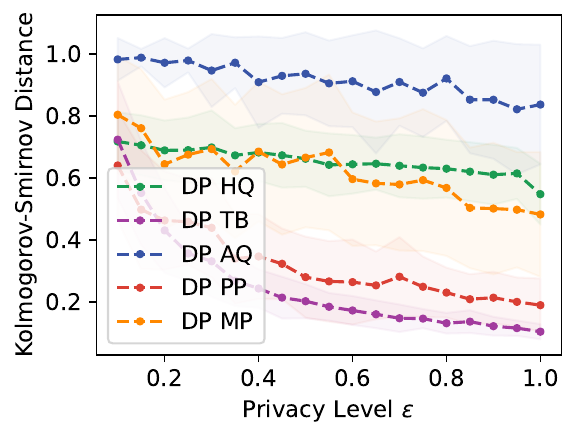}}\hfil
\subfloat[Lognormal$(0, 0.5)$]{\includegraphics[height=2.8cm]{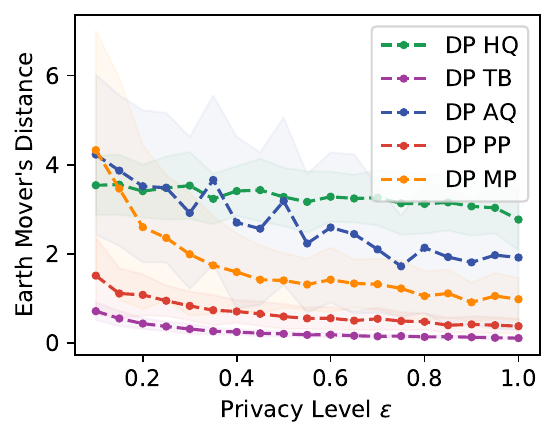}}\hfil
\subfloat[Lognormal$(0, 0.5)$]{\includegraphics[height=2.8cm]{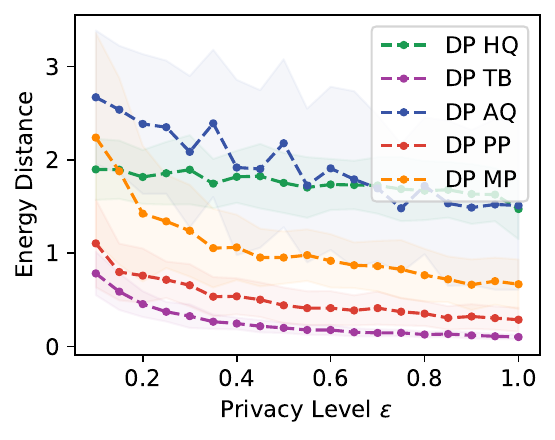}} 

\subfloat[Beta$(10, 2)$]{\includegraphics[height=2.8cm]{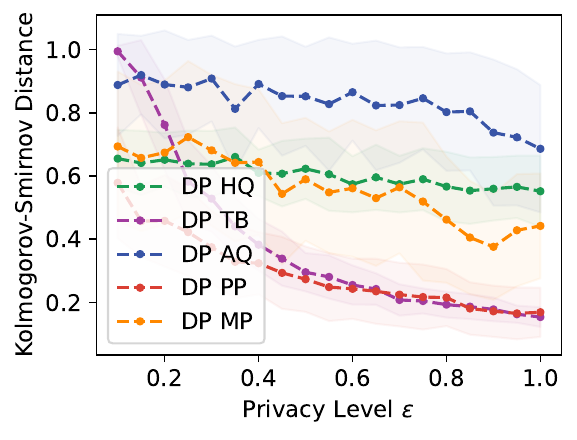}}\hfil
\subfloat[Beta$(10, 2)$]{\includegraphics[height=2.8cm]{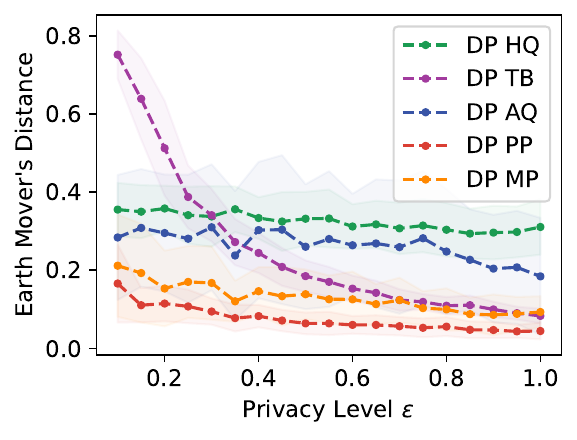}}\hfil
\subfloat[Beta$(10, 2)$]{\includegraphics[height=2.8cm]{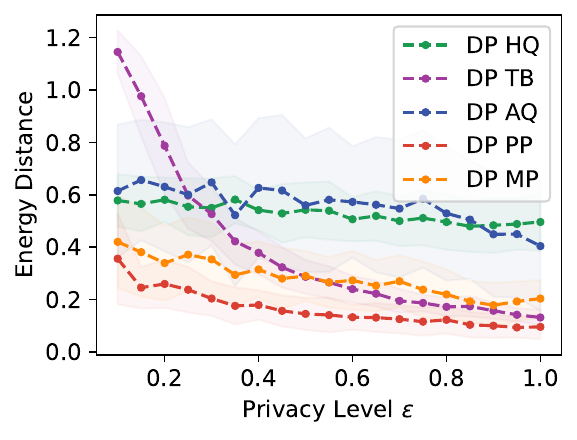}}

\subfloat[Lyft Data]{\includegraphics[height=2.8cm]{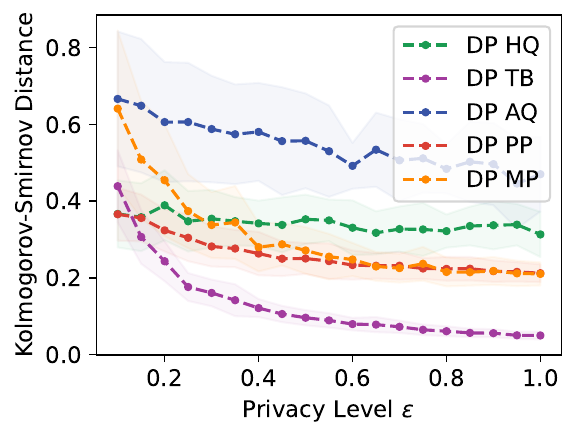}}\hfil
\subfloat[Lyft Data]{\includegraphics[height=2.8cm]{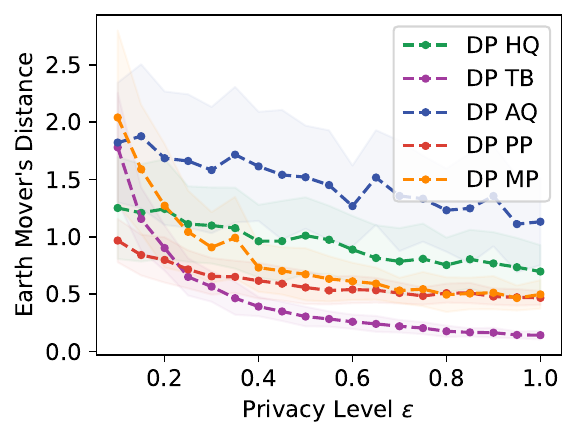}}\hfil
\subfloat[Lyft Data]{\includegraphics[height=2.8cm]{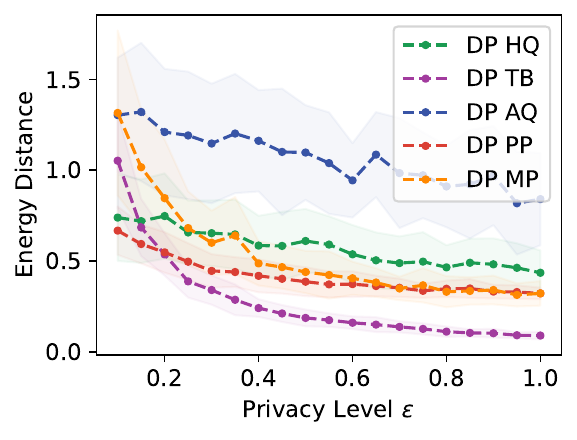}}

\caption{Comparison of distances between different DP CDF methods and the true CDF under three metrics in a newly collected data setting across synthetic distributions and a real-world dataset, where the CDF is updated for a total of $10$ rounds. (a)-(c) Lognormal$(0,0.5)$ distribution with $100$ newly collected samples per round; (d)-(f) Beta$(10,2)$ distribution with $100$ newly collected samples per round; and (g)-(i) \href{https://www.kaggle.com/code/gaborfodor/eda-3d-object-detection-challenge/notebook}{Lyft 3D Object Detection Data} with $320$ newly collected samples per round. Each experiment was repeated $50$ times under pure differential privacy, where $\epsilon$ denotes the overall privacy budget. For all datasets, HQ uses $40$ bins, TB uses $1024$ leaves, AQ runs for $40$ iterations, and PP employs $6$ basis functions. For (a)-(f), MP selects $5$ basis functions from a dictionary of $40$ Legendre atoms, while for (g)-(i), MP selects $6$ basis functions from the same dictionary.}
\label{fig:appendix_new_data}
\end{figure*}

This section provides supplementary experimental results to the main text. We compare differentially private CDF approximation methods on synthetic and real-world datasets, evaluate their accuracy under multiple metrics, and assess their performance in decentralized and dynamic data settings. Figure~\ref{fig:appendix_comparison} shows comparisons across datasets, Figure~\ref{fig:appendix_distance} reports distances to the DP eCDF, Figure~\ref{fig:appendix_decentralized} presents results in decentralized settings, and Figure~\ref{fig:appendix_new_data} illustrates performance with newly collected data.

\subsection{Exploration of Dictionary Compositions}
\label{appendix:dict_compose}

\begin{figure*}[t]
\centering
\subfloat[Lognormal$(0, 0.5)$]{\includegraphics[height=2.5cm]{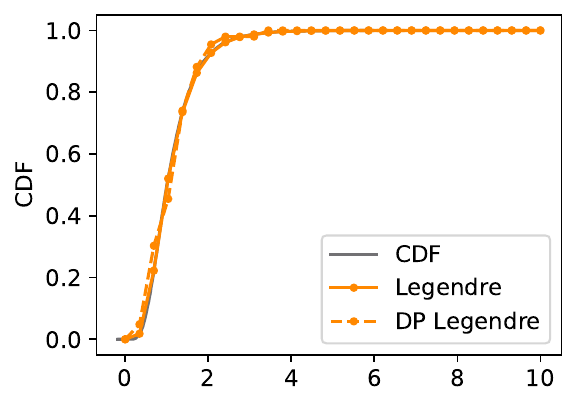}}\hfil
\subfloat[Lognormal$(0, 0.5)$]{\includegraphics[height=2.5cm]{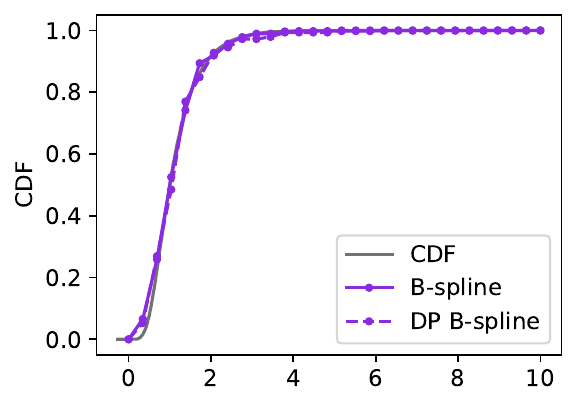}}\hfil
\subfloat[Lognormal$(0, 0.5)$]{\includegraphics[height=2.5cm]{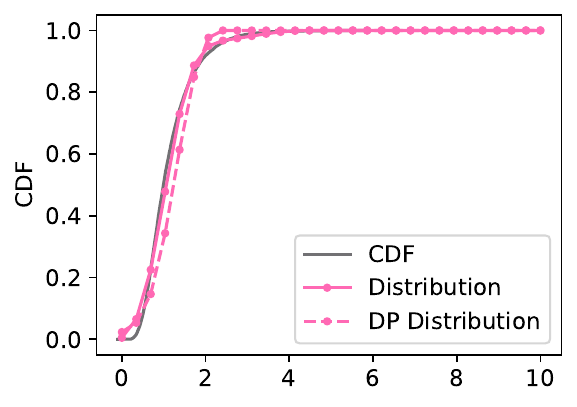}}

\subfloat[Beta$(10, 2)$]{\includegraphics[height=2.5cm]{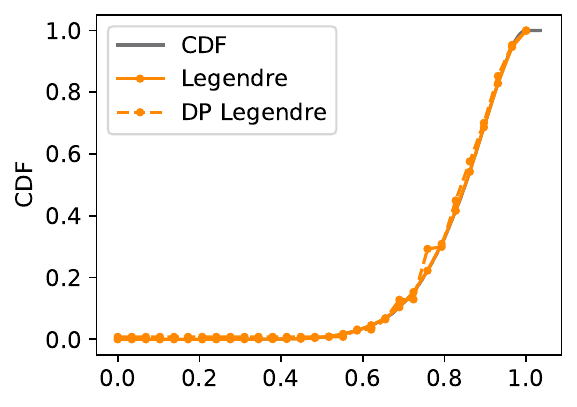}}\hfil
\subfloat[Beta$(10, 2)$]{\includegraphics[height=2.5cm]{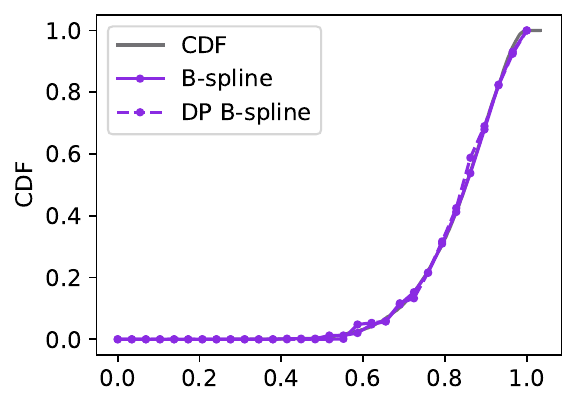}}\hfil
\subfloat[Beta$(10, 2)$]{\includegraphics[height=2.5cm]{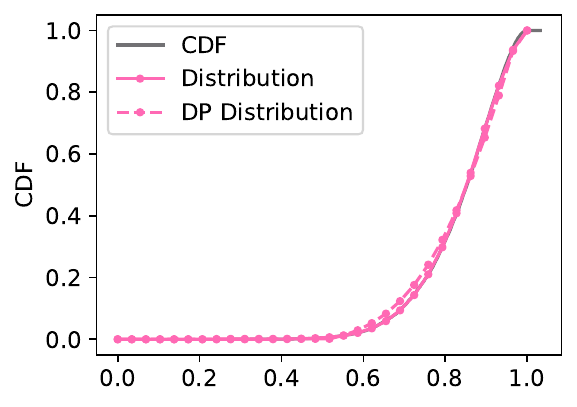}}

\subfloat[Airbnb Data]{\includegraphics[height=2.5cm]{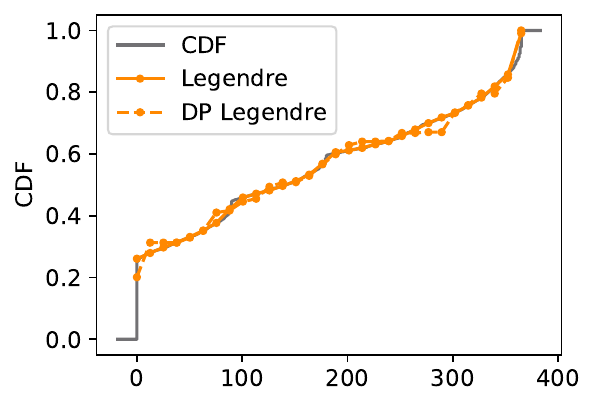}}\hfil
\subfloat[Airbnb Data]{\includegraphics[height=2.5cm]{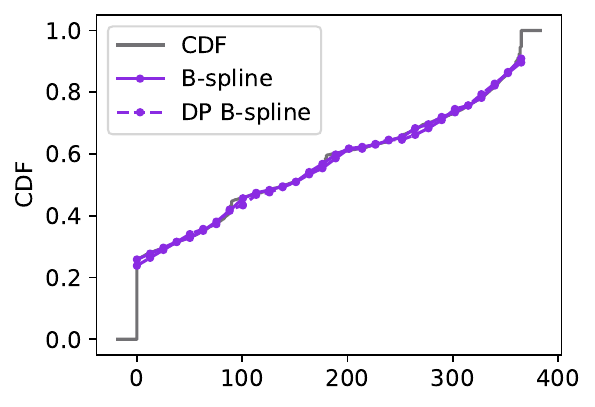}}\hfil
\subfloat[Airbnb Data]{\includegraphics[height=2.5cm]{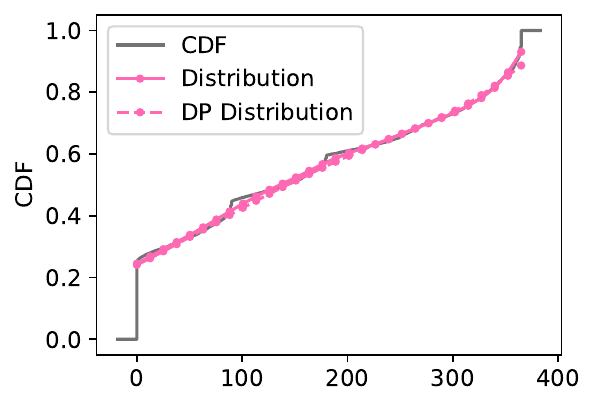}}

\subfloat[Lyft Data]{\includegraphics[height=2.5cm]{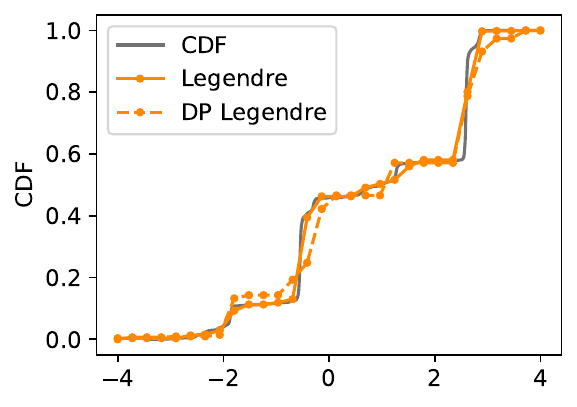}}\hfil
\subfloat[Lyft Data]{\includegraphics[height=2.5cm]{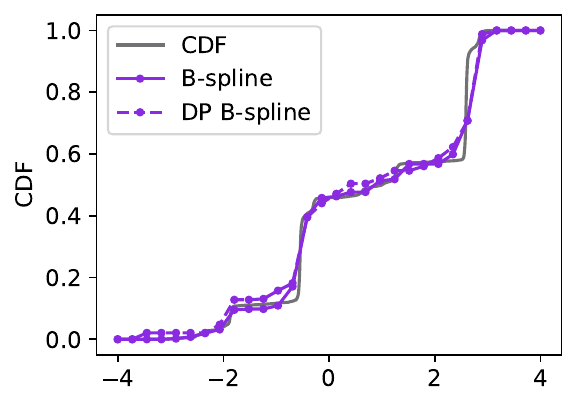}}\hfil
\subfloat[Lyft Data]{\includegraphics[height=2.5cm]{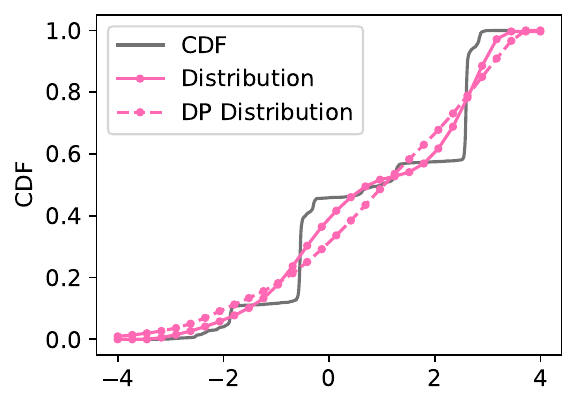}}

\caption{Comparison of CDF reconstruction using different dictionaries across synthetic distributions and real-world datasets. (a)-(c) Lognormal$(0,0.5)$ distribution with $n=10^4$; (d)-(f) Beta$(10,2)$ distribution with $n=10^4$; (g)-(i) U.S. Airbnb Data with $n=22603$; and (j)-(l) Lyft 3D Object Detection Data with $n=31909$. All experiments are conducted with $\epsilon=0.5$, $\delta=n^{-3/2}$, and sparsity level $s=30$. For each dataset, the first column uses a dictionary constructed from $200$ Legendre polynomials, the second column uses a dictionary constructed from $109$ B-spline functions of degrees $0$ and $1$, and the third column uses a dictionary constructed from $400$ normal CDFs with varying means and variances.}
\label{fig:appendix_comparison_dl}
\end{figure*}

\begin{figure*}[ht!]
\centering
\subfloat[Lognormal$(0, 0.5)$]{\includegraphics[height=2.8cm]{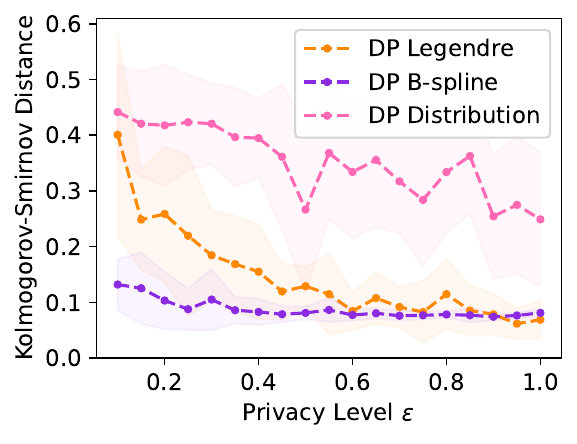}}\hfil
\subfloat[Lognormal$(0, 0.5)$]{\includegraphics[height=2.8cm]{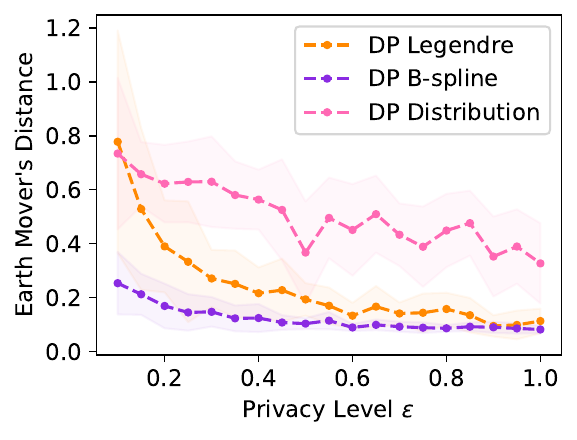}}\hfil
\subfloat[Lognormal$(0, 0.5)$]{\includegraphics[height=2.8cm]{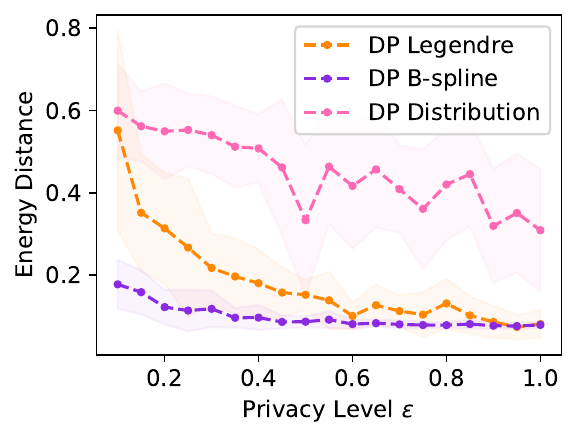}} 

\subfloat[Beta$(10, 2)$]{\includegraphics[height=2.8cm]{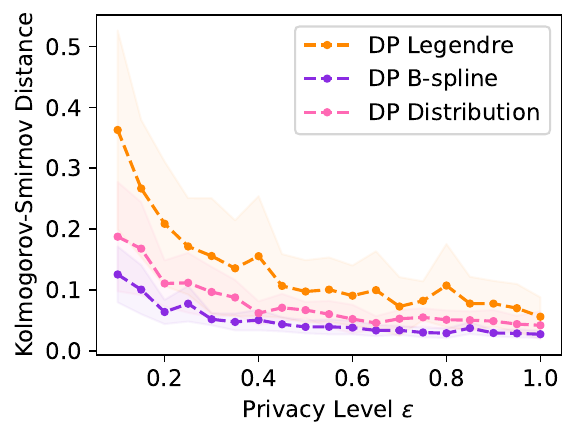}}\hfil
\subfloat[Beta$(10, 2)$]{\includegraphics[height=2.8cm]{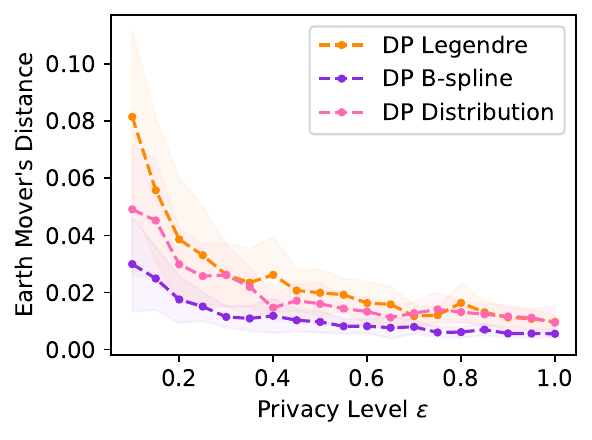}}\hfil
\subfloat[Beta$(10, 2)$]{\includegraphics[height=2.8cm]{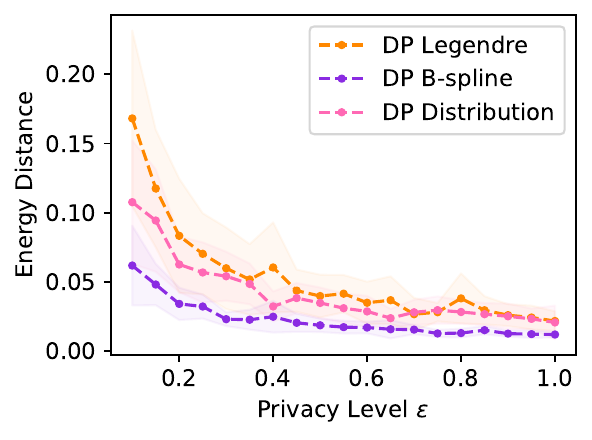}}

\subfloat[Airbnb Data]{\includegraphics[height=2.8cm]{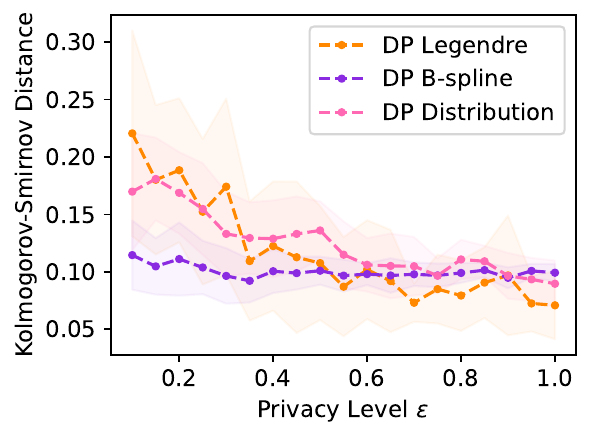}}\hfil
\subfloat[Airbnb Data]{\includegraphics[height=2.8cm]{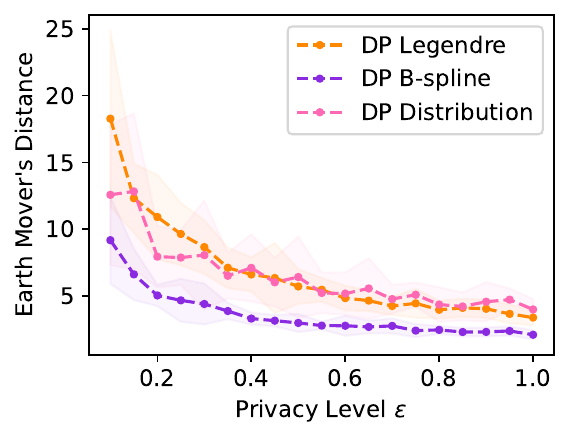}}\hfil
\subfloat[Airbnb Data]{\includegraphics[height=2.8cm]{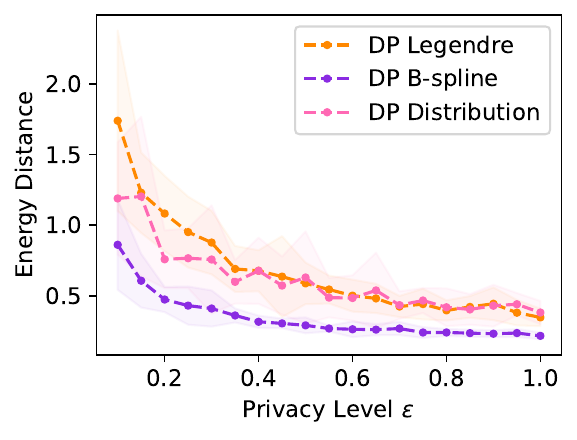}} 

\subfloat[Lyft Data]{\includegraphics[height=2.8cm]{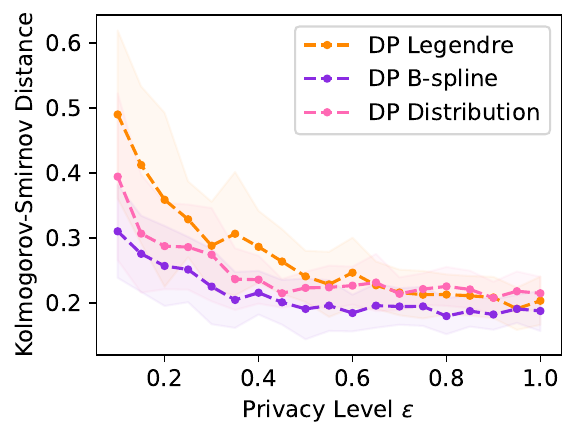}}\hfil
\subfloat[Lyft Data]{\includegraphics[height=2.8cm]{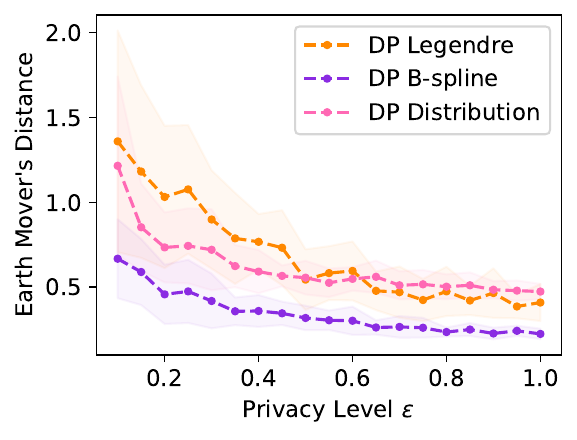}}\hfil
\subfloat[Lyft Data]{\includegraphics[height=2.8cm]{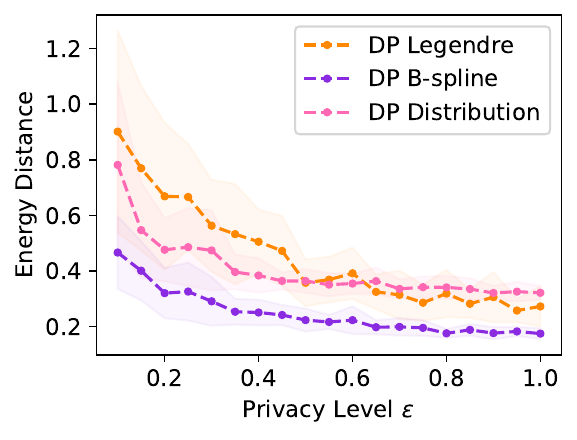}}

\caption{Comparison of the approximation distances between the DP CDFs obtained using different dictionaries and the true CDF across synthetic distributions and real-world datasets. (a)-(c) Lognormal$(0,0.5)$ distribution with $n=10^4$; (d)-(f) Beta$(10,2)$ distribution with $n=10^4$; (g)-(i) U.S. Airbnb Data with $n=22603$; and (j)-(l) Lyft 3D Object Detection Data with $n=31909$. Each experiment was repeated $50$ times with $\delta=n^{-3/2}$ and sparsity level $s=30$.}
\label{fig:appendix_distance_dl}
\end{figure*}

Figure~\ref{fig:appendix_comparison_dl} presents examples of CDF reconstruction, while Figure~\ref{fig:appendix_distance_dl} illustrates the approximation performance using the three different dictionaries. The results are consistent with the analyses presented in the main text.

\end{document}